%% file: main.tex
\pdfoutput=1
\documentclass[manuscript,screen]{acmart}

\AtBeginDocument{%
  \providecommand\BibTeX{{%
    \normalfont B\kern-0.5em{\scshape i\kern-0.25em b}\kern-0.8em\TeX}}}

\setcopyright{acmlicensed}
\copyrightyear{2018}
\acmYear{2018}
\acmDOI{XXXXXXX.XXXXXXX}

\acmISBN{978-1-4503-XXXX-X/18/06}

\usepackage{arydshln}
\usepackage{multirow}
\usepackage{nicematrix}

\usepackage{xcolor,colortbl}
\usepackage{tikz,pgfplots,pgfplotstable}

\pgfplotsset{compat=1.11}

\usepackage{amsmath}

\usepackage{blindtext}
\usepackage{tcolorbox}
\usepackage{graphicx}
\usepackage{enumitem}
\usepackage{booktabs}
\usepackage{pgf-pie}
\usepackage{subfig}

\usepackage{soul}
\newcommand{\hlc}[2][yellow]{{%
    \colorlet{foo}{#1}%
    \sethlcolor{foo}\hl{#2}}%
}

\begin{document}

\title[Diversity's Double-Edged Sword]{Diversity's Double-Edged Sword: Analyzing Race's Effect on Remote Pair Programming Interactions}

\author{Shandler A. Mason}
\email{samason4@ncsu.edu}
\orcid{0000-0003-0587-4527}
\affiliation{%
  \institution{North Carolina State University}
  \streetaddress{1210 Varsity Dr.}
  \city{Raleigh}
  \state{North Carolina}
  \country{USA}
  \postcode{27606}
}
\author{Sandeep Kaur Kuttal}
\email{skuttal@ncsu.edu}
\orcid{0000-0003-0172-7087}
\affiliation{%
  \institution{North Carolina State University}
  \streetaddress{1210 Varsity Dr.}
  \city{Raleigh}
  \state{North Carolina}
  \country{USA}
  \postcode{27606}
}
\renewcommand{\shortauthors}{S. Mason et al.}

\begin{abstract}
  \input{abstract}

\end{abstract}




\received{29 February 2024}
\received[revised]{X X 2023}
\received[accepted]{X X 2023}

\maketitle

\input{introduction}

\input{background}

\input{methodology}

\input{limitations}

\input{results}

\input{discussion}

\input{conclusion}

\begin{acks}
This material is based upon work supported by the Air Force Office of Scientific Research under award number FA9550-21-1-0108 and National Science Foundation under award numbers IIS-2313890 and CCF-2006977. Any opinions, findings, and conclusions or recommendations expressed in this material are those of the authors and do not necessarily reflect the view of the NSF and AFOSR. Finally, we would like to thank Raphael Phillips for his work on labeling the data.

\end{acks}

\bibliographystyle{ACM-Reference-Format}
\bibliography{bibliography}

\end{document}

%% file: abstract.tex
Remote pair programming is widely used in software development, but no research has examined how race affects these interactions. We embarked on this study due to the historical under representation of Black developers in the tech industry, with White developers comprising the majority. Our study involved 24 experienced developers, forming 12 gender-balanced same- and mixed-race pairs. Pairs collaborated on a programming task using the think-aloud method, followed by individual retrospective interviews. Our findings revealed elevated productivity scores for mixed-race pairs, with no differences in code quality between same- and mixed-race pairs. Mixed-race pairs excelled in task distribution, shared decision-making, and role-exchange but encountered communication challenges, discomfort, and anxiety, shedding light on the complexity of diversity dynamics. Our study emphasizes race's impact on remote pair programming and underscores the need for diverse tools and methods to address racial disparities for collaboration.

%% file: introduction.tex
\section{Introduction}

Efficient collaboration within software development teams is paramount, and pair programming stands out as a widely adopted agile practice. In pair programming, two developers collaborate closely, with one actively writing code (the 'driver') and the other reviewing it (the 'navigator') \cite{Dyba2007, Williams2000a, Mcdowell2003}. The two change roles throughout the task. This approach, whether in-person or remote, has a proven history of enhancing productivity, code quality, and self-efficacy \cite{Nagappan2003, Han2010}. However, achieving synergy in pair programming can be especially challenging when working with diverse teams \cite{joshi2009role}. Lack of diversity, in the software industry, can result in the unintended exclusion of certain demographic groups from software products.

Our research is both formative and critical, with a focus on the impact of race\footnote{We acknowledge that developers' identities are shaped by a myriad of factors, including race, ethnicity, gender, socioeconomic status, sexual orientation, age, education, experience, geographic location, and religion. However, we have deliberately narrowed the scope to focus solely on two racial categories.} on pair programming interactions among developers. We chose to concentrate on race, in the United States (US), due to the historical marginalization of Black developers in the tech industry, where White developers predominate \cite{Gitnux_2023}. Furthermore, interracial interactions between these two racial groups, in the US, are complex due to a long history of racial injustices~\cite{pinel1999stigma, shelton2003interpersonal}, including disparities in wealth, education, and incarceration rates ~\cite{dreamsdef, college, prison2017, Gramlich_2019}. These disparities have deep historical roots, dating back to slavery and continuing through discriminatory practices like redlining and Jim Crow (racial segregation) laws ~\cite{Times_2019, boccard1999racial, filer1991voting, king2006jim}. Achieving systematic equality has been a challenge, and even today, a study conducted at Google, which compared pushback from code reviews across various factors such as race, age, and gender, found that White young men face less pushback from peers compared to minorities \cite{murphyhill2022pushback}.

To the best of our knowledge, no prior research has explored the influence of racial dynamics on remote pair programming interactions among professional developers.

Pair programming involves ongoing interaction between partners, fostering collaboration to accomplish tasks and reach decisions. Research indicates that the human brain is predisposed to stereotypes \cite{Spiers2017_Prejudice}, and stereotypes related to race can influence the dynamic interactions between software developers. To investigate the impact of race in both same- and mixed-race pairs, we formulated four research questions:

\begin{itemize}
    \item \textbf{RQ1: How does race affect pair programming dynamics in both same- and mixed-race pairs?}
    Pair programming offers various technical benefits, including increased productivity, improved code quality, and higher self-efficacy \cite{Nagappan2003, Han2010}. Our focus is to examine how race, whether in same- or mixed-race pairs, impacts these technical aspects of pair programming.

    \item \textbf{ RQ2: How does the creativity styles of participants in same-race pairs differ from those in mixed-race pairs during pair programming?}

    Creativity is crucial for individual success \cite{wagner2012, zhao2012, Edelman2010} and for addressing complex, open-ended problems \cite{Brown2009, Liu2004}.  Past research has found that pair programming enhances creativity, potentially leading to superior solutions \cite{Tanja2008,Seo2016,begel2008,Howard2009}. We aimed to investigate if the creative problem-solving process differs between same- and mixed-race pairs.

   \item \textbf{RQ3: How does race influence collaboration dynamics among same- and mixed-race pairs during pair programming?}

   Pair programming decisions can be influenced by human biases, like other decision-making processes, impacting feedback based on partners' demographic perceptions, whether consciously or subconsciously. We investigated leadership style and role-exchange between pairs as research has found differences in these dynamics. Leadership styles, which offer insights into decision-making dynamics, have demonstrated susceptibility to race-related factors in management science literature (e.g., \cite{Gundemir2014, Leadership2009}).  Similarly, gender research has found an impact of gender dynamics on role-exchange within pairs \cite{Kuttal2019}. We aimed to examine whether similar relationships exist concerning racial dynamics.

\item \textbf{RQ4: How does individuals' awareness of their partners' racial backgrounds influence their attitudes and interaction dynamics within same- and mixed-race pairs?}

Race and ethnicity frequently exert a significant influence on individuals and communities, whether through explicit identification or implicit associations \cite{yanow}. 
Research studies have revealed the remarkable speed with which the human brain forms perceptions of other people \cite{Schyns2009} and detects confidence \cite{JIANG20159}. This perception whether positive or negative can affect the racial awareness of a developer. Therefore, we examined individuals' attitude and perceptions during interactions with individuals from same- or mixed-racial backgrounds.

\end{itemize}

To answer these research questions, we conducted think-aloud lab studies and retrospective interviews. The lab studies investigated real-time interactions between 24 Black and White software professionals (gender-balanced), observing their collaborative efforts in same- and mixed-race pairs as they completed a programming task using the think-aloud method. Additionally, we conducted retrospective interviews to capture their experiences and perceptions on working with individuals from same- or mixed-racial backgrounds.

%% file: background.tex
\section{Background and Related Work} 

\subsection{Pair Programming}
Remote or in-person pair programming is a collaborative software development approach involving two developers: a driver and a navigator. The driver actively writes code, controls the keyboard, and oversees task implementation. The navigator reviews the code for errors and improvements, develops strategies, and evaluates decisions \cite{Chong2007}. Pairs can switch roles frequently \cite{Williams2002a, Williams2000a}. This agile methodology is widely recognized in both professional and academic settings \cite{Sommerville10, gregory2021agile}. 

Remote pair programming, also known as distributed pair programming, has been shown to be on par with in-person pair programming \cite{Tsompanoudi2019, Kuttal2019, hanks2005student, hughes2020remote}. Remote pair programming leverages collaborative software, aligning with the global software development industry and the prevalence of remote work due to COVID-19, effectively engaging diverse groups of developers across various geographical locations \cite{aagren2022agile}.

\subsubsection{Benefits of pair programming.}
Pair programming offers increased productivity, knowledge transfer, discussions, and delivers higher-quality code, benefiting both industry professionals and students \cite{sun2015effectiveness}. 
Collaboration in pair programming, backed by research \cite{Zieris2020, Dyba2007, Jones2013, Nagappan2003}, surpasses individual abilities, demonstrating effectiveness in motivation, education, and team communication compared to solo programming \cite{Mcdowell2002, Oviatt2000, Celepkolu2018, Williams2002b, Nosek1998, McDowell2006, Han2010}. This collaborative approach, marked by role-exchange and active interaction, enhances students' understanding of complex concepts and stimulates innovative idea generation \cite{Werner2004}. For professionals, pair programming facilitates project learning and boosts productivity \cite{Zieris2020}. It enhances developers' self-efficacy, making them more adaptable, resilient, and content when tackling programming challenges \cite{Nosek1998}. Developers rate increased creativity as a top benefit of pair programming \cite{begel2008}, and a pair's collective creativity and experience should produce higher quality code compared to an individual \cite{Tanja2008}.

\subsubsection{Challenges in fostering pair programming} Pair programming poses challenges, including the potential to stifle individual exploration \cite{voss2011hippocampal, deci1971effects, disessa2000changing}, steer pairs off tasks \cite{lemov2010teach}, and generate friction between pairs \cite{wray2009pair, bevan2002guidelines, howard2006attitudes, simon2008first}. It introduces the risk of social loafing \cite{balijepally2009two}, pair fatigue, and pair pressure \cite{wray2009pair}, negatively impacting productivity and decision-making. Individuals, particularly children, may face challenges in engaging effectively in collaborative conversations \cite{lemov2010teach}.
Several challenges persist between the driver and navigator, including difficulties for the navigator in keeping pace with the driver's actions and struggling to make effective contributions \cite{plonka2012disengagement}, while the driver is more prone to interruptions \cite{Jones2013}. Differing levels of expertise \cite{Williams2002a, vandegrift2004coupling} and non-compliance with assigned driver/navigator roles \cite{bryant2004double} pose difficulties for pairs.

\subsubsection{Impact of sociodemographic characteristics in pair programming.} Recent studies have revealed insights into communication, collaboration, and coordination challenges among both same- and mixed-gender pairs while pair programming \cite{Kuttal2019, Lott2021}. Pair programming has proven beneficial for female students \cite{werner2004pair} by reducing their levels of frustration \cite{braught2011case}. It plays a role in closing the gender gap in computer science, motivating women to pursue careers in the field \cite{Werner2004, Ruvalcaba2016}. Reports suggest that mixed-gender pairs may face compatibility challenges \cite{katira2005towards}, and that there are observed differences in communication dynamics between same-gender, man and woman pairs \cite{choi2014, aries1987gender}. 

Pair programming research has explored how diverse personal experiences and education levels among students, including varying experience with pair programming and test-driven development, contributes to less comprehension \cite{bowman2019prior}, lower participation \cite{lui2006pair}, and a decreased interest in computer science \cite{thomas2003code, chaparro2005factors}.

Computer science education research has delved into pair programming, spanning K-12 \cite{zhong2016impact, bodaker2023online, denner2014pair, werner2009pair, galdo2022pair} and undergraduate \cite{bowman2021impact, cliburn2003experiences} students. Irrespective of age groups, pair programming heightened students' confidence and enjoyment levels. The nuances of age biases in the pair programming context remain unexplored.

Computer science research has investigated how developers from different geographical location can change the efficiency of teams during collaboration \cite{chen2020incorporating, bass2007collaboration}.

To the best of our knowledge, there is no existing study that has thoroughly explored the influence of participants' racial backgrounds in the context of pair programming. While a few studies have mentioned the diversity of their participant pool \cite{nagappan2003improving, williams2002support, katira2005towards}, only one study has investigated pair programming dynamics involving Latino and White (students) participants within a middle school classroom setting \cite{ruvalcaba2016observations}. 

\subsection{Race in the United States}

Race is a socially constructed concept with multiple established definitions, making it challenging to pinpoint a single, universally accepted meaning~\cite{m2013beyond, delgado2023critical, matthew19s, solorzano2002critical, wellman1993portraits}. In the context of our study, we define race as a multifaceted concept, as defined by \cite{roth2016multiple, bowker2000sorting}, encompassing a participant's self-perception, personal beliefs, external perceptions, survey responses, and physical characteristics. In contemporary United States (US) society, racial categories like ``Black or African American'' and ``White'' are fluid and widely accepted terminology \cite{Jensen_2022}.

\subsubsection{History of Racial Inequalities in the US}
Interracial interactions in the US are complex due to a deep-seated history of racial injustices, particularly between Black and White individuals~\cite{pinel1999stigma, shelton2003interpersonal}. These injustices are exemplified by stark wealth disparities, with the average wealth of Black families being just \$3,600 compared to \$147,000 for White families~\cite{dreamsdef}, lower college graduation rates among Black (22.5\%) and White (36\%) individuals aged 25 and older~\cite{college}, and a six-fold higher likelihood of Black individuals being incarcerated compared to White individuals~\cite{prison2017, Gramlich_2019}. These inequities have significantly exacerbated tensions between racial groups and are deeply embedded in US institutions. 

Slavery, which began in the US in 1619~\cite{Times_2019}, was followed by a history of oppressive measures, including the classification of slaves as three-fifths of a person in the 1788 constitution~\cite{ConstitutionalRights}. Slavery was officially abolished in 1865 with the 13th Amendment~\cite{bogen2023racial}, but subsequent practices like financial redlining~\cite{boccard1999racial}, restrictive voting laws~\cite{filer1991voting}, and Jim Crow legislation~\cite{king2006jim} continued to institutionalize racism. Despite legal equality being established by 1970, it has not been fully accepted in society as evident from the ``Black Lives Matter'' movement~\cite{lebron2023making}.

\subsubsection{Race Affects on Collaboration and the Workplace} Researchers have studied race in collaboration and workplaces in the domain of management, psychology and sociology \cite{edwards2012role, page2007making, mcleod1996ethnic, sommers2006racial, de2003task, moreland2018creating}. Black individuals experience decreased trust \cite{purdie2008social, emerson2015company}, a reduced sense of belonging and acceptance \cite{murphy2009importance, walton2007question}, and concerns about being authentic \cite{shelton2003interpersonal, shelton2006interracial} during collaboration in the school and workplace settings. Black individuals face the additional burden of stereotype threat, fearing the reinforcement of negative group stereotypes \cite{steele1997threat, steele2011whistling, steele2002contending}. The responsibility of debunking these stereotypes can contribute to increased blood pressure \cite{blascovich2001african}, depression \cite{jackson1995composition}, impaired performance and deteriorated memory \cite{lord1985memory, roberson2003stereotype}.

In the workplace, Black individuals grapple with lower career satisfaction and overall well-being compared to their White counterparts in similar positions \cite{feagin2005many, browne2000latinas, feagin1995living}. Black individuals face persistent concerns about discrimination \cite{kaiser2001stop, major2002antecedents} with both overt and subtle forms of racism acting as barriers hindering their success \cite{dipboye2013discrimination, hebl2005promoting, dovidio2001nature}. Additionally, heightened awareness of under-representation in the workplace \cite{ashe1993days} contributes to increased anxiety \cite{bosson2004saying, johns2008stereotype, spencer1999stereotype} and weakened executive functioning \cite{beilock2007stereotype}.

\subsection{Race in Technology}

The study of race in technology often focuses on racial biases present in various technological aspects, including technology designs~\cite{erete2015engaging}, virtual reality~\cite{lopez2019investigating, peck2021evidence}, video games \cite{pace2008can}, autonomous systems~\cite{hosanagar2020human}, facial recognition software~\cite{raji2019actionable, buolamwini2018gender}, Google search engine results~\cite{noble2018algorithms, independent_2015}, and algorithms~\cite{angwin_larson_kirchner_mattu_2016, raghavan2020mitigating}. In contrast, the field of software engineering, which is a specialized area within technology, has conducted minimal research on race. A recent analysis of 376 papers presented at the International Conference on Software Engineering (ICSE) from 2019 to 2021 revealed that only 4\% of these papers described race-related aspects, with none of them analyzing, reflecting upon, or assessing the impact of participants' race in their studies~\cite{dutta2023diversity}. Furthermore, no studies exist that analyze the attitudes, perceptions, and collaborative behaviors of developers in this context.

\subsubsection{The Tech Industry Pipeline Problem}

The issue of underrepresentation of Black developers in the tech industry stems from multifaceted and deeply ingrained systemic problems. According to data from the National Center for Education Statistics (NCES), Black students encounter limited access to advanced math and science courses in high school, which perpetuates educational disparities \cite{barton2003parsing}. Furthermore, the National Science Foundation (NSF) reports that in 2019, despite comprising approximately 13\% of the US population, Black individuals earned only 9\% of Bachelor's degrees in STEM fields \cite{burke2022state}. This educational underrepresentation extends to the professional realm. Industry executives in computing-related fields have acknowledged a substantial decline in the pipeline of Black and Latino developers, from computer science graduates (20\%) to professionals within the software industry (6\%)~\cite{Rankin_2023, alegria2015causes}. In 2014, Black employees at Google accounted for just 2\% of the company's US workforce, with even lower representation in executive and technical roles~\cite{jacobson2014google}. Recent data from the US Equal Employment Opportunity Commission (EEOC) indicates that, as of 2020, Black individuals held a mere 7\% of all computer and mathematical occupations within the tech industry \cite{riccucci2021managing}. Furthermore, a 2020 study by Hired uncovered alarming wage disparities, revealing that Black tech workers, in the US, often receive lower compensation compared to their White counterparts, even when considering factors such as experience and education \cite{Dake_2022}. Tackling these pervasive challenges is vital for promoting diversity, equity, and inclusion within the software industry.

\subsubsection{Race in HCI}

Chen et al. \cite{chen2023and} found that only 3\% of CHI papers, between 2016 and 2021, included data on the race and ethnicity of their participants. Schlesinger et al. \cite{schlesinger2017intersectional} identified a mere 17 CHI papers that directly addressed the topic of race. The current body of HCI literature has consistently emphasized the limited attention given to race within the field of computer science and has called for a more profound exploration of participants' identities and demographics, as evident in works by  \cite{smith2020s, rankin2019straighten, ogbonnaya2020critical, cunningham2023grounds}. Within HCI research, the study of race has spanned various domains, including school interventions \cite{rader2011brick}, gaming habits \cite{dillahunt2009s, rankin2020seat}, health-related topics \cite{grimes2008eatwell}, family communication \cite{tee2009exploring}, and AI literacy \cite{solyst2023would}. However, to the best of our knowledge, there is no existing HCI research that investigates how race influences the collaborative dynamics of developers in pair programming settings.

%% file: methodology.tex
\section{Methodology}

To examine racial interactions within the context of remote pair programming, under controlled conditions, we conducted a think-aloud lab study. During this lab study, participants were tasked with completing programming assignments. To gain deeper insights and triangulate our findings, we conducted retrospective interviews.

Our study design and procedures were approved by our university's Institutional Review Board (IRB). The study was conducted between February 2023 and August 2023, scheduled according to participant availability. Study materials can be accessed at~\cite{GoogleDrive} and details are as follows:

\subsection{Recruitment}
To attract participants, we employed a flyer accompanied by a concise study description that deliberately avoided any reference to race or gender. These materials provided a comprehensive overview of the study, including details about pair programming, the study environment, duration, associated benefits/risks, and compensation. Prospective participants were required to meet minimal criteria: (1) at least 18 years of age; (2) residing in the US; (3) fluent in English; (4) familiar with Java, Python, or C\#; and (5) willing to be video, audio, and screen recorded.

To implement our recruitment strategy, we employed snowball sampling and utilized various online platforms such as LinkedIn, Facebook, Slack, and Discord. Additionally, we conducted a background survey that included questions about race, ethnicity\footnote{Ethnicity refers to a social phenomenon wherein individuals share a unique culture and historical experience \cite{edwards2001race, isajiw1993definition}. One participant self-reported as Hispanic ethnicity.}, and computing experience. The questionnaire was structured with separate, multi-select questions for race and ethnicity, adhering to guidelines from sources like the US Census \cite{Bureau_2022}, prior CHI publications \cite{chen2022collecting, Kuttal2021}, and the NCWIT Guide to Demographic Survey Questions \cite{ncwit}.

\subsubsection{Screening} Out of 557 responses received for the background survey, 77 were incomplete, resulting in 480 completed responses. Unfortunately, 90\% of these responses were identified as spam or bot-generated through manual checking\footnote{For example, responses containing fake names such as ``Gdhdhdjj dhududu Bdhdjdj'', fake emails such as ``kauwjakuwgakuaym397@gmail.com'', and repetitive name variations such as ``B. Bis'', ``R. Bis'', ``I. Bis'',  were categorized as spam.} by the first author. 

After filtering, 48 individuals remained, 40 were deemed eligible for participation through our screening process. We specifically chose participants who met the following inclusion criteria: (1) self-report with one race (either Black or White); (2) self-report as either a man or a woman; (3) individuals who were born and spent majority of their developmental years (up to 18 years old, a time-span with lifelong social and emotional impacts~\cite{barenboim1981development, thompson2006development}) in the US. We established these inclusion criteria for several reasons: (1) to examine interactions between mixed racial groups in a controlled manner and focused only on two races, Black and White; (2) to minimize effects of interactions between two genders in a controlled environment, as we only received background surveys from individuals who self-reported as a man or a woman, we acknowledge that gender is not limited to these binary categories; (3) to compare similar lived experiences within the US \cite{loeber2000stability}; (4) to address the scarcity of research on individuals belonging to intersecting marginalized identities, such as Black women~\cite{SESKO2010356, crenshaw1997mapping, erete2018intersectional, collins2020intersectionality}.

\subsubsection{Sampling} From 40 eligible individuals, we systematically selected 24 participants for our study, considering factors such as availability and the ability to align race, gender, age, experience and education levels between partner. To prevent potential unconscious biases ~\cite{oberai2018unconscious}, we ensured that participants were unaware of their partner and their assigned pairings before the study. We created both same- and mixed-race pairs, including 4 Black-Black pairs, 4 White-White pairs, and 4 Black-White pairs. In each racial category, we ensured a balance between genders, resulting in 2 Man-Man pairs and 2 Woman-Woman pairs. We formed same-gender pairs based on previous research indicating increased communication, satisfaction and rapport \cite{choi2009pair}. We aimed to minimize potential gender-related effects in our study and gain insight into race-related differences.

\subsection{Participants}
Table \ref{participants} provides an overview of the participants' demographics. All participants self-reported as professionals or graduate students in computing-related fields. (21/24) participants held full-time employment or internship experience in the computing industry. (20/24) participants held a minimum of a Bachelor's degree, while (4/24) held a Master's degree. Participants were geographically located in the Southeast (17/24), West (4/24), Midwest (2/24) or Northeast (1/24) regions of the US while participating in the study. Participants self-reported having programming experience with fundamental languages such as Java, Python, C\#, or JavaScript. Each pair was labeled as (P\#-X\#Y\#), where P\# denotes the pair number, X\# indicates the gender of the first or second participant (M for Man, W for Woman), and Y\# represents their race (B for Black, Wh for White).  For example, P1-M1B1 refers to pair 1, participant \#1 who self-reported as Man and Black. P3-W2Wh2 refers to pair 3, participant \#2 who self-reported as Woman and White.

\begin{table}[t]
\caption{Participants general demographics and programming experiences.}
\begin{center}
\scalebox{.75}{
  \begin{tabular}{|c|c|c|c|c|c|c|c|c|c|c|c|c|c|}
     \hline
      \multirow{2}{*}{\textbf{Pair\#}} &
      \multirow{2}{*}{\textbf{ID}} &
      \multirow{2}{*}{\shortstack{\textbf{Self-Reported}\\ \textbf{Gender}}} &
      \multirow{2}{*}{\shortstack{\textbf{Self-Reported}\\ \textbf{Race}}} &
      \multirow{2}{*}{\textbf{Classification}}  &
      \multirow{2}{*}{\textbf{Education}} &
      \multirow{2}{*}{\textbf{Age}} &
      \multirow{2}{*}{\textbf{Location (US)}}  &
      \multicolumn{4}{c|}{\textbf{Experience}} \\
      \cline{9-12}
    & & & & & & & & {\textbf{Prog.}} & {\textbf{Pair Prog.}} & \textbf{TDD.} & \textbf{Industry}\\
        \hline
        P1 & M1B1 & Man & Black & Professional & Bachelor's & 18-23 &  Southeast & \textgreater{}4 yrs. & No & No & 6 mth.-1 yr.\\ \cdashline{2-12}
        & M2B2 & Man & Black \& Hispanic & Professional & Bachelor's & 24-29 & Southeast &\textgreater{}4 yrs. & Yes & No & 1-3 yrs.\\ 
        \hline
        P2 & M1B1 & Man & Black & Professional & Bachelor's & 24-29 &  Southeast & 2 yrs. & Yes & No & 1-3 yrs.\\ \cdashline{2-12}
        & M2B2 & Man & Black & Professional & Bachelor's & 18-23 &  Southeast& \textgreater{}4 yrs. & Yes & Yes & 3-6 mth.\\ 
        \hline
        P3 & W1B1 & Woman & Black & Professional & Bachelor's & 18-23 &  Southeast& 4 yrs. & Yes & No & 6 mth.-1 yr.\\ \cdashline{2-12}
        & W2B2 & Woman & Black & Professional & Bachelor's & \textgreater{}40 &  Southeast& 2 yrs. & No & No & \textgreater{}3 yrs.\\
        \hline
        P4 & W1B1 & Woman & Black & Professional & Bachelor's & 24-29 &  Southeast& 2 yrs. & No & Yes & 1-3 yrs.\\ \cdashline{2-12}
        & W2B2 & Woman & Black & Professional & Bachelor's & 18-23 &  West& \textgreater{}4 yrs. & No & No & 1-3 yrs. \\ 
        \hline
        P5 & M1Wh1 & Man & White & Professional & Bachelor's & 24-29 &  Southeast& 4 yrs. & Yes & Yes & \textgreater{}3 yrs.\\ \cdashline{2-12}
        & M2Wh2 & Man & White & Professional & Bachelor's & 30-40 &  West& \textless{}1 yr. & Yes & No & 6 mth.-1 yr.\\ 
        \hline
        P6 & M1Wh1 & Man & White & Professional & Bachelor's & 24-29 &  West& 2 yrs. & Yes & Yes & \textgreater{}3 yrs.\\ \cdashline{2-12}
        & M2Wh2 & Man & White & Professional & Master's & 30-40 &  Midwest& \textgreater{}4 yrs. & Yes & Yes & \textgreater{}3 yrs.\\ 
         \hline
        P7 & W1Wh1 & Woman & White & Professional & Bachelor's & 24-29 & Northeast& \textless{}1 yr. & Yes & No & 6 mth.-1 yr.\\ \cdashline{2-12}
        & W2Wh2 & Woman & White & Professional & Bachelor's & 24-29 & Midwest&\textless{}1 yr. & Yes & No & None\\ 
        \hline
        P8 & W1Wh1 & Woman & White & PhD student & Bachelor's & 24-29 &  Southeast& \textgreater{}4 yrs. & Yes & No & 1-3 mth.\\ \cdashline{2-12}
        & W2Wh2 & Woman & White & PhD student & Bachelor's & 30-40 &  Southeast& \textgreater{}4 yrs. & Yes & Yes & \textgreater{}3 yrs.\\ 
        \hline
        P9 & M1B1 & Man & Black & PhD student & Bachelor's & 24-29 &  Southeast& \textgreater{}4 yrs. & No & No & 1-3 yrs.\\ \cdashline{2-12}
        & M2Wh2 & Man & White & PhD student & Bachelor's & 24-29 &  West& \textgreater{}4 yrs. & Yes & No & 3-6 mth.\\ 
        \hline
        P10 & M1B1 & Man & Black & Professional & Master's   & 24-29 &  Southeast& \textgreater{}4 yrs. & Yes & Yes & 6 mth.-1 yr.\\ \cdashline{2-12}
        & M2Wh2 & Man & White & Professional & Bachelor's &  \textgreater{}40 &Southeast& 3 yrs. & No & Yes & 1-3 yrs.\\
        \hline
        P11 & W1B1 & Woman & Black & PhD student & Master's  & 18-23 &  Southeast& \textgreater{}4 yrs. & No & Yes & 1-3 yrs.\\ \cdashline{2-12}
        & W2Wh2 & Woman & White & PhD student & Master's  & 24-29 &  Southeast& \textgreater{}4 yrs. & Yes & Yes & 1-3 yrs.\\ 
        \hline
        P12 & W1B1 & Woman & Black & PhD student & Bachelor's & 24-29 &  Southeast& \textgreater{}4 yrs. & Yes & Yes & None \\ \cdashline{2-12}
        & W2Wh2 & Woman & White & PhD student & Bachelor's & 18-23 &  Southeast& \textgreater{}4 yrs. & No & No & None\\ 
    \hline
  \end{tabular}}
  \label{participants}
  \vspace*{-10pt}
\end{center}
\end{table}

\subsection{Study Design}
We conducted a controlled lab study with Zoom, a virtual collaboration tool~\cite{archibald2019using}, to establish our controlled study environment for participants to pair program. Participants were instructed to log in from their laptops in their respective working environments, allowing us to control the study conditions. The virtual setup enabled participants from various geographical locations across the US to participate, facilitating the recruitment of minority participants. Each participant received a \$30 Amazon gift card as compensation for their participation in our 1.5-2 hour study. Participants were not provided with information regarding the study's emphasis on race-related effects before their participation.

\subsubsection{Before the Programming Tasks}
We used a script to ensure uniformity across each pair's session. We initiated the study by instructing participants to enable their video and audio for the study's entirety. All participants adhered to the instructions by upholding the continuous operation of video and audio throughout the study's duration. We continued the study by having participants complete a consent form\footnote{Participants might have been aware of the examination of race-related effects during the study. The inclusion of the phrase  ``to understand diverse pairs'' in the study's consent form could have introduced a potential expectancy bias \cite{miller2004expectancy}, by providing participants with information about the study's focus.} and a self-efficacy questionnaire. 

We explained the study procedures and provided tutorials on the necessary concepts needed to complete the study. Our instructional sequence began with a 3-minute video tutorial introducing pair programming, outlining the driver and navigator roles. Participants then watched a 3-minute video tutorial on test-driven development (TDD) principles and a 1-minute video of a think-aloud example. Each participant practiced think-aloud by engaging in a sample task, counting the number of windows in their home. We continued with a live demonstration of Replit covering code location, test case creation, code execution, and task details. Replit served as the collaborative development environment where participants performed their programming tasks together~\cite{cooper2020exploring}. Replit was set up for test-driven development. Test-driven development is a well-established approach that requires writing test cases first then refactoring code~\cite{beck2022test}. Each participant received login credentials to access this programming environment. 

We explicitly instructed participants to use pair programming, test-driven development, and think-aloud methods during the programming tasks. Participants were not mandated to produce a specific number of test cases, and the assignments of the driver/navigator roles were open-ended. This approach empowered participants to independently determine their roles and strategies for solving tasks with their partners.

\subsubsection{Programming Tasks} 

Participants completed the programming tasks using the think-aloud method by verbally expressing their ideas and emotions as they complete the task~\cite{Lewis1982}. After the tutorials, participants practiced pair programming, test-driven development, and think aloud by completing a 10-minute ``Simple Task'' aimed at fostering pair jelling, a practice known to enhance compatibility and productivity in pair programming~\cite{Jones2013}. This task was particularly essential as our participants did not know each other beforehand and required some time to familiarize themselves before proceeding to the main task.  This task involved writing Java code to validate password length and login information.

Following the ``Simple Task,'' participants engaged in the ``Main Task'' for 40 minutes, with the entire session being audio, video, and screen recorded. We allocated 40 minutes for the task, considering the study's duration of 1.5-2 hours, with the intention to prevent pair fatigue \cite{wray2009pair}. Additional time may have altered scores.
In the ``Main Task,'' participants implemented a Tic-Tac-Toe game. We chose the Tic-Tac-Toe game, as our task, due to its inherent simplicity, eliminating the need for extensive task requirements or explanations to the participants. It involves two players, X and O, taking turns placing marks on a 3x3 grid until a player achieves three consecutive placements or the game ends in a tie. Participants were provided with a sample 3x3 grid where player X's marks were placed repeatedly each turn without declaring a winner. Additionally, we provided participants with sample user stories (high-level scenarios) and acceptance criteria (the conditions to fulfill the game) typically employed in test-driven development~\cite{santos2020increasing}. The acceptance criteria encompassed player swapping (taking turns) and verifying if a winner was determined through vertical, horizontal, or diagonal placements.

\subsubsection{After the Programming Tasks:} Following the 40-minute task, we provided participants with a 5-minute break to prevent fatigue. Subsequently, participants filled out post-questionnaires adopted from ~\cite{Kuttal2019, Kuttal2021, Robe2022} regarding their self-efficacy and overall experience with pair programming and test-driven development. These questionnaires were utilized to triangulate the results of our data analysis.

\begin{table}[t]
\caption{Semi-structured retrospective interview questions used by all interviewers.}
\label{tbl:sample int. ques.}
\scalebox{.75}{
\begin{NiceTabular}{|p{13cm}|}[hvlines, first-col]
 & \textbf{Sample Questions}\\
\Block{4-0}{General Pair Prog.} 
 & Did you as a pair think about who would be the driver or navigator? \\
 & How did you decide when to switch driver/navigator roles?\\
 & In the future, when doing pair programming, when do you want to be the driver? the navigator?\\
 & How did that partnership feel?\\
\Block{4-0}{Race Related} & Would you prefer partnering with someone of your same racial group or someone from a different racial group?\\
 & Did you feel that there were any benefits to working with someone of the same/different racial group?\\
 & Do you think communication/collaboration would have been easier or more difficult with a partner from the same racial group? with a partner from a different racial group?\\
 & What do you think if you were partnering with someone from the same/different racial group, would you have performed better? Performed the same? Performed worse?
\CodeAfter
\SubMatrix\{{2-1}{5-1}.[left-xshift=2mm]
\SubMatrix\{{6-1}{9-1}.[left-xshift=2mm]
\end{NiceTabular}}
\end{table}

\subsubsection{Retrospective Interview} 
Table \ref{tbl:sample int. ques.} provides a sample of our semi-structured retrospective interview questions. Each participant engaged in an interview, which was video and audio recorded, within separate Zoom breakout rooms. The questions covered a wide range of topics, from sociodemographic aspects such as education level and geographic location, general pair programming inquiries, and specific inquiries about race. We recognized the importance of discussing race openly to delve deeper into interracial interactions during remote pair programming. Our objectives for these interview questions were as follows: (1) to gain insights into participants' emotions during the task; (2) to make connections about race beyond survey responses.

We carefully crafted the semi-structured interview questions with input from a professor who specializes in  Human Development and Family Studies and conducts research on racial influences in society. We incorporated the professor's recommendations to enhance the sensitivity and comfort of the interviews for participants.  To create a more comfortable environment, we matched interviewers with interviewees based on the participant's self-reported race. For instance, our first author, a Black woman, conducted interviews with all the Black participants, while two other researchers, a White man and a White woman, interviewed all the White participants. We took into consideration the concept of social desirability, as previous research has demonstrated that the interviewer's role can significantly impact participants' comfort levels and the depth of their responses, including the sharing of more details and personal stories~\cite{bergen2020everything, kaushal2014social}.

\subsection{Data Analysis}
We conducted both quantitative and qualitative analyses to address our research questions (RQs). For RQ1, we employed quantitative methods to assess productivity, code quality, changes in self-efficacy before and after the task, and gathered post-questionnaire feedback on participants' pair programming preferences. We opted for a scoring rubric, as a measurement for productivity and code quality, drawing parallels with the grading methodology employed in academic settings \cite{stevens2023introduction, becker2003grading}. Due to our small sample size, we utilized descriptive statistical analyses, employing mean and standard deviation for group comparisons \cite{fisher2009understanding}.

For qualitative analysis of the data pertaining to RQ2, RQ3, and RQ4, we leveraged the integrated functionalities within Zoom to capture video, audio, and screen recordings for each participant.  Subsequently, two researchers manually corrected terminology and phrasing inaccuracies. We first transcribed the recordings, breaking them down into individual utterances. Each utterance was then manually annotated with relevant information, including timestamps, the speaker's identity, their self-reported race/gender, and their role within the pair programming session (i.e., the person driving). The code sets we used to code the developers' utterances, for our qualitative analysis, can be found in Table \ref{creativity} and  \ref{tab:leadership codeset} presented throughout our paper. An essential step in qualitative research is the process of coding, which entails identifying key concepts or phenomena within the data and marking their occurrences \cite{auerbach2003qualitative}. 

For interview responses, we employed thematic analysis to categorize qualitative data into themes aligned with the research questions and adopted an iterative, open-coding approach to scrutinize pair behavior. This involved coding the developers' retrospective interview responses to identify significant concepts and phenomena, following established guidelines in the field \cite{Seaman1999, Kitchenham2002}.

To assess inter-rater reliability, we used the Jaccard measure \cite{Jaccard1901}. Initially, two researchers independently coded 20\% of the task transcripts and interviews, achieving an 85\% agreement on the coded data. The remaining transcripts were divided between the two researchers for independent coding.

%% file: limitations.tex
\section{Limitations \& Future Work}
While our sample size consisted of 8 same-race and 4 mixed-race pairs, which might be considered limited for making broad generalizations or conducting extensive quantitative analyses, it serves as an initial step in understanding racial interactions within pair programming. Other factors, such as sample selection, may have affected the results, but we controlled for gender and assigned pairings based on race, age, prior experience, education levels, and availability in an effort to minimize any potential influence from participant classification on our results. Recruiting participants from marginalized groups, such as Black men, Black women, and White women, presented challenges for this study, such as constantly advertising recruitment materials and receiving bot-generated surveys, resulting in our study taking 6 months to complete. Additionally, we focused on examining interactions between two genders, specifically men and women, as participants self-reported their genders, and our responses exclusively fell within these two categories. Our results may be minimally influenced by gender-related effects, as we did not include mixed-gender pairs (e.g., M1Wh1 - W2Wh2) or mixed-gender, race pairs (e.g., M1Wh1 - W2B2) in our study. During retrospective interviews, we couldn't always achieve both racial and gender matching due to the availability of interviewers. It's important to note that our findings are not solely attributed to race, as we acknowledge that other factors like participants' personalities{\footnote{Hannay et al. found that personality test results did not strongly correlate with the performance of developer pairs \cite{hannay2009effects}.}}, skill levels, programming language preferences, exposure to diversity and inclusion initiatives, and knowledge of racial differences may have influenced our results. A follow-up study with a larger and more diverse sample size could enhance the generalizability of the research findings. Further research could evaluate a self-reported measure regarding participants' engagement with diversity, inclusion, and race. Future studies could delve into additional dimensions of individuals' identity, including gender, sexual orientation, and disability, as these factors intersect with race, potentially shaping pair programming dynamics.
Furthermore, our study employed a single, straightforward task, a game of Tic-Tac-Toe, while real-world industry programming challenges can be considerably more complex.

%% file: results.tex
\section{Results}

\subsection{RQ1: How does race affect pair programming dynamics in both same- and mixed-race pairs?}

Pair programming is a well-established agile software development method known to significantly influence productivity, code quality, self-efficacy, and the overall experience of developer pairs. We analyzed same- and mixed-race participant pairs interactions and evaluated four factors: (1) productivity - how each participant progressed on the task; (2) code quality - accuracy of written code and tests; (3) self-efficacy - a participant's belief in their capability to complete the task; (4) overall experience - what participant's gained from pair programming with test-driven development.

For our quantitative analysis, refer to Table \ref{tab:comparing 3 groups prod,cq,self}, we used descriptive statistical analyses (mean, standard deviation) to compare three groups (4 - same-race (BB), 4 - same-race (WhWh), 4 - mixed-race (BWh) participant pairs). Throughout RQ1, we present our hypotheses based on our quantitative results, which need to be tested with a statistically significant population to draw concrete conclusions. We analyzed:

  \begin{table}[t]
    \caption{Comparing technical aspects across 4 same-race (BB), 4 same-race (WhWh), and 4 mixed-race (BWh) pairs.}
    \scalebox{.75}{
    \begin{tabular}{|c||c|cc||cc||cc|}
    \hline
    & \textbf{} & \multicolumn{2}{c||}{\textbf{Same-Race (BB)}} & \multicolumn{2}{c||}{\textbf{Same-Race (WhWh)}} & \multicolumn{2}{c|}{\textbf{Mixed-Race (BWh)}}  \\ \hline
    & Samples & \multicolumn{1}{c|}{Mean}  & SD  & \multicolumn{1}{c|}{Mean}  & SD  & \multicolumn{1}{c|}{Mean} & SD   \\ \hline
    Productivity  & 4 & \multicolumn{1}{c|}{28} & 12.36 & \multicolumn{1}{c|}{19.25} & 17.75 & \multicolumn{1}{c|}{32.5} & 19.36 \\ \hline
    Code Quality  & 4 & \multicolumn{1}{c|}{17.5} & 16.77 & \multicolumn{1}{c|}{14} & 9.25  & \multicolumn{1}{c|}{17.5} & 4.33 \\ \hline
    Self-Efficacy & 8  & \multicolumn{1}{c|}{6.5}  & 7.58  & \multicolumn{1}{c|}{-4.75} & 12.09 & \multicolumn{1}{c|}{1.5}  & 3.51 \\ \hline
    \end{tabular}}
    \label{tab:comparing 3 groups prod,cq,self}
\end{table}

\subsubsection{Productivity}
Productivity was assessed based on the completion of the task within the 40-minute time frame. We scored each pair's work out of 100 points, equally weighing the progress on test cases and code. We assigned 10 points each to {\tt vertical}, {\tt horizontal}, {\tt diagonal}, {\tt tie}, and {\tt taking turns} test cases. Additionally, equal points were awarded for code written in each of the {\tt vertical}, {\tt horizontal}, and {\tt diagonal} categories. An additional 5 bonus points were awarded for pairs showcasing innovation through code that went beyond the instructions. The grading criteria is available at \cite{GoogleDrive}.

Based on our study participants, mixed-race (BWh) pairs had the highest average score (32.5) compared to same-race (BB) (28), and same-race (WhWh) (19.25) pairs (refer Table \ref{tab:comparing 3 groups prod,cq,self}). However, the scores for mixed-race (BWh) pairs showed a greater standard deviation (SD) of 19.36, indicating that the results were more dispersed. The elevated variability observed among mixed-race (BWh) pairs may be attributed to (3/4) pairs (P9, P10, P11) possessing industry experience, while (1/4) pairs (P12) lacked such experience, resulting in limited progress on the tasks. According to our quantitative results, White participants demonstrated higher productivity in mixed-race (BWh) pairs than in same-race (WhWh) pairs.

The mixed-race (BWh) pairs tended to employ a `brute force' approach in task completion. This approach involved exploring numerous potential solutions, even if they were not entirely correct or optimal. Moreover, in the event of a failed solution attempt, they promptly made another attempt to resolve the problem \cite{guo2014local, higson2019bayesian, michie1989brute}. Such an approach significantly enhanced the pairs productivity. For example, during the Main Task, P10-M2Wh2 stated, \textit{``This is kind of brute force, but we just want to check each one of these [cells].''} 

In the same-race (BB) pairs, the driver primarily worked on the Main Task, while the navigator assumed a relaxed and disengaged stance (as elaborated in Section 5.3.2). Consequently, progress through the task was facilitated primarily by one partner working independently. 

The mean productivity score of the same-race (WhWh) pairs was influenced by the behaviors observed in the same-race (WhWh) women pairs. For example, P7 (a same-race (WhWh) women pair) spent roughly 8 minutes having a repetitive discussion on the same idea for switching between players (X or O). As recognized by P7-W1Wh1, \textit{``\dots we're talking about the same thing over and over.''}

We investigated whether race has an effect on productivity. \textit{\textbf{We hypothesize that (H1) pair productivity will increase for mixed-race pairs compared to same-race pairs.}}

\subsubsection{Code Quality}

We evaluated the quality and completeness of the code produced by same- and mixed-race participant pairs. Each pair's work was assessed on a scale of 100 points, with 60 points for correct code and 40 points for correct test cases. For correct code written, we assigned 15 points each for {\tt vertical}, {\tt horizontal}, {\tt diagonal}, and {\tt swap}. We allocated 8 points for each test case in {\tt vertical}, {\tt horizontal}, {\tt diagonal}, {\tt tie}, and {\tt taking turns} categories. Additionally, we allocated 5 bonus points for any code that was innovative and went beyond the instructions. The grading criteria are available at \cite{GoogleDrive}.

Our results showed that mixed-race (BWh) pairs and same-race (BB) pairs had the same average code quality score (17.5), while same-race (WhWh) pairs had the lowest (14) (refer Table \ref{tab:comparing 3 groups prod,cq,self}). The low code quality averages across all groups stemmed from (10/12) pairs producing 0 test cases and (9/12) pairs producing none or only 1 correct method of code. The scores among same-race (BB) pairs were more dispersed with a standard deviation (SD) of 16.77. The variability for same-race (BB) pairs is due to (1/4) pairs producing entirely incorrect code, resulting in a score of 0. According to our results, code quality scores were comparable between same- and mixed-race pairs.

Despite employing a `brute force' approach (as discussed in 5.1.1), the mixed-race (BWh) pairs overlooked the importance of code quality. For example, during the Main Task, P9 applied `brute force' approach and acknowledged the resulting lower code quality, as P9-M2Wh2 commented, \textit{``I don't know if everything here actually works.''} In same-race (BB) pairs, the disengaged navigator showed little concern for code quality. In same-race (WhWh) pairs, repetitive discussions (as discussed in 5.1.1) and lower self-efficacy (as discussed in 5.1.3) limited progress, thereby limiting the code quality.

We investigated whether race has an effect on code quality. \textit{\textbf{We hypothesize that (H2) there will be no differences in code quality between same- and mixed-race pairs.}}

\begin{figure}[h]
\caption{Participants self-efficacy before and after the task. Red circles represent a decrease, green circles represent an increase, and black circles represent no change in self-efficacy scores.}
\centering
\noindent\scalebox{.75}{%
    \begin{tikzpicture}
    \begin{axis} [
    ybar = 2pt,
    x=0.19cm, 
    bar width = 4pt, 
    width=18cm, 
    height=6.5cm,
    ymin = 12, 
    ymax = 65, 
    xmax=dummy,
    xticklabel style={rotate=45}, 
    xtick pos=left,
    ytick pos=left,
    symbolic x coords={P1-M1B1, dummy, dummy, P1-M2B2, dummy, dummy, P2-M1B1, dummy, dummy, P2-M2B2, dummy, dummy, P3-W1B1, dummy, dummy, P3-W2B2, dummy, dummy, P4-W1B1, dummy, dummy, P4-W2B2, dummy, dummy, P5-M1Wh1, dummy, dummy, P5-M2Wh2, dummy, dummy, P6-M1Wh1, dummy, dummy, P6-M2Wh2, dummy, dummy, P7-W1Wh1, dummy, dummy, P7-W2Wh2, dummy, dummy, P8-W1Wh1, dummy, dummy, P8-W2Wh2, dummy, dummy, P9-M1B1, dummy, dummy, P9-M2Wh2, dummy, dummy, P10-M1B1, dummy, dummy, P10-M2Wh2, dummy, dummy, P11-W1B1, dummy, dummy, P11-W2Wh2, dummy, dummy, P12-W1B1, dummy, dummy, P12-W2Wh2, dummy},
    xtick=data,
    enlarge y limits=0.045,
    enlarge x limits = {abs=0.38cm}, 
    tick label style={font=\scriptsize}, 
    xlabel={\hspace{0.2cm} \textbf{Same-Race (BB)} \hspace{2.5cm} \textbf{Same-Race (WhWh)} \hspace{2.5cm} \textbf{Mixed-Race (BWh)}},
    ylabel={Self-Efficacy Scores},
    xlabel style={font=\scriptsize},
    ylabel style={font=\scriptsize},
    legend image code/.code={
        \draw [#1] (0cm,-0.1cm) rectangle (0.2cm,0.1cm); },
    legend style={
        at={(1.009,1)},
        anchor=north west,
        font=\scriptsize,
    },
    ]
    \addplot[fill=black!40] coordinates {(P1-M1B1, 52) (P1-M2B2, 54) (P2-M1B1, 42) (P2-M2B2, 47) (P3-W1B1, 47) (P3-W2B2, 40) (P4-W1B1, 42) (P4-W2B2, 41) (P5-M1Wh1, 53) (P5-M2Wh2, 48) (P6-M1Wh1, 42) (P6-M2Wh2, 57) (P7-W1Wh1, 47) (P7-W2Wh2, 47) (P8-W1Wh1, 45) (P8-W2Wh2, 53) (P9-M1B1, 57) (P9-M2Wh2, 52) (P10-M1B1, 51) (P10-M2Wh2, 55) (P11-W1B1, 51) (P11-W2Wh2, 56) (P12-W1B1, 46) (P12-W2Wh2, 39)};
    \addlegendentry{Before Task}
    \addplot[fill=blue!40] coordinates {(P1-M1B1, 54) (P1-M2B2, 63) (P2-M1B1, 39) (P2-M2B2, 45) (P3-W1B1, 54) (P3-W2B2, 46) (P4-W1B1, 61) (P4-W2B2, 55) (P5-M1Wh1, 56) (P5-M2Wh2, 49) (P6-M1Wh1, 51) (P6-M2Wh2, 57) (P7-W1Wh1, 38) (P7-W2Wh2, 45) (P8-W1Wh1, 13) (P8-W2Wh2, 47) (P9-M1B1, 57) (P9-M2Wh2, 56) (P10-M1B1, 51) (P10-M2Wh2, 58) (P11-W1B1, 54) (P11-W2Wh2, 59) (P12-W1B1, 40) (P12-W2Wh2, 44)};
    \addlegendentry{After Task}
    \end{axis}
    \draw [dashed] (4.65, -0.2) -- (4.65, 5.1);
    \draw [dashed] (9.22, -0.2) -- (9.22, 5.1);
    \draw (2.6,3.5) [draw=green] ellipse (0.2 and 0.5);
    \draw (3.2,2.9) [draw=green] ellipse (0.2 and 0.5);
    \draw (3.75,3.5) [draw=green] ellipse (0.2 and 1);
    \draw (4.35,3.3) [draw=green] ellipse (0.2 and 0.8);
    \draw (7.2,2.8) [draw=red] ellipse (0.2 and 0.6); 
    \draw (7.8,3.15) [draw=red] ellipse (0.2 and 0.3); 
    \draw (8.44,1.65) [draw=red] ellipse (0.2 and 1.45); 
    \draw (8.95,3.5) [draw=red] ellipse (0.2 and 0.4); 
    \draw (9.48,4) ellipse (0.2 and 0.2); 
    \draw (10.05,3.75) [draw=green] ellipse (0.2 and 0.4); 
    \draw (10.65,3.5) ellipse (0.2 and 0.2); 
    \draw (11.2,3.9) [draw=green] ellipse (0.2 and 0.4); 
    \end{tikzpicture}}
\label{fig:self efficacy}
\end{figure}
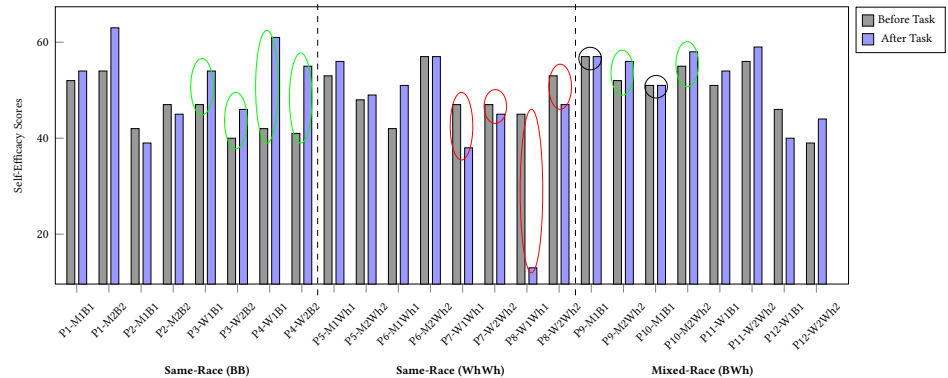

\subsubsection{Self-Efficacy} Self-efficacy reflects participants' confidence in their ability to succeed. One notable outcome of pair programming is its positive influence on participant morale, as demonstrated in prior research \cite{Nosek1998}. To quantify this impact, we assessed self-efficacy using a 7-point Likert scale with 9 questions administered through surveys conducted at both the beginning and end of the task. 

We employed the well-established Computer Programming Self-Efficacy Scale (CPSES) developed by Ramalingam and Wiedenbeck \cite{ramalingam1998development} and translated by Altun and Mazman \cite{altun2012programlamaya}. This widely accepted scale gauges short-term changes in individuals' self-efficacy, considering factors such as magnitude, strength, and generality \cite{yukselturk2017investigation, tsai2019developing}. Hence, our questions focused on problem-solving and navigating challenges during pair programming, factors known to impact self-efficacy \cite{compeau1995computer}. Fig. \ref{fig: self efficacy questions}. outlines the questions (Q1-Q9) and ratings (1-7) used to measure participants' self-efficacy. Rating 1 indicated the lowest confidence level (not confident at all) while 7 signified the highest confidence level (absolutely confident). We aggregated the scores from the individual questions to generate a total score for each participant in both the pre and post self-efficacy questionnaires. It's important to note that participants' self-efficacy may have been influenced by their interactions with their programming partner.

\begin{figure}[t]
\centering
\caption{Self-efficacy questionnaire utilizing a 7-point Likert scale consisting of 9 questions administered both before and after the task \cite{ramalingam1998development, compeau1995computer}.}
\vspace{5pt}
\scalebox{.95}{
    \begin{minipage}{.5\textwidth}
      \scriptsize
       \rule{215pt}{0.3pt} \\
      \textbf{1 } Not confident at all \hspace{11.5pt} \textbf{4 } 50/50 \hspace{33pt} 
      \textbf{7 } Absolutely confident  \\
      \textbf{2 } Mostly not confident \hspace{7pt}  \textbf{5 } Fairly confident \hspace{10pt} \\
      \textbf{3 } Slightly confident \hspace{15pt} \textbf{6 } Mostly confident\\
      \rule{215pt}{0.3pt} \\
      \textbf{Q1: } I could come up with a suitable strategy for a given programming project in a \\short time. \\
      \rule{215pt}{0.3pt} \\
      \textit{I could find a way}\dots \\
      \textbf{Q2: } of overcoming the problem if I got stuck at a point, while working on a \\programming project. \\
      \textbf{Q3: } to concentrate on my program, even when there were distractions around me. \\
      \textbf{Q4: } of motivating myself to program, even if problem area was of no interest to me. \\
      \rule{215pt}{0.3pt} \\
    \end{minipage}
    \begin{minipage}{.5\textwidth}
      \scriptsize
      \rule{215pt}{0.3pt} \\
      \textit{I could write a}\dots \\
      \textbf{Q5: } program that someone else could comprehend and add features to at a later date. \\
      \textbf{Q6: } small Java program given a small problem that is familiar to me. \\
      \rule{215pt}{0.3pt} \\
      \textit{I could complete a programming project}\dots \\
      \textbf{Q7: } if I could call someone for help if I got stuck. \\
      \textbf{Q8: } if someone showed me how to solve the problem first. \\
      \textbf{Q9: } once someone else helped me get started. \\
      \rule{215pt}{0.3pt} \\
   \end{minipage}
}
\label{fig: self efficacy questions}
\end{figure}

We compared the differences in self-efficacy scores for each participant by analyzing their self-reported levels before and after the task. Specifically, we examined the self-efficacy score variations among three groups: same-race (BB), same-race (WhWh), and mixed-race (BWh) pairs. A higher score reflects a more favorable attitude towards pair programming. The results of this comparison are summarized in Table \ref{tab:comparing 3 groups prod,cq,self}.

Based on our study participants, same-race (BB) pairs had the highest average change in self-efficacy (6.5, refer to Table \ref{tab:comparing 3 groups prod,cq,self}). The same-race (BB) woman pairs had a large increase in average self-efficacy (P3: +6.5, P4: +16.5) (refer to Fig. \ref{fig:self efficacy}. green circles). For example, P4-W1B1 increased by 19 and P4-W2B2 increased by 14 which is the highest self-efficacy increase among all the pairs. P3-W1B1 mentioned her reasoning for a lower confidence level at the beginning of the study, \textit{``We definitely would have benefited if we had a little bit more confidence, I think, in our skill set. I think we both kind of doubted ourselves a little bit. I know I did initially because I tend to not want to over compensate.''} We investigated whether a partner's perceived race, during pair programming, effects an individual's self-efficacy, \textit{\textbf{we hypothesize that (H3) an individual's self-efficacy will increase for same-race (BB) pairs compared to same-race (WhWh) and mixed-race (BWh) pairs.}}

Based on our study participants, same-race (WhWh) pairs had the lowest average change in self-efficacy (-4.75, refer to Table \ref{tab:comparing 3 groups prod,cq,self}). Both same-race (WhWh) woman pairs average self-efficacy decreased (P7: -5.5, P8: -19) (refer to Fig. \ref{fig:self efficacy}. red circles). For example, P7-W1Wh1 decreased by 9 and P7-W2Wh2 decreased by 2. When asked about self-efficacy, P8-W1Wh1 said, \textit{``That ended up really hurting my confidence, because I felt that I was performing poorly, not only in front of a researcher, but also in front of a peer.''} Our findings contradict pair programming studies that reported woman having increased self-efficacy while working together~\cite{Kuttal2019, choi2013evaluating, choi2015comparative, mcdowell2006pair, ying2019their}. \textit{\textbf{We hypothesize that (H4) an individual's self-efficacy will decrease for same-race (WhWh) women pairs compared to same-race (BB) and mixed-race (BWh) pairs.}}

Based on our study participants, mixed-race (BWh) pairs average change in self-efficacy scores (1.5) were the most approximate (SD = 3.51), in reference to Table \ref{tab:comparing 3 groups prod,cq,self}. The main pattern we observed was between the mixed-race (BWh) man pairs. The Black man participants' self-efficacy remained constant (P9-M1B1: 0, P10-M1B1: 0) whereas the White man participants' self-efficacy increased (P9-M2Wh2: +4, P10-M2Wh2: +3) (refer to Fig. \ref{fig:self efficacy}. black and green circles). There was no pattern across the mixed-race (BWh) woman pairs. \textit{\textbf{We hypothesize that (H5) there will be no difference in an individual's self-efficacy for mixed-race (BWh) pairs compared to same-race (BB) and same-race (WhWh) pairs.}}

\pgfplotstableread[col sep=&, header=true]{
Question&v1&v2&v3
Q1&4.125 & 4.125&4
Q2&4.125 &4 &4.25
Q3& 4&3.875 &3.875
Q4&4.375 &4.5 &4.625
Q5&4.5 &4.25 &4.375
Q6&4.5 &3.75 &3.25
Q7&4.75 &3.625 &3.75
Q8& 2.125&2.875 &2.5
Q9&1.625 &2.5 &1.875
Q10&2.25 & 2.25&1.875
Q11&2.125 &3 &2
Q12& 2.25&3.25 &2.25
Q13& 2.25& 3.625&2.5
Q14& 2.125&2.75 &2
Q15& 3.875&3.375 &3.125
Q16&4.25 &3.75 &3.5
Q17&4.25 &3.5 &3.625
Q18&4.125 & 3.5&3.5
Q19&4.5 &4.25 &4.25
}\ppdata

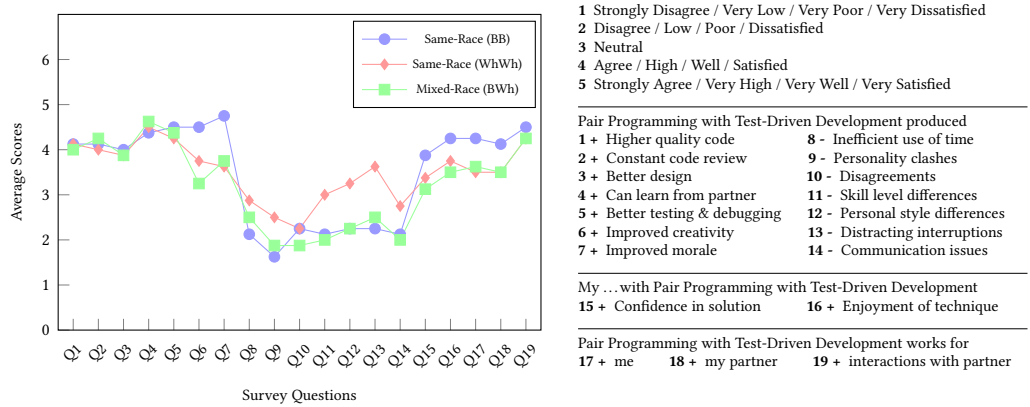
\begin{figure}[h]
\caption{Pair programming preferences questionnaire with 1 (Lowest) - 5 (Highest) scale and 19 questions regarding pair programming with test-driven development.}
\vspace{5pt}
\begin{minipage}[c]{8cm}
\centering
    \begin{tikzpicture}
        \begin{axis}[
        xtick=data,
        xticklabels from table={\ppdata}{Question},  
        width=8cm, 
        height=8cm,
        bar width=10mm, y=6mm,
        xtick pos=left,
        ytick pos=left,
        enlarge x limits = {abs=0.2cm},
        xlabel={Survey Questions},
        xlabel style={font=\scriptsize},
        ylabel style={font=\scriptsize},
        ylabel={Average Scores},
        ytick={0,1,2,3,4,5,6},
        ymin=0,
        ymax=7,
        xticklabel style={rotate=45, font=\scriptsize},
        yticklabel style={font=\scriptsize},
        legend style={at={(0.8,0.98)}, anchor=north, font=\tiny},
        tick label style={font=\scriptsize},
        nodes near coords align={vertical},
        ]
        \addplot [blue!40,mark=*]  table [ybar, y=v1, x expr=\coordindex,] {\ppdata};
        \addplot [red!40,mark=diamond*] table [y=v2, x expr=\coordindex] {\ppdata};
        \addplot [green!40,mark=square*] table [y=v3, x expr=\coordindex] {\ppdata};
        \legend{\strut Same-Race (BB), \strut Same-Race (WhWh), \strut Mixed-Race (BWh)}
        \end{axis}
    \end{tikzpicture}
    \end{minipage}%
    \begin{minipage}{7\textwidth}
      \scriptsize
      \textbf{1 } Strongly Disagree / Very Low / Very Poor / Very Dissatisfied \\
      \textbf{2 } Disagree / Low / Poor / Dissatisfied \\
      \textbf{3 } Neutral \\
      \textbf{4 } Agree / High / Well / Satisfied \\
      \textbf{5 } Strongly Agree / Very High / Very Well / Very Satisfied \\
      \rule{170pt}{0.3pt} \\
      Pair Programming with Test-Driven Development produced \\
      \textbf{1 + } Higher quality code \hspace{24.5pt} \textbf{8 - } Inefficient use of time\\
      \textbf{2 + } Constant code review  \hspace{20pt} \textbf{9 - } Personality clashes \\
      \textbf{3 + } Better design \hspace{40pt} \textbf{10 - } Disagreements\\
      \textbf{4 + } Can learn from partner  \hspace{16pt} \textbf{11 - } Skill level differences \\
      \textbf{5 + } Better testing \& debugging \hspace{6.5pt} \textbf{12 - } Personal style differences\\
      \textbf{6 + } Improved creativity \hspace{24.5pt} \textbf{13 - } Distracting interruptions \\
      \textbf{7 + } Improved morale \hspace{31pt} \textbf{14 - } Communication issues\\
      \rule{170pt}{0.3pt} \\
      My \dots with Pair Programming with Test-Driven Development \\
      \textbf{15 + } Confidence in solution \hspace{14pt} \textbf{16 + } Enjoyment of technique\\
      \rule{170pt}{0.3pt} \\
      Pair Programming with Test-Driven Development works for \\
      \textbf{17 + } me \hspace{10pt} \textbf{18 + } my partner \hspace{10pt} \textbf{19 + } interactions with partner \\ \\ \\
   \end{minipage}
\label{fig: pp post study}
\end{figure}

\subsubsection{Overall Experience} We measured participants overall experience using pair programming with test-driven development. We collected participants pair programming with test-driven development preferences individually using a questionnaire adapted from Kuttal et. al \cite{Kuttal2019, Kuttal2021} and Robe et. al \cite{Robe2022}. These responses helped to triangulate our findings in RQ1. 

In Fig. \ref{fig: pp post study}. each trend line represents a group and the points are the average response scores for same-race (BB), same-race (WhWh), mixed-race (BWh) participant pairs per question. Participants rated using values 1 - 5 with 1 as the lowest response (Strongly Disagree / Very Low / Very Poor / Very Dissatisfied) and 5 as the highest response (Strongly Agree / Very High / Very Well / Very Satisfied). Whether a higher or lower value is preferred is indicated beside the question numbers with a (+) or (-) sign. For questions 1-5, all three groups favorably ranked pair programming.

 \begin{itemize}
   
    \item \textbf{\textit {Same-race (BB) participant pairs preferred pair programming:}} The pair programming questionnaire results reinforced our quantitative self-efficacy results for same-race (BB) pairs. The same-race (BB) participants (blue, dotted line in Fig. \ref{fig: pp post study}.) had the highest average scores for questions 6 - ``improved creativity,'' 7 - ``improved morale,'' 15 - ``confidence in solution,'' 16 - ``enjoyment of technique,'' 17 - ``me,'' 18 - ``my partner,'' and 19 - ``interactions with partner.'' These questions preferred higher values because they were benefits of using pair programming with test-driven development. Same-race (BB) participants had the lowest average scores for questions 8 - ``inefficient use of time'' and 9 - ``personality clashes'' which preferred lower values. P3-W2B2 mentioned why she prefers pair programming, \textit{``Sometimes 2 heads are better than one [when you] look at something\dots somebody else can look at it for a minute and automatically see the error.''} Based on our study participants, same-race (BB) participants enjoyed pair programming with test driven development which increased their self-efficacy and morale in completing the task.
    \item \textbf{\textit{Same-race (WhWh) participant pairs preferred solo programming:}} The pair programming questionnaire results reinforced our quantitative self-efficacy results for same-race (WhWh) pairs. The same-race (WhWh) participants (red, diamond line in Fig. \ref{fig: pp post study}.) had the highest average scores for questions 8 - ``inefficient use of time,'' 9 - ``personality clashes,'' 11 - ``skill level differences,'' 12 - ``personal style differences,'' 13 - ``distracting interruptions,'' and 14 - ``communication issues.'' These questions preferred lower values because they were downsides of using pair programming with test-driven development. P5-M2Wh2 talked about his preference for solo programming because of skill level differences, \textit{``If I don't know this stuff and I'm trying to learn while we're trying to actually get work done or vice versa, trying to get someone else caught up that can be kind of detrimental sometimes. Where if you were just working by yourself, you could just knock it out [but instead] you're getting dragged down, and vice versa, you can also be the person that's dragging someone else down. Then you're feeling bad because some people do, unfortunately, show their frustrations when you're not [on] the same level as they are, and then working with someone that's getting frustrated with you that's not fun. I would rather work by myself.''} Based on our study participants, same-race (WhWh) participants ranked pair programming with test-driven development unfavorably which negatively impacted their self-efficacy. 
    
    \item \textbf{\textit {Mixed-race (BWh) participant pairs benefited from diversity of ideas from their partners:}} Based on our study participants, mixed-race (BWh) pairs (green, square line in Fig. \ref{fig: pp post study}.) had the lowest average score for question 6 - ``improved creativity.'' However, five participants from mixed-race pairs noted having different perspectives as a benefit to completing the task with a partner from a different racial group. P12-W1B1 discussed the advantages of working with a partner from a different racial group, \textit{``They see things that you don't see, and you see things that they don't see\dots so communication within [different] people is always a good thing.''} These results aligns with prior work on diversity in higher education has found that diversity enriches perspectives by exposing students to a broader range of viewpoints \cite{gurin1999new, smith2000benefits}.
 \end{itemize}

\subsection{RQ2: How does the creativity styles of participants in same-race pairs differ from those in mixed-race pairs during pair programming?}
To compare same- and mixed-race participant pairs, we used the established Osborn-Parnes Creative Problem Solving Process \cite{Osborn1957, Page1988, Kavitha2013}, which consists of four stages (see Table \ref{creativity}). This framework guided our systematic analysis of creative problem-solving dynamics, in racial pairings, during pair programming. For qualitative analysis, we utilized a code set from \cite{Kuttal2020} and open-coded participant behavior during the task, triangulating results through interview responses. We observed that in both same- and mixed-race pairs, both the driver and navigator made equal contributions in terms of {\tt clarify} and {\tt idea}.

\begin{table}[b]
\centering
\caption{Osborn-Parnes Creative Problem Solving Process creativity stages and definitions \cite{Isakesen2004, Kuttal2020}.}
\label{creativity}
\scalebox{.75}{
\begin{tabular}{|l|l|}
\hline
\textbf{\begin{tabular}[c]{@{}l@{}}Creativity\\ Stages\end{tabular}} & \textbf{Definitions} \\ \hline
Clarify   & Identify the goal, wish, or challenge.                       \\ \hline
Idea      & Generate ideas on how to solve the challenge.                \\ \hline
Develop   & Evaluate, strengthen, and select solutions for best ``fit.'' \\ \hline
Implement & Support implementation of the selected solution(s).          \\ \hline
\end{tabular}}
\end{table}

\begin{figure}[h]
\centering
\caption{Same-race (BB) vs. same-race (WhWh) vs. mixed-race (BWh) pairs total develop and implement frequencies for driver (blue) and navigator (orange).\\}
\label{creativity circles}
\noindent\scalebox{.4}{%
\begin{tikzpicture}[
    pie-label/.style = {text width=54mm, align=left, 
                        anchor=south west, inner xsep=0pt}
                    ]
    \pie[xshift=4cm, color={blue!40, orange!40}, font={\huge}]{79.59/, 20.41/}
    \hspace{1cm}
    \pie[xshift=14cm, color={blue!40, orange!40}, font={\huge}]{73.75/, 26.25/}
    \hspace{.8cm}
    \pie[xshift=24cm,text=legend, color={blue!40, orange!40}, font={\huge}]{48.85/{Driver}, 51.15/{Navigator}}
\end{tikzpicture}}
\label={ \\ \scriptsize\textbf{Same-Race (BB)} \hspace{2.7cm} \textbf{Same-Race (WhWh)} \hspace{2.5cm} \textbf{Mixed-Race (BWh)}}
\end{figure}
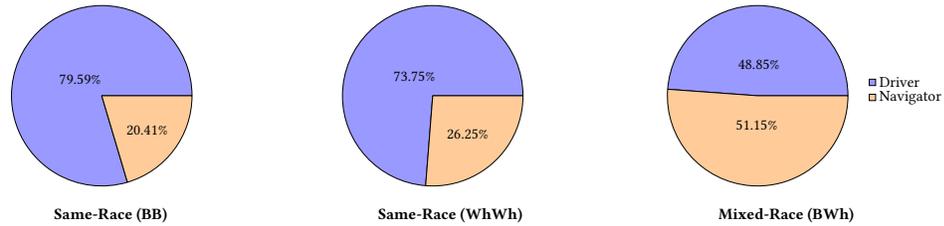

 \begin{itemize}

    \item \textbf{\textit{Same-race (BB) and (WhWh) participant pairs drivers made majority of the contribution to the develop and implement stages:}} In (7/8) same-race pairs, the primary contributor to the {\tt develop} and {\tt implement} phases was the driver. The driver had roughly (BB) 79.59\% and (WhWh) 73.75\% of the {\tt develop} and {\tt implement} frequencies compared to about (BB) 20.41\% and (WhWh) 26.25\% for the navigator (refer to Fig. \ref{creativity circles}.). For example, P2-M2B2, the driver, contributed to the {\tt develop} stage for 15 instances. P2-M1B1, the navigator, contributed to the {\tt develop} stage for 7 instances. P2-M2B2 contributed to the {\tt implement} stage for 49 instances while P2-M1B1 contributed one-sixth of that amount.  When asked why the navigator lacked contribution to the stages, P6-M1Wh1 (navigator) said, \textit{``I felt very outclassed that I did not really know what I was doing. He was moving faster than I could, so when I was trying to sit there and kind of understand the logic, he was quickly moving beyond \dots he seemed to jump right in and get to it.''}

     \item \textbf{\textit{Mixed-race (BWh) participant pairs the driver and navigator distributed the tasks evenly in the develop and implement stages:}} All (4/4) mixed-race pairs evenly distributed the work, between driver and navigator, in the {\tt develop} and {\tt implement} stages. The driver (blue in Fig. \ref{creativity circles}.) contributed about 48.85\% while the navigator (orange in Fig. \ref{creativity circles}.) contributed about 51.15\%  to the {\tt develop} and {\tt implement} stages. For example, P9-M1B1 (driver) contributed 64 instances and P9-M2Wh2 (navigator) contributed 63 instances to the {\tt develop} and {\tt implement} stages. P12-W1B1 mentioned the even task distribution, \textit{``In [the task] she identified the problems pretty easily and I was the person [who] was like, oh yeah, we have to import the array list so everybody has their shrink.''}  
    
\end{itemize}

\vspace{-10pt}

\subsection{RQ3: How does race influence collaboration dynamics among same- and mixed-race pairs during pair programming?}

To understand the key dynamics and factors that influence collaboration between developers in same- and mixed-race pairs, and how these dynamics impact software development outcome, we analyzed:

\begin{table}[b]
    \centering
    \caption{Leadership styles and definitions \cite{cuadrado2012gender, Kuttal2019, adeliyi2021investigating, bass1996transformational}.}
    \scalebox{.75}{
    \begin{tabular}{|p{2.2cm}|p{10.5cm}|}
        \hline
        \textbf{Leadership Styles} & \textbf{Definitions}\\
        \hline
        \hlc[orange!40]{Authoritative} & When P1 dominates the interaction/makes decisions alone and P2 follows/doesn't contribute any input.\\
        \hline
        \hlc[orange!40]{Democratic} & When P1 shares the decisions with the other P2 or encourages P2 to take initiative. \\
        \hline
        Laissez-faire & When P1 says do what you think is right and gives all decision making to P2. \\
        \hline
        \hlc[orange!40]{Paternalistic}& When P1 teaches/instructs P2 or leads in a ‘fatherly/motherly’ way. \\
        \hline
        Transformational & When P1 changes, transforms, or redirects thoughts and P2 starts thinking the same way as P1. \\
        \hline
    \end{tabular}}
    \label{tab:leadership codeset}
\end{table}

\begin{figure}[b]
\caption{Percentage of time participants spent using each leadership style.}
\centering
\noindent\scalebox{.75}{%
    \begin{tikzpicture}
        \begin{axis}[
        name=MyAxis,
        xbar stacked,
        width=10cm, 
        height=10cm, 
        bar width=5.5pt,
        enlarge y limits=0.045,
        xtick pos=left,
        ytick pos=left,
        xlabel={Overall Interaction},
        ylabel={\hspace{0.5cm} Same-Race (BB) \hspace{1cm} Same-Race (WhWh)  
        \hspace{1cm} Mixed-Race (BWh) \hspace{1cm} },
        xmin=0, 
        xmax=100,
        legend pos=outer north east,
        legend style={font=\footnotesize},
        tick label style={font=\footnotesize}, 
        ylabel style={font=\footnotesize},
        xticklabel={$\pgfmathprintnumber{\tick}\%$},
        symbolic y coords={P1-M1B1, P1-M2B2, dummy, P2-M1B1, P2-M2B2, dummy, P3-W1B1, P3-W2B2, dummy, P4-W1B1, P4-W2B2, dummy, P5-M1Wh1, P5-M2Wh2, dummy, P6-M1Wh1, P6-M2Wh2, dummy, P7-W1Wh1, P7-W2Wh2, dummy, P8-W1Wh1, P8-W2Wh2, dummy, P9-M1B1, P9-M2Wh2, dummy, P10-M1B1, P10-M2Wh2, dummy, P11-W1B1, P11-W2Wh2, dummy, P12-W1B1, P12-W2Wh2},
        ytick=data,
        ]
        \addplot[color=black, fill=green!40]  coordinates {(69.74,P1-M1B1) (50,P1-M2B2) (81.25,P2-M1B1) (81.82,P2-M2B2) (60.47,P3-W1B1) (45.61,P3-W2B2) (51.47,P4-W1B1) (22.92,P4-W2B2) (53.52,P5-M1Wh1) (40,P5-M2Wh2) (27.27,P6-M1Wh1) (55.77,P6-M2Wh2) (22.5,P7-W1Wh1) (36.54,P7-W2Wh2) (16.67,P8-W1Wh1) (31.82,P8-W2Wh2) (47.5,P9-M1B1) (34.55,P9-M2Wh2) (41.38,P10-M1B1) (51.72,P10-M2Wh2) (31.11,P11-W1B1) (35.71,P11-W2Wh2) (40.68,P12-W1B1) (41.38,P12-W2Wh2) };
        \addplot[color=black, fill=blue!40]  coordinates {(25,P1-M1B1) (36,P1-M2B2) (12.5,P2-M1B1) (6.06,P2-M2B2) (24.42,P3-W1B1) (36.84,P3-W2B2) (30.88,P4-W1B1) (45.83,P4-W2B2) (11.27,P5-M1Wh1) (20,P5-M2Wh2) (59.09,P6-M1Wh1) (30.77,P6-M2Wh2) (37.50,P7-W1Wh1) (30.77,P7-W2Wh2) (50,P8-W1Wh1) (15.91,P8-W2Wh2) (42.5,P9-M1B1) (34.55,P9-M2Wh2) (51.72,P10-M1B1) (41.38,P10-M2Wh2) (64.44,P11-W1B1) (61.9,P11-W2Wh2) (49.15,P12-W1B1) (48.28,P12-W2Wh2)};
        \addplot[color=black, fill=black!40]  coordinates {(0,P1-M1B1) (4,P1-M2B2) (0,P2-M1B1) (0,P2-M2B2) (1.16,P3-W1B1) (0,P3-W2B2) (0,P4-W1B1) (4.17,P4-W2B2) (0,P5-M1Wh1) (0,P5-M2Wh2) (0,P6-M1Wh1) (0,P6-M2Wh2) (0,P7-W1Wh1) (0,P7-W2Wh2) (0,P8-W1Wh1) (0,P8-W2Wh2) (0,P9-M1B1) (9.09,P9-M2Wh2) (3.45,P10-M1B1) (3.45,P10-M2Wh2) (0,P11-W1B1) (0,P11-W2Wh2) (1.69,P12-W1B1) (0,P12-W2Wh2)};
        \addplot[color=black, fill=red!40]  coordinates {(1.32,P1-M1B1) (2,P1-M2B2) (3.13,P2-M1B1) (3.03,P2-M2B2) (12.79,P3-W1B1) (15.79,P3-W2B2) (14.71,P4-W1B1) (22.92,P4-W2B2) (29.58,P5-M1Wh1) (23.33,P5-M2Wh2) (13.64,P6-M1Wh1) (13.46,P6-M2Wh2) (30,P7-W1Wh1) (26.92,P7-W2Wh2) (33.33,P8-W1Wh1) (52.27,P8-W2Wh2) (5,P9-M1B1) (16.36,P9-M2Wh2) (0,P10-M1B1) (0,P10-M2Wh2) (2.22,P11-W1B1) (0,P11-W2Wh2) (3.39,P12-W1B1) (3.45,P12-W2Wh2)};
        \addplot[color=black, fill=orange!40]  coordinates {(3.95,P1-M1B1) (8,P1-M2B2) (3.13,P2-M1B1) (9.09,P2-M2B2) (1.16,P3-W1B1) (1.75,P3-W2B2) (2.94,P4-W1B1) (4.17,P4-W2B2) (5.63,P5-M1Wh1) (16.67,P5-M2Wh2) (0,P6-M1Wh1) (0,P6-M2Wh2) (10,P7-W1Wh1) (5.77,P7-W2Wh2) (0,P8-W1Wh1) (0,P8-W2Wh2) (5,P9-M1B1) (5.45,P9-M2Wh2) (3.45,P10-M1B1) (3.45,P10-M2Wh2) (2.22,P11-W1B1) (2.38,P11-W2Wh2) (5.08,P12-W1B1) (6.9,P12-W2Wh2)};
        \legend{\strut Authoritative, \strut Democratic, \strut Laissez-faire, \strut Paternalistic, \strut Transformational}
        \end{axis}
    \draw [dashed] (-2,2.85) -- (9.5,2.85);
    \draw [dashed] (-2,5.55) -- (9.5,5.55);
    \end{tikzpicture}%
}    
\label{fig:leadership figure}
\end{figure}

\subsubsection{Effect of Leadership Style}
Leadership style is an important aspect to consider during remote pair programming. Management research has explored how race impacts self-perception, how others perceive and treat you, and evaluations all of which can influence an individual's leadership style~\cite{ospina2009critical, case1997african, lomotey1993african}. We coded the task for leadership styles and used interview responses to triangulate our results. Table ~\ref{tab:leadership codeset} is the code set we used to label leadership styles which is inspired by various works~\cite{cuadrado2012gender, Kuttal2019, adeliyi2021investigating, bass1996transformational}. Fig. ~\ref{fig:leadership figure}. illustrates the percentage of time each participant spent using a certain leadership style.

 \begin{itemize}
    \item \textbf{\textit {Same-race (BB) and (WhWh) participant pairs had a solo decision-making partner:}} In (7/8) same-race pairs, one partner displayed an {\tt authoritative} style (green in Fig. \ref{fig:leadership figure}.) more than the other partner. The authoritative partner made majority of the coding decisions alone during the task. For example, in P6 (same-race pair), P6-M2Wh2 was about 30\% more authoritative than P6-M1Wh1 (refer to Fig. \ref{fig:leadership figure}.). \textit{``I'm just gonna do this [add a variable], so that I don't have to use self [keyword] and hopefully, that does it,''} is an example dialogue of how P6-M2Wh2 made task design decisions alone. The more authoritative partner took control over the task implicitly or explicitly.  Diversity research, in the workplace, suggests that same-race dyads tend to have higher-quality leader-member exchanges, influencing more democratic decision-making \cite{bauer1996development, scandura1986managers, roberson20016, randolph2016diversity}. However, our results are based on same-race dyads operating at the same power level, warranting further investigation with a larger population.

    \item \textbf{\textit {Same-race (WhWh) and (BB) participant pairs educated each other during the task:}} Same-race (WhWh) pairs, irrespective of gender, and same-race (BB) women pairs educated one another. On average, (4/4) same-race (WhWh) pairs displayed a strong {\tt paternalistic} approach (red in Fig. \ref{fig:leadership figure}.): P5 - 26\%, P6 - 14\%, P7 - 28\%, P8 - 43\%. Also, on average, both same-race (BB) woman pairs had a strong {\tt paternalistic} style: P3 - 15\%, P4 - 19\% (red in Fig. \ref{fig:leadership figure}.). The pairs used this style when instructing their partner on programming syntax. For example, P5-M1Wh1 instructed on writing the method to switch players between X and O, \textit{``So leave that method, the getPlayer one, leave that the way it is, create another method right below it\dots perfect, and then go ahead and put a parameter in there, so setPlayer and in those parentheses do, like char\dots there you go, and then just do\dots lowercase c-h-a-r, maybe newMark\dots I think it's just one [equal sign]\dots''} while P5-M2Wh2 followed the directions. P6-M2Wh2 confirmed the paternalistic style we observed, \textit{``If you're coaching them through syntactical things sometimes it's like here's a simpler ways of writing what you wrote or hey you're missing the class name, or you're missing the new keyword.''} Irrespective of race, women exhibited paternalistic tendencies consistent with prior research on gender and leadership styles, where women tend to prioritize understanding and concern for others' feelings \cite{eagly1990gender}. 
    
 \item \textbf{\textit {Mixed-race (BWh) participant pairs shared decision-making:}} In all (4/4) mixed-race pairs, both partners spent roughly an equal percentage of the interaction employing the {\tt authoritative} (green in Fig. \ref{fig:leadership figure}.) and {\tt democratic} (blue in Fig. \ref{fig:leadership figure}.) styles but were slightly more democratic. For example, both participants in P12 (mixed-race pair) spent approximately an equal amount of time using both styles ({\tt authoritative} - 41\%, {\tt democratic} - 48\%). The mixed-race pairs were authoritative during the beginning and end of the task due to the pressure of the time limit. Mixed-race pairs transitioned between leadership styles by explicitly releasing control of the task.  Management research studies discovered that mixed-race entrepreneurial teams shared decisions more democratically, in the workplace, while performing tasks~\cite{bass1990bass, bass2008handbook}.

    \item \textbf{\textit {Mixed-race (BWh) women participant pairs were authoritative during idea generation and democratic during coding, while men participant pairs exhibited the reverse pattern: }}
    We observed that men and women mixed-race pairs transitioned between leadership styles using different patterns. Mixed-race men pairs (P9 and P10) exhibited a {\tt democratic} style during discussion of ideas and an {\tt authoritative} style while developing code. Mixed-race women pairs (P11 and P12) showed an {\tt authoritative} style while brainstorming ideas and shifted to a {\tt democratic} style while writing code. P11-W2Wh2 verified the democratic trends we observed, \textit{``I personally felt like it [the partnership] was pretty democratic.''} This behavior parallels findings in gender studies, where women tended to prefer the role of navigator when they knew how to complete a task, while men preferred the role of driver under similar circumstances \cite{Kuttal2019}.

 \end{itemize}

\subsubsection{Role-exchange dynamics}The well-established pair programming roles are driver (writes code) and navigator (guides with verbal suggestions). The frequency and flexibility of how pairs exchange roles can determine individuals level of engagement in the programming task~\cite{bryant2006pair, zhong2017investigating}.

 \begin{itemize}
    \item \textbf{\textit {Navigator of same-race (BB) and (WhWh) participant pairs tended to disengage from the task:}} Navigator disengagement is described as the navigator failing to grasp the task and letting the driver continue programming alone. The disengaged navigator loses the advantages of pair programming when they stop participating and lack concern in completing the task~\cite{plonka2012disengagement}. Five participants from same-race pairs mentioned experiencing navigator disengagement. P1-M1B1 (driver) talked about this phenomenon, \textit{``I feel like since I was talking so much, I wasn't getting a lot of reciprocation other than just like, yeah, you're right, you can do this.''} P2-M1B1 (navigator) expressed his disengagement during the task, \textit{``There [were] definitely times when I was kind of, I didn't know honestly what [was] going on.''} The observed disengagement in same-race pairs is reminiscent of the study by Chong et al. \cite{chong2007social} that found less knowledgeable developers tended to remain disengaged or adopted a passive approach to tasks.

    \item \textbf{\textit {Mixed-race (BWh) participant pairs explicitly discussed roles:}} In (4/4) mixed-race participant pairs, partners communicated explicitly and strictly adhered to the driver/navigator roles. The discussion of roles took place at the beginning, middle, and end of the interaction. We observed mixed-race (BWh) pairs discussing roles about six times more than same-race (BB) and (WhWh) pairs. P9-M2Wh2 confirmed our observation, \textit{``We went pretty explicit, I said. Hey, do you want to be a driver? He said. Sure, and he was a driver. I was navigator\dots I think we stuck pretty close to the roles.''} In contrast, P6-M2Wh2 highlighted the lack of discussion in same-race pairs, \textit{``For the most part we had one person do most of the driving\dots I did a little bit\dots I wasn't necessarily navigating [or] driving. He did most of the driving and then he was also navigating as well.''}

 \end{itemize}

\subsection{RQ4: How does individuals’ awareness of their partners’ racial backgrounds influence their attitudes and interaction dynamics within same- and mixed-race pairs?}

To explore the influence of individuals' awareness of their partners' racial backgrounds on their attitudes and behaviors, we conducted retrospective interviews and examined their perspectives regarding interactions with individuals from same or mixed racial backgrounds.

\begin{itemize}
  \item \textbf{\textit{Same-race (BB) and (WhWh) participant pairs communicated better while mixed-race (BWh) participant pairs limited communication:}} 
 Effective communication involves active listening, clear articulation of ideas, open sharing of thoughts, and approachability. A survey of Microsoft professional developers identified good communication as a top-ten desirable quality in a pair programming partner~\cite{begel2008pair}. 

Same-race (BB) and (WhWh) participants (5/16 mentioned explicitly) frequently demonstrated enhanced communication, attributed to the increased comfort levels resulting from shared demographics. P2-M1B1 underscored this aspect, emphasizing that a deeper sense of comfort facilitated open and effective communication with his partner. He stated, \textit{``I do like partnering [with] people who are the same demographic or same race, just because [I may] be able to communicate certain things or we might just have similar ideas because of similar experiences.''} 

Conversely, (6/8) mixed-race (BWh) participants frequently encountered communication challenges stemming from concerns about inadvertently causing offense or misunderstanding due to racial dynamics. P11-W2Wh2, a participant in a mixed-race pair, candidly expressed how race considerations occasionally hindered her ability to communicate openly. She noted, \textit{``It would be naive to say no [race had no impact on communication] because\dots I felt like there were times where I didn't fully communicate [because] it could have came across like I was overstepping.''} Hill et al. \cite{koglerhill2000} found that same-race relationships typically offer greater psychosocial support, highlighting potential challenges in communication dynamics for mixed-race pairs.

    \item \textbf{\textit {Same-race (BB) and (WhWh) participant pairs were vulnerable with each other and mixed-race (BWh) participant pairs showed empathy towards their partner:}} 
Vulnerability, characterized by participants opening up and sharing personal aspects of their personality with their partner~\cite{schroeder2009vulnerability}, manifested differently within same- and mixed-race participant pairs.

In the case of  (4/8) same-race pairs, vulnerability was more pronounced as participants felt a sense of camaraderie rooted in shared experiences. P4-W1B1, a participant in a same-race pair, described this vulnerability, stating, \textit{``It's more just like we've all been through the same stuff like we're just trying to get this done. So let's just try to knock this out, and there's more of a jovial tone\dots probably from the same [racial group]. I feel like there's maybe just greater ease, and I could reveal all of my personality.''}

Conversely, (3/4) mixed-race pairs demonstrated empathy towards each other, marked by heightened awareness and consideration of their partner's feelings and reactions. Three participants, in mixed-race pairs, mentioned exercising extra caution while working with partners from different racial backgrounds. For instance, P11-W2Wh2, a participant in a mixed-race pair, expressed the need for heightened sensitivity, explaining, \textit{``You sort of have to be more careful about what you say especially when it comes to people who are already in marginalized groups.''}

The empathy and carefulness shown by mixed-race pairs reflect findings in counseling psychology, where in mixed-race counselor/client dynamics, the counselors' culturally-sensitive responses impact client reactions and engagement~\cite{pomales1986effects}.
    
    \item \textbf{\textit {Participants had a cultural rapport working with same-race partners but lacked experience with other races:}} 
    The concept of cultural rapport, often characterized by an unspoken understanding, familiarity with culturally specific terms, and the exchange of non-verbal cues~\cite{johnson1999ties, merriam2001power}, played a significant role in shaping interactions within (4/8) same-race participant pairs. These pairs tended to share a natural cultural rapport with their partners, which translated into increased comfort levels during collaboration. P3-W1B1 articulated this phenomenon explaining, \textit{``I think in general, I do find it easier to work in groups or collaborate with other people from [the] same racial or ethnic background just because there's some stuff that you sort of get, even if it's stuff that's not directly related to the task. Building cultural rapport with somebody does go a long way in making you feel more comfortable and able to do the task that you're working on.''}

    On the contrary, (9/24) participants often remarked on their limited collaborative experience with individuals from different racial backgrounds which often resulted in uncomfortable situations. For instance, P1-M1B1 admitted to his limited experience collaborating with individuals from different racial groups, stating, \textit{``I grew up in Black areas only I always went to all Black schools, everything so there would just be a level of uneasiness working with different races.''} Similarly, P9-M2Wh2 recounted a personal anecdote highlighting the impact of limited cross-cultural exposure, sharing, \textit{``For example, when I was young I did not encounter Black people at all right until college started\dots I guess here's a good anecdote. I remember when I first became a TA, I associated lynching (the unauthorized public hangings of Black individuals in the US from the 1890s to 1940s ~\cite{wood2011lynching}) with like werewolves right? It's from a game I would play. I was telling a professor I have a really good joke\dots I don't remember the full context. I think I was just going to make fun of [a college] team [by writing] oh, don't lynch me for this! but then [the professor] was like you may not want to use that word. I'm like, oh, I didn't think of this.''} Lack of experience with diverse groups can indeed pose challenges in group dynamics as  a study on diverse group dynamics revealed that individuals initially relied on perceived stereotypes of others, which, in turn, led to difficulties in group development \cite{enayati2002research}.

    \item \textbf{\textit {Participants felt comfortable working with same-race partners but had anxiety working with other races:}}  
    The concept of comfort levels, often associated with the ``feel-good'' effect in collaborations, pertains to the extent to which both partners feel at ease during their interactions and how this comfort can potentially influence their performance~\cite{muller2004empirical}. (7/24) participants highlighted the advantages of working with partners of the same racial background, emphasizing the higher level of comfort and the sense of natural connection they experienced. P3-W1B1, a participant in a same-race pair, articulated her reasons for feeling more at ease during the task, stating, \textit{``I will say there could have just been some natural comfort being with someone I'm slightly more familiar with\dots I think that could have maybe unknowingly affected how I talked, how I joked, or you know things of that nature. I think there's benefit in me, being not as nervous, or on edge, or trying to present a certain way, I was able to be myself more likely.''}

Anxiety, particularly in the workplace, can lead to fatigue, reduced focus, and decreased overall performance~\cite{haslam2005anxiety}. While many software companies have recognized the importance of supporting mental health programs, including Employee Assistance Programs (EAPs), Mental Health Benefits, and Mental Health Training, the issue of anxiety continues to affect employees \cite{attridge2019global}. 

(4/24)  participants specifically pointed out discomfort and anxiety as drawbacks that had a detrimental impact on their productivity in the workplace. P1-M1B1 shared his experiences of anxiety while working with developers from different racial backgrounds, explaining, \textit{``I don't have that much exposure to working with someone that's not Black so it'd be some anxiety from just getting used to that format.''}

In psychology, research on same-race Black therapeutic pairs has shown that shared comfort and culturally grounded language can accelerate the formation of strong bonds~\cite{goode2016unspoken, cabral2011racial}.

    \item \textbf{\textit {Participants related more to same-race partners but ``sugar coat'' while working with other races:}} Relatability, an attribute characterized by simplicity in comprehension and the ability to empathize with others~\cite{aish2018people}, played a role in how participants expressed uncertainty during collaborative tasks. (10/24) participants often demonstrated uncertainty by conveying incomplete thoughts/statements or by explicitly expressing confusion about code bugs or programming syntax. For instance, P4-W1B1 inquired of P4-W2B2, \textit{``What's that?''} in reference to \verb|Scanner in = new Scanner(System.in);| within the code. P4-W2B2 responded, \textit{``I don't know what that means. I'm going to look it up. I'm going to look up the scanner.''}

Notably, same-race participant pairs exhibited roughly triple the amount of uncertainty expressions compared to mixed-race participant pairs. This pattern may be attributed to the greater comfort levels experienced when collaborating with individuals of the same racial background. P3-W1B1, a participant in a same-race pair, spoke about openly acknowledging uncertainty with her partner at the outset of the task, stating, \textit{``I probably knew that I was going to take on that driver role a little bit more since we both expressed uncertainty.''}

Three participants also discussed a tendency to employ indirect communication, often referred to as ``sugar coating''~\cite{sices2009sugar}, when collaborating with developers from different racial backgrounds. P2-M1B1 elaborated on this phenomenon, explaining, \textit{``Someone who's similar to me demographically\dots I think it would just be a lot faster because I could be super open with them, super transparent. I wouldn't have to sugar coat anything. I could just say, hey, I'm having this problem, and I don't know what to do, and they could really level with me and relate to me.''} This tendency to ``sugar coat'' can be attributed to the findings of a social psychology study involving 40 (Black-White) pairs in which the lack of prior social contact between partners resulted in a natural inclination to be reserved and cautious \cite{ickes1984compositions}.

\vspace{-0.2pc} 
\end{itemize}

While participants tended to identify more with individuals of the same racial background, they often moderated their communication style while collaborating with individuals from different racial backgrounds, highlighting the complex interplay between relatability, comfort, and communication dynamics in collaborative settings.

%% file: discussion.tex
\begin{table}[]
\centering
\caption{Summary of our study findings and implications. The positive (green), neutral (gray), and negative (pink) aspects of diversity are highlighted.}
\label{tbl:summary of results}
\scalebox{0.85}{
\begin{tabular}{|cc|cc|c|c|}
\hline
\multicolumn{2}{|c|}{} &
  \multicolumn{2}{c|}{\textbf{Same-Race}} &
  \textbf{Mixed-Race} &
   \\ \cline{3-5}
\multicolumn{2}{|c|}{\multirow{-2}{*}{}} &
  \multicolumn{1}{c|}{\textbf{(BB)}} &
  \textbf{(WhWh)} &
  \textbf{(BWh)} &
  \multirow{-2}{*}{\textbf{Implications}} \\ \hline
\multicolumn{1}{|c|}{} &
  \textit{Productivity} &
  \multicolumn{1}{c|}{More} &
  Less &
  \cellcolor[HTML]{9AFF99}More \textit{\textbf{(H1)}}&
  Facilitator Agent \\ \cline{2-6} 
\multicolumn{1}{|c|}{} &
  \textit{Code Quality} &
  \multicolumn{1}{c|}{-- }&
  -- &
  \cellcolor[HTML]{C0C0C0} \textit{\textbf{(H2)}}&
  -- \\ \cline{2-6} 
\multicolumn{1}{|c|}{} &
  \textit{Self-Efficacy} &
  \multicolumn{1}{c|}{Increased \textit{\textbf{(H3)}}} &
  Decreased \textit{\textbf{(H4)}}&
  \cellcolor[HTML]{C0C0C0} \textit{\textbf{(H5)}}&
  Facilitator Agent / Methodology \\ \cline{2-6} 
\multicolumn{1}{|c|}{\multirow{-4}{*}{\textbf{RQ1}}} &
  \textit{Overall Experience} &
  \multicolumn{1}{c|}{\begin{tabular}[c]{@{}c@{}}Preferred pair\\ programming\end{tabular}} &
  \begin{tabular}[c]{@{}c@{}}Preferred solo\\ programming\end{tabular} &
  \cellcolor[HTML]{9AFF99}Diversity of ideas &
  Facilitator Agent / Methodology \\ \specialrule{.15em}{.07em}{.07em}   
\multicolumn{1}{|c|}{\textbf{RQ2}} &
  \textit{Creativity} &
  \multicolumn{1}{c|}{\begin{tabular}[c]{@{}c@{}}Distributed task\\ unevenly\end{tabular}} &
  \begin{tabular}[c]{@{}c@{}}Distributed task\\ unevenly\end{tabular} &
  \cellcolor[HTML]{9AFF99}\begin{tabular}[c]{@{}c@{}}Distributed task\\ evenly\end{tabular} &
  Facilitator Agent  / Methodology \\ \specialrule{.15em}{.07em}{.07em}
\multicolumn{1}{|c|}{} &
  \textit{Leadership Style} &
  \multicolumn{1}{c|}{\begin{tabular}[c]{@{}c@{}}Solo \\ decision-making\end{tabular}} &
  \begin{tabular}[c]{@{}c@{}}Solo \\ decision-making\end{tabular} &
  \cellcolor[HTML]{9AFF99}\begin{tabular}[c]{@{}c@{}}Shared\\  decision-making\end{tabular} &
  Facilitator Agent / Methodology / Workshops \\ \cline{2-6} 
\multicolumn{1}{|c|}{\multirow{-3}{*}{\textbf{RQ3}}} &
  \textit{Role-Exchange} &
  \multicolumn{1}{c|}{\begin{tabular}[c]{@{}c@{}}Disengaged \\ navigator\end{tabular}} &
  \begin{tabular}[c]{@{}c@{}}Disengaged \\ navigator\end{tabular} &
  \cellcolor[HTML]{9AFF99}\begin{tabular}[c]{@{}c@{}}Explicit discussion \\ of roles\end{tabular} &
  Facilitator Agent / Methodology / Workshops \\ \specialrule{.15em}{.07em}{.07em}
\multicolumn{1}{|c|}{} &
  \textit{Communication} &
  \multicolumn{1}{c|}{More} &
  More &
  \cellcolor[HTML]{FFCCC9}Less &
  Remote Work / Asynchronous Interviews \\ \cline{2-6} 
\multicolumn{1}{|c|}{} &
  \textit{Vulnerability} &
  \multicolumn{1}{c|}{More} &
  More &
  \cellcolor[HTML]{9AFF99}More Empathy &
  \begin{tabular}[c]{@{}c@{}}VR / Video Games / Workshops / \\ Courses / Facilitator Agent\end{tabular} \\ \cline{2-6} 
\multicolumn{1}{|c|}{} &
  \textit{Cultural Rapport} &
  \multicolumn{1}{c|}{High} &
  High &
  \cellcolor[HTML]{FFCCC9}\begin{tabular}[c]{@{}c@{}}Lack experience / \\ Uncomfortable\end{tabular} &
  Facilitator Agent / Methodology / Video Games \\ \cline{2-6} 
\multicolumn{1}{|c|}{} &
  \textit{Comfort Levels} &
  \multicolumn{1}{c|}{High} &
  High &
  \cellcolor[HTML]{FFCCC9}High Anxiety &
  Remote Work / Asynchronous Interviews \\ \cline{2-6} 
\multicolumn{1}{|c|}{\multirow{-7}{*}{\textbf{RQ4}}} &
  \textit{Relatability} &
  \multicolumn{1}{c|}{More} &
  More &
  \cellcolor[HTML]{FFCCC9}Less &
  Video Games / Workshops \\ \hline
\end{tabular}}
\end{table}

\section{Discussions and Implications}

Table \ref{tbl:summary of results} summarizes key differences in remote pair programming between same- and mixed-race pairs and their implications for tools, methods, and racial awareness. We formulated hypotheses (\textit{\textbf{H1-H5}}) that require further testing with a statistically significant sample size to generalize our claims.

Our results offer valuable insights into various scenarios that may arise in pair programming. One notable benefit of pair programming is the ability to remain focused on tasks and adhere to established techniques. For example, P8 demonstrated strict adherence to test-driven development principles, as per the requirements. Throughout the Main Task, P8-W2Wh2 stated, \textit{``In pure test-driven development I think we do one test at a time\dots If it were me I would write [it another way] but for pure test-driven development, I think we go to the [tests]\dots Well, in pure test-driven development we could just have it return true.''} Conversely, a drawback of pair programming may be heightened anxiety when coding in front of another individual. For example, P8-W1Wh1 experienced nervousness when coding in front of her partner, during the Main Task, \textit{``I'm feeling very stressed.''}

\subsection{Diversity's Double-Edged Sword of Pair Programming}
\subsubsection{Positive Aspects:} 

In the realm of mixed-race (BWh) pair programming, a series of beneficial practices emerged. Participants were more productive (RQ1, \textit{\textbf{H1}}), had different perspectives (RQ1), effectively distributed tasks (RQ2), shared decision-making responsibilities (RQ3), and frequently switched roles (RQ3). These practices allowed developers to fully harness the potential of pair programming. One contributing factor to this success may be the heightened attentiveness and commitment to collaboration observed in diverse teams, as articulated by participant P11-W2Wh2 during an interview, \textit{``I try to be cognizant and more aware of the situation at hand.''}

Diversity also fostered empathy among developers, particularly in the context of racial awareness during collaborative efforts (RQ4). P11-W2Wh2 offered valuable insights into this racial awareness, explaining, \textit{``I do try to\dots especially when working with women or people who identify as non-binary, Black, Asian, or any other race that’s not White, [I make an effort] to be very cognizant of my words. As someone who identifies as a woman and is also White, I recognize that I have certain privileges and a degree of authority in the situation. I am cognizant not to overstep [boundaries] and I make sure I give them the space to be themselves and also communicate about the problem.''} This increased racial awareness influenced nuanced attitudes and dynamics, observed in both same- and mixed-race pairs.

Furthermore, developers recognize the advantages of diversity as one participant, P8-W1Wh1, expressed a preference for a pair programming partner from a different racial group. When asked about the reasoning behind this preference, P8-W1Wh1 remarked, \textit{``I assume, if you had 2 people from different backgrounds, racial or ethnic or otherwise, you would potentially have more access to a diverse thought process for problem-solving.''} A study in school administration also highlighted the importance of diversity in instructor and staff teams to provide students with a broad spectrum of perspectives and backgrounds for a richer educational environment~\cite{ngounou2019value}.

\subsubsection{Negative Aspects:} While diversity can offer numerous benefits, it also poses certain challenges, as evidenced in RQ4. These challenges encompass impediments to effective communication, the cultivation of discomfort and anxiety, and a potential reduction in participants' ability to establish connections with one another. P1-M1B1 expressed, \textit{``Anxiety [causes me to] preemptively respond, by not talking or communicating as much.''}

One notable challenge that arises in diverse teams is an elevated level of stress, as voiced by four participants who feel compelled to continually prove themselves as mentioned by P1-M2B2, \textit{``A lot of times you find yourself trying to prove that you know how to code.''}

Additionally, lack of diversity may result in the departure of some minority individuals from the workplace. Participant P1-M2B2 spoke on a personal anecdote, \textit{``I didn't like the fact that they [company administration] didn't really see everything that I could bring to the table. So I ended up quitting.''}

\subsubsection{Neutral Aspects:}
Diversity is inconsequential in terms of code quality (RQ1, \textit{\textbf{H2}}), and self-efficacy (RQ1, \textit{\textbf{H5}}). However, same-race (BB) pairs exhibited higher self-efficacy levels (RQ1, \textit{\textbf{H3}}). Four participants attributed their heightened self-efficacy to a natural rapport and connection with their partners, a high degree of comfort, and a sense of inclusion in their collaborative efforts.

Moreover, trusting a partner proved to be a cornerstone of pair programming. Partners leaned on each other's advice and contributions, with trust predominantly vested in individuals with more experience, higher skill levels, or established relationships. For example, P6-M1Wh1 expressed trust in his partner skills, noting, \textit{``I felt like [P6-M2Wh2] knew what he was doing a lot\dots he seemed to jump right in and get to it\dots so I was fine with letting him go with his thing\dots I think we have the same general concept of what to do, his was just better and more advanced.''}

\subsubsection{Do our results apply to in-person pair programming?}

It's worth noting that the technical, creative, collaborative, and attitude differences we identified between same- and mixed-race participant pairs in the context of remote pair programming may indeed have relevance to in-person pair programming scenarios. In fact, these distinctions might be even more pronounced during face-to-face interactions, given the immediate and intense nature of in-person collaborations.

\subsubsection{What sociodemographic factors have a potential influence on our results?}

Individual sociodemographic factors can intersect and influence study outcomes. Notably, gender distinctions emerged, revealing that men pairs (average score: 34.3) demonstrated nearly double the productivity scores of women pairs (average score: 18.8). This disparity extended to their attitudes towards pair programming with test-driven development, as men reported a higher level of enjoyment (higher values for Q6, Q7, Q15-Q19, refer to Fig. \ref{fig: pp post study}.) compared to women. As observed in RQ1, same-race (WhWh) woman pairs experienced a decline in average self-efficacy, and they exhibited a paternalistic leadership style (RQ3). Previous pair programming studies have indicated that female students generally enjoy pair programming. In contrast, our findings suggest that women professional developers exhibited a degree of indifference toward this method. Further investigations are needed to understand and explore the reasons behind this disparity. Nuances in leadership styles emerged among men and women pairs, revealing that men tended to adopt a more authoritative approach, while women exhibited an equal distribution between authoritative and democratic styles (refer to Fig. \ref{fig:leadership figure}.). These findings align with previous research \cite{Kuttal2019, Lott2021}.

Personal experiences and education levels, particularly prior exposure to pair programming and test-driven development, can shape outcomes due to increased comfort and proficiency. Additionally, personal experience influences self-efficacy and leadership styles. In same-race (BB) pairs, an increase in self-efficacy was associated with more authoritative partners for (5/8) participants, while in mixed-race (BWh) pairs, (4/8) participants with increased self-efficacy assumed more democratic roles.

Although age can be a potential influencing factor, with older partners often assuming leadership roles \cite{leman2015groups, leman2005children}, our findings revealed no age influence on results, warranting further investigation.

Geographical location, influencing exposure to diversity, inclusion, and race, adds a layer of complexity that could potentially impact results. According to our findings, differences in geographic location had no influence on the results. We conjecture as all study participants were located in the US, suggesting minimal geographical impact on our results.

\subsection{Creating Tools, Interfaces, and Methodologies That Foster Racial Diversity}

Tools offer an efficient and effective means to educate individuals about race. In this context, we explore various tools that can enhance the experience of collaborating with individuals from diverse racial backgrounds. These implications support RQ1 - RQ4 (refer to Table \ref{tbl:summary of results}). We recommend:

 \begin{itemize}
\item \textit{\textbf{Empathetic facilitator and conversational agents for fostering inclusive collaboration:} }
Developing empathetic facilitator and conversational agents for pair programming holds promise for improving collaboration and fostering experiences of working with individuals from diverse racial backgrounds. Facilitator agents, are software systems designed to assist and support users in various tasks, such as decision-making and problem-solving \cite{ikeda2017generating}. The findings in Table \ref{tbl:summary of results} suggest that certain dimensions in pair programming, such as productivity, self-efficacy, creativity, leadership style, and role exchange, may be detectable by facilitator agents. These agents could provide empathetic and motivational feedback to help individual developers become proficient collaborators, encouraging them to adopt different perspectives.

Conversational agents have shown promise in pair programming contexts, as demonstrated in recent studies~\cite{Kuttal2021, Robe2022, Robe2020}. By creating conversational agents with the added feature of possessing characteristics of individuals from diverse racial groups, individuals can practice, learn to collaborate effectively, and educate one another. These agents can help increase comfort levels among individuals from diverse racial backgrounds during in-person collaborations, ultimately contributing to more inclusive and harmonious work environments to build cultural rapport. 

 \item \textbf{\textit{Leveraging video games for inclusive collaboration education:}} Video games has the potential to entertain and educate players about the significance of diversity and inclusion in collaborative ventures.
 Video games have demonstrated their effectiveness as tools for engaging, educating, and training individuals across various technical and social domains~\cite{squire2008video, janarthanan2012serious}. However, the gaming industry is grappling with concerns regarding the portrayal of Black characters and the stark underrepresentation of Black developers, comprising only 2\% of the game development community \cite{Browne_2020}. While some strides have been made in using games to explore Black history and address racial bias, there remains a significant gap in fostering collaboration among individuals from diverse racial backgrounds.
An innovative solution involves infusing diversity within video games by featuring characters from diverse racial and cultural backgrounds. These games can seamlessly integrate cultural elements, offering players valuable cultural insights during collaborative game-play without disrupting the overall gaming experience. 

  \item \textit{\textbf{Using virtual reality to increase racial empathy for collaboration:}} 
Virtual reality (VR) is a transformative technology that enhances our understanding and empathy for diverse racial experiences. VR has been used to create impactful simulations, like a VR film addressing racism's impact on Black individuals \cite{VRfilmRace}. Recently, VR enables embodied perspective-taking, allowing users to immerse themselves in various racial backgrounds, fostering empathy \cite{theriault2021bodyswapping}. VR experiences and games like 1000 Cut journey \cite{VHIL}, Traveling While Black \cite{Felix}, I Am A Man \cite{YouTube_2021}, and Greenwood Avenue \cite{YouTube_2020}, further aim to deepen empathy for Black individuals. VR has the potential to cultivate empathy by allowing individuals from both Black and White (or any) racial backgrounds to experience each other's perspectives when collaborating. 

  \item \textbf{\textit{Developing HCI methodologies infusing diversity of remote collaborative software:} }
   The absence of racially diverse teams in the tech industry can have detrimental effects on the development of software products and services. Therefore, as HCI researchers, it is crucial that we develop methodologies specifically tailored to identifying and mitigating racial biases in software, similar to the successful GenderMag approach, which addresses gender biases in software. GenderMag achieves this through personas and cognitive walkthroughs, offering software practitioners valuable insights into gender-based biases and user experiences \cite{Gendermag, Burnett2010, Burnett2016, Burnett2018}. 

   To systematically address racial biases in software, a ``race lens'' HCI methodology is needed. Such a methodology, which we might call ``RaceMag,'' could systematically uncover and rectify racial biases in software products. Our research has identified several factors that a race-focused methodology should consider, including leadership style, self-efficacy, role-exchange, communication patterns, and rapport building. A ``RaceMag'' approach can use these factors as facets of personas so that remote collaborative products can be analyzed for racial bias. By implementing ``RaceMag'', software development teams can actively strive to develop racially inclusive remote software products, even within industries characterized by unequal racial workforce representation.

 \end{itemize}

\subsection{Integrating Racial Awareness in Education and Workplaces}
Promoting racial awareness in K-12 education, university campuses, and workplaces is essential for fostering a harmonious, equitable, and prosperous society. Embracing diversity and reducing bias through inclusivity and open dialogue are crucial steps in bridging racial divides. These implications support RQ3 and RQ4 (refer to Table \ref{tbl:summary of results}). We recommend:

 \begin{itemize}
  \item \textbf{\textit{Promoting collaborative racial awareness through mandatory courses:} }
Our study revealed challenges faced by mixed-race participant pairs, such as lower comfort levels and increased anxiety during collaboration. While K-12 schools and universities strive to instill racial awareness and inclusivity \cite{washington2020twice}, these efforts often fall short in addressing collaboration difficulties due to race-related biases. For instance, P1-M2B2 shared the impact of these biases on performance, stating, \textit{``A lot of times you're automatically seen as less than, and there's not really anything you can do about it\dots it can definitely affect your performance.''}
To tackle this challenge, we propose the introduction of compulsory course content in schools explicitly designed to nurture collaboration among individuals from diverse racial backgrounds. Additionally, we recommend developing a collaboration activity called `Who are You?' inspired by the `Who am I?' game found in the race awareness game for teachers~\cite{RaceAwareness}. `Who am I?' provides an engaging platform for students to reflect upon and inquire about their own race, ethnicity, and various aspects of identity in a thoughtful and intriguing manner. We believe that `Who are You?' can offer similar benefits for peers. These educational initiatives are poised to play a crucial role in elevating comfort levels and fostering cultural rapport among students.

\item \textbf{\textit{Promoting collaborative racial awareness through workshops:} }
While workplace leaders are increasingly recognizing the importance of racial awareness and diversity, our analysis highlighted ongoing challenges in seamless collaboration among mixed racial groups. As expressed by P10-M1B1, a participant in a mixed-race pair, partnerships can initially feel tense, \textit{``At first, I was trying to feel out my partner. It was a bit strained on my end.''}
To address these challenges, we recommend implementing workshops specifically designed to create understanding and appreciation of cultural differences among individuals from various races, cultures, and genders. These workshops can take various forms, such as informal movie nights featuring race-related films, followed by reflective discussions. By fostering open dialogues about race, organizations can significantly improve collaboration among diverse racial groups in the workplace.

   \item \textit{\textbf{Enhancing minority recruitment with asynchronous technical interviews:}} 
   Asynchronous technical interviews offer a promising solution to improving the recruitment process for minority candidates. Traditional interviews often create anxiety and confusion due to unclear expectations~\cite{behroozi2020debugging}. In contrast, asynchronous interviews have demonstrated benefits, such as enhanced communication, reduced stress, and the preservation of technical problem-solving skills and code quality \cite{BehrooziFSE2022}. These advantages are likely to apply to minority candidates, contributing to more inclusive and equitable hiring practices.

\item \textit{\textbf{Remote work benefits for marginalized groups to alleviate anxiety:}}
Remote work environments can also alleviate anxiety, as individuals have the option to turn off their cameras. As participant P2-M2B2 noted remote work, \textit{``seems a little bit more protective.''} Recent studies have highlighted the substantial advantages of remote and hybrid work structures in the software industry, particularly for professionals from marginalized groups, including caregivers~\cite{ralph2020pandemic, santos2022grounded}, individuals with disabilities~\cite{santos2023benefits}, and LGBTQIA+ individuals~\cite{ford2019remote}. Remote work empowers them to openly express their identities and provides autonomy in their interactions, granting greater control over their work experiences.

 \end{itemize}

%% file: conclusion.tex
\section{Conclusion}

Our study represents a pioneering effort of investigating race dynamics in Human-Computer Interaction (HCI) and Software Engineering (SE) literature, specifically within the context of pair programming. We examine both same- and mixed-race pairs of professional developers, aiming to capture the voices of minority developers, including Black men, Black women, and White women, within the context of collaborative coding.
Through a comprehensive analysis, combining qualitative and quantitative methods, with a focus on race, our study reveals the following key findings:
\begin{itemize}

   \item \textbf{\textit{RQ1: Technical Aspects:}} Mixed-race pairs demonstrated higher productivity, while code quality was consistent across all same- and mixed-race pairs. Same-race (BB) pairs exhibited increased self-efficacy, while same-race (WhWh) pairs reported a decrease. Meanwhile, mixed-race pairs demonstrated the advantages of diverse perspectives, resulting in a variety of ideas.

\item \textbf{\textit{RQ2: Creativity: }}In same-race pairs, creative problem solving predominantly originated from the driver, while mixed-race pairs excelled in evenly distributing these intellectual contributions.

\item \textit{\textbf{RQ3: Collaboration: }}Same-race pairs often displayed imbalanced decision-making and role exchange dynamics, where one partner assumed an authoritative role while the other disengaged from the task. Same-race (BB) women pairs and same-race (WhWh) pairs, regardless of gender, frequently took on instructional roles within their partnerships. In contrast, mixed-race pairs excelled in leadership through collaborative decision-making and active role-exchange.

\item \textit{\textbf{RQ4: Attitude: }}Awareness of partners' racial backgrounds significantly influenced interaction dynamics. Same-race pairs reported fluid communication, high comfort levels, rapport, and a sense of connection. Mixed-race pairs, however, faced communication challenges and occasional anxiety, but they exhibited empathy.

\end{itemize}

These results have several implications for the HCI and SE field, including the design of empathetic agents, the development of HCI methodologies, applications in virtual reality (VR) and video games, and the enhancement of coursework and workshops rooted in inclusive collaboration.
Our study underscores the nuanced impact of diversity in pair programming interactions and highlights its potential for both positive and negative effects.  As Maya Angelou beautifully stated, \textit{``We all should know that diversity makes for a rich tapestry, and we must understand that all the threads of the tapestry are equal in value no matter their color.''} Our research strives to contribute to weaving this rich tapestry in the world of technology and collaboration.

%% file: main.bbl

\begin{thebibliography}{270}


\ifx \showCODEN    \undefined \def \showCODEN     #1{\unskip}     \fi
\ifx \showDOI      \undefined \def \showDOI       #1{#1}\fi
\ifx \showISBNx    \undefined \def \showISBNx     #1{\unskip}     \fi
\ifx \showISBNxiii \undefined \def \showISBNxiii  #1{\unskip}     \fi
\ifx \showISSN     \undefined \def \showISSN      #1{\unskip}     \fi
\ifx \showLCCN     \undefined \def \showLCCN      #1{\unskip}     \fi
\ifx \shownote     \undefined \def \shownote      #1{#1}          \fi
\ifx \showarticletitle \undefined \def \showarticletitle #1{#1}   \fi
\ifx \showURL      \undefined \def \showURL       {\relax}        \fi
\providecommand\bibfield[2]{#2}
\providecommand\bibinfo[2]{#2}
\providecommand\natexlab[1]{#1}
\providecommand\showeprint[2][]{arXiv:#2}

\bibitem[dre({[n.\,d.]})]%
        {dreamsdef}
 \bibinfo{year}{[n.\,d.]}\natexlab{}.
\newblock
\newblock
\urldef\tempurl%
\url{https://ips-dc.org/wp-content/uploads/2019/01/IPS_RWD-Report_FINAL-1.15.19.pdf}
\showURL{%
\tempurl}


\bibitem[col({[n.\,d.]})]%
        {college}
 \bibinfo{year}{[n.\,d.]}\natexlab{}.
\newblock
\newblock
\urldef\tempurl%
\url{https://www.census.gov/content/dam/Census/library/publications/2016/demo/p20-578.pdf}
\showURL{%
\tempurl}


\bibitem[pri({[n.\,d.]})]%
        {prison2017}
 \bibinfo{year}{[n.\,d.]}\natexlab{}.
\newblock
\newblock
\urldef\tempurl%
\url{https://bjs.ojp.gov/content/pub/pdf/p17.pdf}
\showURL{%
\tempurl}


\bibitem[Con({[n.\,d.]})]%
        {ConstitutionalRights}
 \bibinfo{year}{[n.\,d.]}\natexlab{}.
\newblock
\newblock
\urldef\tempurl%
\url{https://www.crf-usa.org/black-history-month/the-constitution-and-slavery}
\showURL{%
\tempurl}


\bibitem[Goo({[n.\,d.]})]%
        {GoogleDrive}
 \bibinfo{year}{[n.\,d.]}\natexlab{}.
\newblock
\newblock
\urldef\tempurl%
\url{https://drive.google.com/drive/folders/1Sb8k9_2AzJnGK97s0vL8e-2fJ8gaPu5w?usp=drive_link}
\showURL{%
\tempurl}


\bibitem[VHI({[n.\,d.]})]%
        {VHIL}
 \bibinfo{year}{[n.\,d.]}\natexlab{}.
\newblock
\newblock
\urldef\tempurl%
\url{https://stanfordvr.com/1000cut/}
\showURL{%
\tempurl}


\bibitem[ind(2015)]%
        {independent_2015}
 \bibinfo{year}{2015}\natexlab{}.
\newblock \bibinfo{title}{Google's algorithm shows prestigious job ads to men, but not to women}.
\newblock
\newblock
\urldef\tempurl%
\url{https://www.independent.co.uk/tech}
\showURL{%
\tempurl}


\bibitem[Tim(2019)]%
        {Times_2019}
 \bibinfo{year}{2019}\natexlab{}.
\newblock
\newblock
\urldef\tempurl%
\url{https://www.nytimes.com/interactive/2019/08/14/magazine/1619-america-slavery.html?mtrref=undefined&amp;gwh=AA520C829192CC8AF8894579A780882E&amp;gwt=pay&amp;assetType=PAYWALL}
\showURL{%
\tempurl}


\bibitem[You(2020)]%
        {YouTube_2020}
 \bibinfo{year}{2020}\natexlab{}.
\newblock
\newblock
\urldef\tempurl%
\url{https://www.youtube.com/watch?v=fMxU4D0pO6Q}
\showURL{%
\tempurl}


\bibitem[Gen(2020)]%
        {Gendermag}
 \bibinfo{year}{2020}\natexlab{}.
\newblock \bibinfo{title}{GenderMag}.
\newblock
\newblock
\urldef\tempurl%
\url{http://gendermag.org/}
\showURL{%
\tempurl}


\bibitem[You(2021)]%
        {YouTube_2021}
 \bibinfo{year}{2021}\natexlab{}.
\newblock
\newblock
\urldef\tempurl%
\url{https://www.youtube.com/watch?v=GkvxHnC7Zzo}
\showURL{%
\tempurl}


\bibitem[Adeliyi et~al\mbox{.}(2021)]%
        {adeliyi2021investigating}
\bibfield{author}{\bibinfo{person}{Adeola Adeliyi}, \bibinfo{person}{Michel Wermelinger}, \bibinfo{person}{Karen Kear}, {and} \bibinfo{person}{Jon Rosewell}.} \bibinfo{year}{2021}\natexlab{}.
\newblock \showarticletitle{Investigating Remote Pair Programming In Part-Time Distance Education}. In \bibinfo{booktitle}{\emph{Proceedings of the 2021 Conference on United Kingdom \& Ireland Computing Education Research}}. \bibinfo{pages}{1--7}.
\newblock


\bibitem[{\AA}gren et~al\mbox{.}(2022)]%
        {aagren2022agile}
\bibfield{author}{\bibinfo{person}{Pernilla {\AA}gren}, \bibinfo{person}{Eli Knoph}, {and} \bibinfo{person}{Richard Berntsson~Svensson}.} \bibinfo{year}{2022}\natexlab{}.
\newblock \showarticletitle{Agile software development one year into the COVID-19 pandemic}.
\newblock \bibinfo{journal}{\emph{Empirical Software Engineering}} \bibinfo{volume}{27}, \bibinfo{number}{6} (\bibinfo{year}{2022}), \bibinfo{pages}{121}.
\newblock


\bibitem[Aish et~al\mbox{.}(2018)]%
        {aish2018people}
\bibfield{author}{\bibinfo{person}{Nir Aish}, \bibinfo{person}{Philip Asare}, {and} \bibinfo{person}{Elif~Eda Miskio{\u{g}}lu}.} \bibinfo{year}{2018}\natexlab{}.
\newblock \showarticletitle{People like me: Providing relatable and realistic role models for underrepresented minorities in STEM to increase their motivation and likelihood of success}. In \bibinfo{booktitle}{\emph{2018 IEEE integrated STEM education conference (ISEC)}}. IEEE, \bibinfo{pages}{83--89}.
\newblock


\bibitem[Alegria and Branch(2015)]%
        {alegria2015causes}
\bibfield{author}{\bibinfo{person}{Sharla~N Alegria} {and} \bibinfo{person}{Enobong~Hannah Branch}.} \bibinfo{year}{2015}\natexlab{}.
\newblock \showarticletitle{Causes and Consequences of Inequality in the STEM: Diversity and its Discontents}.
\newblock \bibinfo{journal}{\emph{International Journal of Gender, Science and Technology}} \bibinfo{volume}{7}, \bibinfo{number}{3} (\bibinfo{year}{2015}), \bibinfo{pages}{321--342}.
\newblock


\bibitem[ALTUN and MAZMAN(2012)]%
        {altun2012programlamaya}
\bibfield{author}{\bibinfo{person}{Arif ALTUN} {and} \bibinfo{person}{Sacide~G{\"u}zin MAZMAN}.} \bibinfo{year}{2012}\natexlab{}.
\newblock \showarticletitle{Programlamaya ili{\c{s}}kin {\"o}z yeterlilik alg{\i}s{\i} {\"o}l{\c{c}}e{\u{g}}inin T{\"u}rk{\c{c}}e formunun g{\"u}venirlik ve ge{\c{c}}erlik {\c{c}}al{\i}{\c{s}}mas{\i}}.
\newblock \bibinfo{journal}{\emph{Journal of Measurement and Evaluation in Education and Psychology}} \bibinfo{volume}{3}, \bibinfo{number}{2} (\bibinfo{year}{2012}), \bibinfo{pages}{297--308}.
\newblock


\bibitem[Angwin et~al\mbox{.}(2016)]%
        {angwin_larson_kirchner_mattu_2016}
\bibfield{author}{\bibinfo{person}{Julia Angwin}, \bibinfo{person}{Jeff Larson}, \bibinfo{person}{Lauren Kirchner}, {and} \bibinfo{person}{Surya Mattu}.} \bibinfo{year}{2016}\natexlab{}.
\newblock \bibinfo{title}{Machine bias}.
\newblock
\newblock
\urldef\tempurl%
\url{https://www.propublica.org/article/machine-bias-risk-assessments-in-criminal-sentencing}
\showURL{%
\tempurl}


\bibitem[Anonymous(2019)]%
        {VRfilmRace}
\bibfield{author}{\bibinfo{person}{Anonymous}.} \bibinfo{year}{2019}\natexlab{}.
\newblock \bibinfo{title}{The Reality of Racism Comes to Life in VR Film}.
\newblock
\newblock
\urldef\tempurl%
\url{https://news.columbia.edu/news/reality-racism-comes-life-vr-film}
\showURL{%
\tempurl}


\bibitem[Anonymous(2023)]%
        {RaceAwareness}
\bibfield{author}{\bibinfo{person}{Anonymous}.} \bibinfo{year}{2023}\natexlab{}.
\newblock \bibinfo{title}{Who Am I? Race Awareness Game}.
\newblock
\newblock
\urldef\tempurl%
\url{https://www.commonsense.org/education/reviews/who-am-i-race-awareness-game#:~:text=Who%20Am%20I%3F%20is%20a,that%20make%20sense%20to%20themselves.}
\showURL{%
\tempurl}


\bibitem[Archibald et~al\mbox{.}(2019)]%
        {archibald2019using}
\bibfield{author}{\bibinfo{person}{Mandy~M Archibald}, \bibinfo{person}{Rachel~C Ambagtsheer}, \bibinfo{person}{Mavourneen~G Casey}, {and} \bibinfo{person}{Michael Lawless}.} \bibinfo{year}{2019}\natexlab{}.
\newblock \showarticletitle{Using zoom videoconferencing for qualitative data collection: perceptions and experiences of researchers and participants}.
\newblock \bibinfo{journal}{\emph{International journal of qualitative methods}}  \bibinfo{volume}{18} (\bibinfo{year}{2019}), \bibinfo{pages}{1609406919874596}.
\newblock


\bibitem[Aries(1987)]%
        {aries1987gender}
\bibfield{author}{\bibinfo{person}{Elizabeth Aries}.} \bibinfo{year}{1987}\natexlab{}.
\newblock \bibinfo{booktitle}{\emph{Gender and communication.}}
\newblock \bibinfo{publisher}{Sage Publications, Inc}.
\newblock


\bibitem[Ashe and Rampersad(1993)]%
        {ashe1993days}
\bibfield{author}{\bibinfo{person}{Arthur Ashe} {and} \bibinfo{person}{A Rampersad}.} \bibinfo{year}{1993}\natexlab{}.
\newblock \bibinfo{title}{Days ofgrace: A memoir}.
\newblock
\newblock


\bibitem[Attridge(2019)]%
        {attridge2019global}
\bibfield{author}{\bibinfo{person}{Mark Attridge}.} \bibinfo{year}{2019}\natexlab{}.
\newblock \bibinfo{title}{A global perspective on promoting workplace mental health and the role of employee assistance programs}.
\newblock , \bibinfo{numpages}{622--629}~pages.
\newblock


\bibitem[Auerbach and Silverstein(2003)]%
        {auerbach2003qualitative}
\bibfield{author}{\bibinfo{person}{Carl Auerbach} {and} \bibinfo{person}{Louise~B Silverstein}.} \bibinfo{year}{2003}\natexlab{}.
\newblock \bibinfo{booktitle}{\emph{Qualitative data: An introduction to coding and analysis}}. Vol.~\bibinfo{volume}{21}.
\newblock \bibinfo{publisher}{NYU press}.
\newblock


\bibitem[B.~A.~Kitchenham and Rosenberg(2002)]%
        {Kitchenham2002}
\bibfield{author}{\bibinfo{person}{D.~C.~Hoaglin B.~A.~Kitchenham, S. L.~Pfleeger} {and} \bibinfo{person}{J. Rosenberg}.} \bibinfo{year}{2002}\natexlab{}.
\newblock \showarticletitle{"Preliminary Guidelines for Empirical Research in Software Engineering"}. In \bibinfo{booktitle}{\emph{IEEE Transactions on Software Engineering}}, Vol.~\bibinfo{volume}{28}. \bibinfo{pages}{721--734}.
\newblock


\bibitem[Balijepally et~al\mbox{.}(2009)]%
        {balijepally2009two}
\bibfield{author}{\bibinfo{person}{VenuGopal Balijepally}, \bibinfo{person}{RadhaKanta Mahapatra}, \bibinfo{person}{Sridhar Nerur}, {and} \bibinfo{person}{Kenneth~H Price}.} \bibinfo{year}{2009}\natexlab{}.
\newblock \showarticletitle{Are two heads better than one for software development? The productivity paradox of pair programming}.
\newblock \bibinfo{journal}{\emph{MIS quarterly}} (\bibinfo{year}{2009}), \bibinfo{pages}{91--118}.
\newblock


\bibitem[Barenboim(1981)]%
        {barenboim1981development}
\bibfield{author}{\bibinfo{person}{Carl Barenboim}.} \bibinfo{year}{1981}\natexlab{}.
\newblock \showarticletitle{The development of person perception in childhood and adolescence: From behavioral comparisons to psychological constructs to psychological comparisons}.
\newblock \bibinfo{journal}{\emph{Child development}} (\bibinfo{year}{1981}), \bibinfo{pages}{129--144}.
\newblock


\bibitem[Barton(2003)]%
        {barton2003parsing}
\bibfield{author}{\bibinfo{person}{Paul~E Barton}.} \bibinfo{year}{2003}\natexlab{}.
\newblock \showarticletitle{Parsing the Achievement Gap: Baselines for Tracking Progress. Policy Information Report.}
\newblock  (\bibinfo{year}{2003}).
\newblock


\bibitem[Bass et~al\mbox{.}(1996)]%
        {bass1996transformational}
\bibfield{author}{\bibinfo{person}{Bernard~M Bass}, \bibinfo{person}{Bruce~J Avolio}, {and} \bibinfo{person}{Leanne Atwater}.} \bibinfo{year}{1996}\natexlab{}.
\newblock \showarticletitle{The transformational and transactional leadership of men and women}.
\newblock \bibinfo{journal}{\emph{Applied psychology}} \bibinfo{volume}{45}, \bibinfo{number}{1} (\bibinfo{year}{1996}), \bibinfo{pages}{5--34}.
\newblock


\bibitem[Bass and Bass(2008)]%
        {bass2008handbook}
\bibfield{author}{\bibinfo{person}{Bernard~M Bass} {and} \bibinfo{person}{Ruth Bass}.} \bibinfo{year}{2008}\natexlab{}.
\newblock \bibinfo{booktitle}{\emph{Handbook of leadership: Theory, research, and application}}.
\newblock \bibinfo{publisher}{Free Press}.
\newblock


\bibitem[Bass and Stogdill(1990)]%
        {bass1990bass}
\bibfield{author}{\bibinfo{person}{Bernard~M Bass} {and} \bibinfo{person}{Ralph~Melvin Stogdill}.} \bibinfo{year}{1990}\natexlab{}.
\newblock \bibinfo{booktitle}{\emph{Bass \& Stogdill's handbook of leadership: Theory, research, and managerial applications}}.
\newblock \bibinfo{publisher}{Simon and Schuster}.
\newblock


\bibitem[Bass et~al\mbox{.}(2007)]%
        {bass2007collaboration}
\bibfield{author}{\bibinfo{person}{Matthew Bass}, \bibinfo{person}{James~D Herbsleb}, {and} \bibinfo{person}{Christian Lescher}.} \bibinfo{year}{2007}\natexlab{}.
\newblock \showarticletitle{Collaboration in global software projects at siemens: An experience report}. In \bibinfo{booktitle}{\emph{International Conference on Global Software Engineering (ICGSE 2007)}}. IEEE, \bibinfo{pages}{33--39}.
\newblock


\bibitem[Bauer and Green(1996)]%
        {bauer1996development}
\bibfield{author}{\bibinfo{person}{Talya~N Bauer} {and} \bibinfo{person}{Stephen~G Green}.} \bibinfo{year}{1996}\natexlab{}.
\newblock \showarticletitle{Development of leader-member exchange: A longitudinal test}.
\newblock \bibinfo{journal}{\emph{Academy of management journal}} \bibinfo{volume}{39}, \bibinfo{number}{6} (\bibinfo{year}{1996}), \bibinfo{pages}{1538--1567}.
\newblock


\bibitem[Beck(2022)]%
        {beck2022test}
\bibfield{author}{\bibinfo{person}{Kent Beck}.} \bibinfo{year}{2022}\natexlab{}.
\newblock \bibinfo{booktitle}{\emph{Test driven development: By example}}.
\newblock \bibinfo{publisher}{Addison-Wesley Professional}.
\newblock


\bibitem[Becker(2003)]%
        {becker2003grading}
\bibfield{author}{\bibinfo{person}{Katrin Becker}.} \bibinfo{year}{2003}\natexlab{}.
\newblock \showarticletitle{Grading programming assignments using rubrics}. In \bibinfo{booktitle}{\emph{Proceedings of the 8th annual conference on Innovation and technology in computer science education}}. \bibinfo{pages}{253--253}.
\newblock


\bibitem[Begel and Nagappan(2008a)]%
        {begel2008}
\bibfield{author}{\bibinfo{person}{Andrew Begel} {and} \bibinfo{person}{Nachi Nagappan}.} \bibinfo{year}{2008}\natexlab{a}.
\newblock \showarticletitle{{Pair programming: What's in it For Me?}}. In \bibinfo{booktitle}{\emph{{ESEM '08: Proceedings of the Second ACM-IEEE International Symposium on Empirical Software Engineering and Measurement}} (\bibinfo{edition}{{ESEM '08: Proceedings of the Second ACM-IEEE International Symposium on Empirical Software Engineering and Measurement}} ed.)}. \bibinfo{publisher}{ACM}, \bibinfo{pages}{120--128}.
\newblock
\showISBNx{978-1-59593-971-5}


\bibitem[Begel and Nagappan(2008b)]%
        {begel2008pair}
\bibfield{author}{\bibinfo{person}{Andrew Begel} {and} \bibinfo{person}{Nachiappan Nagappan}.} \bibinfo{year}{2008}\natexlab{b}.
\newblock \showarticletitle{Pair programming: what's in it for me?}. In \bibinfo{booktitle}{\emph{Proceedings of the Second ACM-IEEE international symposium on Empirical software engineering and measurement}}. \bibinfo{pages}{120--128}.
\newblock


\bibitem[Behroozi et~al\mbox{.}(2022)]%
        {BehrooziFSE2022}
\bibfield{author}{\bibinfo{person}{Mahnaz Behroozi}, \bibinfo{person}{Chris Parnin}, {and} \bibinfo{person}{Chris Brown}.} \bibinfo{year}{2022}\natexlab{}.
\newblock \showarticletitle{Asynchronous Technical Interviews: Reducing the Effect of Supervised Think-Aloud on Communication Ability}. In \bibinfo{booktitle}{\emph{Proceedings of the 30th ACM Joint European Software Engineering Conference and Symposium on the Foundations of Software Engineering}} (Singapore, Singapore) \emph{(\bibinfo{series}{ESEC/FSE 2022})}. \bibinfo{publisher}{Association for Computing Machinery}, \bibinfo{address}{New York, NY, USA}, \bibinfo{pages}{294–305}.
\newblock
\showISBNx{9781450394130}
\urldef\tempurl%
\url{https://doi.org/10.1145/3540250.3549168}
\showDOI{\tempurl}


\bibitem[Behroozi et~al\mbox{.}(2020)]%
        {behroozi2020debugging}
\bibfield{author}{\bibinfo{person}{Mahnaz Behroozi}, \bibinfo{person}{Shivani Shirolkar}, \bibinfo{person}{Titus Barik}, {and} \bibinfo{person}{Chris Parnin}.} \bibinfo{year}{2020}\natexlab{}.
\newblock \showarticletitle{Debugging hiring: What went right and what went wrong in the technical interview process}. In \bibinfo{booktitle}{\emph{Proceedings of the ACM/IEEE 42nd International Conference on Software Engineering: Software Engineering in Society}}. \bibinfo{pages}{71--80}.
\newblock


\bibitem[Beilock et~al\mbox{.}(2007)]%
        {beilock2007stereotype}
\bibfield{author}{\bibinfo{person}{Sian~L Beilock}, \bibinfo{person}{Robert~J Rydell}, {and} \bibinfo{person}{Allen~R McConnell}.} \bibinfo{year}{2007}\natexlab{}.
\newblock \showarticletitle{Stereotype threat and working memory: mechanisms, alleviation, and spillover.}
\newblock \bibinfo{journal}{\emph{Journal of Experimental Psychology: General}} \bibinfo{volume}{136}, \bibinfo{number}{2} (\bibinfo{year}{2007}), \bibinfo{pages}{256}.
\newblock


\bibitem[Bergen and Labont{\'e}(2020)]%
        {bergen2020everything}
\bibfield{author}{\bibinfo{person}{Nicole Bergen} {and} \bibinfo{person}{Ronald Labont{\'e}}.} \bibinfo{year}{2020}\natexlab{}.
\newblock \showarticletitle{“Everything is perfect, and we have no problems”: detecting and limiting social desirability bias in qualitative research}.
\newblock \bibinfo{journal}{\emph{Qualitative health research}} \bibinfo{volume}{30}, \bibinfo{number}{5} (\bibinfo{year}{2020}), \bibinfo{pages}{783--792}.
\newblock


\bibitem[Bevan et~al\mbox{.}(2002)]%
        {bevan2002guidelines}
\bibfield{author}{\bibinfo{person}{Jennifer Bevan}, \bibinfo{person}{Linda Werner}, {and} \bibinfo{person}{Charlie McDowell}.} \bibinfo{year}{2002}\natexlab{}.
\newblock \showarticletitle{Guidelines for the use of pair programming in a freshman programming class}. In \bibinfo{booktitle}{\emph{Proceedings 15th Conference on Software Engineering Education and Training (CSEE\&T 2002)}}. IEEE, \bibinfo{pages}{100--107}.
\newblock


\bibitem[Bipp et~al\mbox{.}(2008)]%
        {Tanja2008}
\bibfield{author}{\bibinfo{person}{Tanja Bipp}, \bibinfo{person}{Andreas Lepper}, {and} \bibinfo{person}{Doris Schmedding}.} \bibinfo{year}{2008}\natexlab{}.
\newblock \showarticletitle{Pair Programming in Software Development Teams - An Empirical Study of Its Benefits}.
\newblock \bibinfo{journal}{\emph{Inf. Softw. Technol.}} \bibinfo{volume}{50}, \bibinfo{number}{3} (\bibinfo{date}{Feb.} \bibinfo{year}{2008}), \bibinfo{pages}{231--240}.
\newblock
\showISSN{0950-5849}
\urldef\tempurl%
\url{https://doi.org/10.1016/j.infsof.2007.05.006}
\showDOI{\tempurl}


\bibitem[Blascovich et~al\mbox{.}(2001)]%
        {blascovich2001african}
\bibfield{author}{\bibinfo{person}{Jim Blascovich}, \bibinfo{person}{Steven~J Spencer}, \bibinfo{person}{Diane Quinn}, {and} \bibinfo{person}{Claude Steele}.} \bibinfo{year}{2001}\natexlab{}.
\newblock \showarticletitle{African Americans and high blood pressure: The role of stereotype threat}.
\newblock \bibinfo{journal}{\emph{Psychological science}} \bibinfo{volume}{12}, \bibinfo{number}{3} (\bibinfo{year}{2001}), \bibinfo{pages}{225--229}.
\newblock


\bibitem[Boccard et~al\mbox{.}(1999)]%
        {boccard1999racial}
\bibfield{author}{\bibinfo{person}{Nicolas Boccard}, \bibinfo{person}{Yves Zenou}, {et~al\mbox{.}}} \bibinfo{year}{1999}\natexlab{}.
\newblock \bibinfo{booktitle}{\emph{Racial discrimination and Redlining in cities}}.
\newblock \bibinfo{type}{{T}echnical {R}eport}. \bibinfo{institution}{Universit{\'e} catholique de Louvain, Center for Operations, Research \& Econometrics}.
\newblock


\bibitem[Bodaker and Rosenberg-Kima(2023)]%
        {bodaker2023online}
\bibfield{author}{\bibinfo{person}{Liat Bodaker} {and} \bibinfo{person}{Rinat~B Rosenberg-Kima}.} \bibinfo{year}{2023}\natexlab{}.
\newblock \showarticletitle{Online pair-programming: Elementary school children learning scratch together online}.
\newblock \bibinfo{journal}{\emph{Journal of Research on Technology in Education}} \bibinfo{volume}{55}, \bibinfo{number}{5} (\bibinfo{year}{2023}), \bibinfo{pages}{799--816}.
\newblock


\bibitem[Bogen(2023)]%
        {bogen2023racial}
\bibfield{author}{\bibinfo{person}{David~S Bogen}.} \bibinfo{year}{2023}\natexlab{}.
\newblock \showarticletitle{From Racial Discrimination to Separate but Equal: The Common Law Impact of the Thirteenth Amendment}.
\newblock \bibinfo{journal}{\emph{Ohio Northern University Law Review}} \bibinfo{volume}{38}, \bibinfo{number}{1} (\bibinfo{year}{2023}), \bibinfo{pages}{3}.
\newblock


\bibitem[Bosson et~al\mbox{.}(2004)]%
        {bosson2004saying}
\bibfield{author}{\bibinfo{person}{Jennifer~K Bosson}, \bibinfo{person}{Ethan~L Haymovitz}, {and} \bibinfo{person}{Elizabeth~C Pinel}.} \bibinfo{year}{2004}\natexlab{}.
\newblock \showarticletitle{When saying and doing diverge: The effects of stereotype threat on self-reported versus non-verbal anxiety}.
\newblock \bibinfo{journal}{\emph{Journal of experimental social psychology}} \bibinfo{volume}{40}, \bibinfo{number}{2} (\bibinfo{year}{2004}), \bibinfo{pages}{247--255}.
\newblock


\bibitem[Bowker and Star(2000)]%
        {bowker2000sorting}
\bibfield{author}{\bibinfo{person}{Geoffrey~C Bowker} {and} \bibinfo{person}{Susan~Leigh Star}.} \bibinfo{year}{2000}\natexlab{}.
\newblock \bibinfo{booktitle}{\emph{Sorting things out: Classification and its consequences}}.
\newblock \bibinfo{publisher}{MIT press}.
\newblock


\bibitem[Bowman et~al\mbox{.}(2019)]%
        {bowman2019prior}
\bibfield{author}{\bibinfo{person}{Nicholas~A Bowman}, \bibinfo{person}{Lindsay Jarratt}, \bibinfo{person}{KC Culver}, {and} \bibinfo{person}{Alberto~Maria Segre}.} \bibinfo{year}{2019}\natexlab{}.
\newblock \showarticletitle{How prior programming experience affects students' pair programming experiences and outcomes}. In \bibinfo{booktitle}{\emph{Proceedings of the 2019 ACM Conference on innovation and technology in computer science education}}. \bibinfo{pages}{170--175}.
\newblock


\bibitem[Bowman et~al\mbox{.}(2021)]%
        {bowman2021impact}
\bibfield{author}{\bibinfo{person}{Nicholas~A Bowman}, \bibinfo{person}{Lindsay Jarratt}, \bibinfo{person}{KC Culver}, {and} \bibinfo{person}{Alberto~M Segre}.} \bibinfo{year}{2021}\natexlab{}.
\newblock \showarticletitle{The impact of pair programming on college students’ interest, perceptions, and achievement in computer science}.
\newblock \bibinfo{journal}{\emph{ACM Transactions on Computing Education}} \bibinfo{volume}{21}, \bibinfo{number}{3} (\bibinfo{year}{2021}), \bibinfo{pages}{1--19}.
\newblock


\bibitem[Braught et~al\mbox{.}(2011)]%
        {braught2011case}
\bibfield{author}{\bibinfo{person}{Grant Braught}, \bibinfo{person}{Tim Wahls}, {and} \bibinfo{person}{L~Marlin Eby}.} \bibinfo{year}{2011}\natexlab{}.
\newblock \showarticletitle{The case for pair programming in the computer science classroom}.
\newblock \bibinfo{journal}{\emph{ACM Transactions on Computing Education (TOCE)}} \bibinfo{volume}{11}, \bibinfo{number}{1} (\bibinfo{year}{2011}), \bibinfo{pages}{1--21}.
\newblock


\bibitem[Brown(2009)]%
        {Brown2009}
\bibfield{author}{\bibinfo{person}{Tim Brown}.} \bibinfo{year}{2009}\natexlab{}.
\newblock \bibinfo{booktitle}{\emph{Change by Design: How Design Thinking Transforms Organizations and Inspires Innovation}}.
\newblock \bibinfo{publisher}{HarperBusiness}.
\newblock
\showISBNx{9780061766084}


\bibitem[Browne(2000)]%
        {browne2000latinas}
\bibfield{author}{\bibinfo{person}{Irene Browne}.} \bibinfo{year}{2000}\natexlab{}.
\newblock \bibinfo{booktitle}{\emph{Latinas and African American women at work: Race, gender, and economic inequality}}.
\newblock \bibinfo{publisher}{Russell Sage Foundation}.
\newblock


\bibitem[Browne(2020)]%
        {Browne_2020}
\bibfield{author}{\bibinfo{person}{Ryan Browne}.} \bibinfo{year}{2020}\natexlab{}.
\newblock \bibinfo{title}{The 150 billion video game industry grapples with a murky track record on diversity}.
\newblock
\newblock
\urldef\tempurl%
\url{https://www.cnbc.com/2020/08/14/video-game-industry-grapples-with-murky-track-record-on-diversity.html}
\showURL{%
\tempurl}


\bibitem[Bryant(2004)]%
        {bryant2004double}
\bibfield{author}{\bibinfo{person}{Sallyann Bryant}.} \bibinfo{year}{2004}\natexlab{}.
\newblock \showarticletitle{Double trouble: Mixing qualitative and quantitative methods in the study of extreme programmers}. In \bibinfo{booktitle}{\emph{2004 IEEE symposium on visual languages-human centric computing}}. IEEE, \bibinfo{pages}{55--61}.
\newblock


\bibitem[Bryant et~al\mbox{.}(2006)]%
        {bryant2006pair}
\bibfield{author}{\bibinfo{person}{Sallyann Bryant}, \bibinfo{person}{Pablo Romero}, {and} \bibinfo{person}{Benedict Du~Boulay}.} \bibinfo{year}{2006}\natexlab{}.
\newblock \showarticletitle{Pair programming and the re-appropriation of individual tools for collaborative software development}.
\newblock \bibinfo{journal}{\emph{Frontiers in Artificial Intelligence and Applications}}  \bibinfo{volume}{137} (\bibinfo{year}{2006}), \bibinfo{pages}{55}.
\newblock


\bibitem[Buolamwini and Gebru(2018)]%
        {buolamwini2018gender}
\bibfield{author}{\bibinfo{person}{Joy Buolamwini} {and} \bibinfo{person}{Timnit Gebru}.} \bibinfo{year}{2018}\natexlab{}.
\newblock \showarticletitle{Gender shades: Intersectional accuracy disparities in commercial gender classification}. In \bibinfo{booktitle}{\emph{Conference on fairness, accountability and transparency}}. PMLR, \bibinfo{pages}{77--91}.
\newblock


\bibitem[Bureau(2022)]%
        {Bureau_2022}
\bibfield{author}{\bibinfo{person}{US~Census Bureau}.} \bibinfo{year}{2022}\natexlab{}.
\newblock \bibinfo{title}{About the topic of race}.
\newblock
\newblock
\urldef\tempurl%
\url{https://www.census.gov/topics/population/race/about.html}
\showURL{%
\tempurl}


\bibitem[Burke et~al\mbox{.}(2022)]%
        {burke2022state}
\bibfield{author}{\bibinfo{person}{Amy Burke}, \bibinfo{person}{Abigail Okrent}, \bibinfo{person}{Katherine Hale}, {and} \bibinfo{person}{Nancy Gough}.} \bibinfo{year}{2022}\natexlab{}.
\newblock \showarticletitle{The State of US Science \& Engineering 2022. National Science Board Science \& Engineering Indicators. NSB-2022-1.}
\newblock \bibinfo{journal}{\emph{National Science Foundation}} (\bibinfo{year}{2022}).
\newblock


\bibitem[Burnett et~al\mbox{.}(2010)]%
        {Burnett2010}
\bibfield{author}{\bibinfo{person}{Margaret Burnett}, \bibinfo{person}{Scott~D. Fleming}, \bibinfo{person}{Shamsi Iqbal}, \bibinfo{person}{Gina Venolia}, \bibinfo{person}{Vidya Rajaram}, \bibinfo{person}{Umer Farooq}, \bibinfo{person}{Valentina Grigoreanu}, {and} \bibinfo{person}{Mary Czerwinski}.} \bibinfo{year}{2010}\natexlab{}.
\newblock \showarticletitle{{Gender Differences and Programming Environments: Across Programming Populations}}. In \bibinfo{booktitle}{\emph{{Proceedings of the 2010 ACM-IEEE International Symposium on Empirical Software Engineering and Measurement}}} \emph{(\bibinfo{series}{ESEM '10})}. \bibinfo{publisher}{ACM}, \bibinfo{pages}{28:1--28:10}.
\newblock
\showISBNx{978-1-4503-0039-1}


\bibitem[Burnett et~al\mbox{.}(2016)]%
        {Burnett2016}
\bibfield{author}{\bibinfo{person}{Margaret Burnett}, \bibinfo{person}{Simone Stumpf}, \bibinfo{person}{Jamie Macbeth}, \bibinfo{person}{Stephann Makri}, \bibinfo{person}{Laura Beckwith}, \bibinfo{person}{Irwin Kwan}, \bibinfo{person}{Anicia Peters}, {and} \bibinfo{person}{William Jernigan}.} \bibinfo{year}{2016}\natexlab{}.
\newblock \showarticletitle{GenderMag: A Method for Evaluating Software’s Gender Inclusiveness}.
\newblock \bibinfo{journal}{\emph{Interacting with Computers}}  \bibinfo{volume}{forthcoming} (\bibinfo{date}{01} \bibinfo{year}{2016}).
\newblock
\urldef\tempurl%
\url{https://doi.org/10.1093/iwc/iwv046}
\showDOI{\tempurl}


\bibitem[Cabral and Smith(2011)]%
        {cabral2011racial}
\bibfield{author}{\bibinfo{person}{Raquel~R Cabral} {and} \bibinfo{person}{Timothy~B Smith}.} \bibinfo{year}{2011}\natexlab{}.
\newblock \showarticletitle{Racial/ethnic matching of clients and therapists in mental health services: a meta-analytic review of preferences, perceptions, and outcomes.}
\newblock \bibinfo{journal}{\emph{Journal of counseling psychology}} \bibinfo{volume}{58}, \bibinfo{number}{4} (\bibinfo{year}{2011}), \bibinfo{pages}{537}.
\newblock


\bibitem[Case(1997)]%
        {case1997african}
\bibfield{author}{\bibinfo{person}{Karen~I Case}.} \bibinfo{year}{1997}\natexlab{}.
\newblock \showarticletitle{African American othermothering in the urban elementary school}.
\newblock \bibinfo{journal}{\emph{The Urban Review}} \bibinfo{volume}{29}, \bibinfo{number}{1} (\bibinfo{year}{1997}), \bibinfo{pages}{25--39}.
\newblock


\bibitem[Celepkolu and Boyer(2018)]%
        {Celepkolu2018}
\bibfield{author}{\bibinfo{person}{Mehmet Celepkolu} {and} \bibinfo{person}{Kristy~Elizabeth Boyer}.} \bibinfo{year}{2018}\natexlab{}.
\newblock \showarticletitle{Thematic Analysis of Students’ Reflections on Pair Programming in CS1}. In \bibinfo{booktitle}{\emph{Proceedings of the 49th ACM Technical Symposium on Computer Science Education}} (Baltimore, Maryland, USA) \emph{(\bibinfo{series}{SIGCSE ’18})}. \bibinfo{publisher}{Association for Computing Machinery}, \bibinfo{address}{New York, NY, USA}, \bibinfo{pages}{771–776}.
\newblock
\showISBNx{9781450351034}
\urldef\tempurl%
\url{https://doi.org/10.1145/3159450.3159516}
\showDOI{\tempurl}


\bibitem[Chaparro et~al\mbox{.}(2005)]%
        {chaparro2005factors}
\bibfield{author}{\bibinfo{person}{Edgar~Acosta Chaparro}, \bibinfo{person}{Aybala Yuksel}, \bibinfo{person}{Pablo Romero}, {and} \bibinfo{person}{Sallyann Bryant}.} \bibinfo{year}{2005}\natexlab{}.
\newblock \showarticletitle{Factors Affecting the Perceived Effectiveness of Pair Programming in Higher Education.}. In \bibinfo{booktitle}{\emph{PPIG}}. \bibinfo{pages}{2}.
\newblock


\bibitem[Chen et~al\mbox{.}(2020)]%
        {chen2020incorporating}
\bibfield{author}{\bibinfo{person}{Liang Chen}, \bibinfo{person}{Yongjian Ye}, \bibinfo{person}{Angyu Zheng}, \bibinfo{person}{Fenfang Xie}, \bibinfo{person}{Zibin Zheng}, {and} \bibinfo{person}{Michael~R Lyu}.} \bibinfo{year}{2020}\natexlab{}.
\newblock \showarticletitle{Incorporating geographical location for team formation in social coding sites}.
\newblock \bibinfo{journal}{\emph{World Wide Web}}  \bibinfo{volume}{23} (\bibinfo{year}{2020}), \bibinfo{pages}{153--174}.
\newblock


\bibitem[Chen et~al\mbox{.}(2022)]%
        {chen2022collecting}
\bibfield{author}{\bibinfo{person}{Yiqun~T Chen}, \bibinfo{person}{Angela~DR Smith}, \bibinfo{person}{Katharina Reinecke}, {and} \bibinfo{person}{Alexandra To}.} \bibinfo{year}{2022}\natexlab{}.
\newblock \showarticletitle{Collecting and reporting race and ethnicity data in HCI}. In \bibinfo{booktitle}{\emph{CHI Conference on Human Factors in Computing Systems Extended Abstracts}}. \bibinfo{pages}{1--8}.
\newblock


\bibitem[Chen et~al\mbox{.}(2023)]%
        {chen2023and}
\bibfield{author}{\bibinfo{person}{Yiqun~T Chen}, \bibinfo{person}{Angela~DR Smith}, \bibinfo{person}{Katharina Reinecke}, {and} \bibinfo{person}{Alexandra To}.} \bibinfo{year}{2023}\natexlab{}.
\newblock \showarticletitle{Why, when, and from whom: considerations for collecting and reporting race and ethnicity data in HCI}. In \bibinfo{booktitle}{\emph{Proceedings of the 2023 CHI Conference on Human Factors in Computing Systems}}. \bibinfo{pages}{1--15}.
\newblock


\bibitem[Choi(2013)]%
        {choi2013evaluating}
\bibfield{author}{\bibinfo{person}{Kyungsub~S Choi}.} \bibinfo{year}{2013}\natexlab{}.
\newblock \showarticletitle{Evaluating gender significance within a pair programming context}. In \bibinfo{booktitle}{\emph{2013 46th Hawaii International Conference on System Sciences}}. IEEE, \bibinfo{pages}{4817--4825}.
\newblock


\bibitem[Choi(2015a)]%
        {choi2014}
\bibfield{author}{\bibinfo{person}{Kyungsub~Stephen Choi}.} \bibinfo{year}{2015}\natexlab{a}.
\newblock \showarticletitle{A comparative analysis of different gender pair combinations in pair programming}.
\newblock \bibinfo{journal}{\emph{Behaviour \& Information Technology}} \bibinfo{volume}{34}, \bibinfo{number}{8} (\bibinfo{year}{2015}), \bibinfo{pages}{825--837}.
\newblock


\bibitem[Choi(2015b)]%
        {choi2015comparative}
\bibfield{author}{\bibinfo{person}{Kyungsub~Stephen Choi}.} \bibinfo{year}{2015}\natexlab{b}.
\newblock \showarticletitle{A comparative analysis of different gender pair combinations in pair programming}.
\newblock \bibinfo{journal}{\emph{Behaviour \& Information Technology}} \bibinfo{volume}{34}, \bibinfo{number}{8} (\bibinfo{year}{2015}), \bibinfo{pages}{825--837}.
\newblock


\bibitem[Choi et~al\mbox{.}(2009)]%
        {choi2009pair}
\bibfield{author}{\bibinfo{person}{Kyungsub~Steve Choi}, \bibinfo{person}{Fadi~P Deek}, {and} \bibinfo{person}{Il Im}.} \bibinfo{year}{2009}\natexlab{}.
\newblock \showarticletitle{Pair dynamics in team collaboration}.
\newblock \bibinfo{journal}{\emph{Computers in Human Behavior}} \bibinfo{volume}{25}, \bibinfo{number}{4} (\bibinfo{year}{2009}), \bibinfo{pages}{844--852}.
\newblock


\bibitem[Chong and Hurlbutt(2007a)]%
        {Chong2007}
\bibfield{author}{\bibinfo{person}{Jan Chong} {and} \bibinfo{person}{Tom Hurlbutt}.} \bibinfo{year}{2007}\natexlab{a}.
\newblock \showarticletitle{The Social Dynamics of Pair Programming}. In \bibinfo{booktitle}{\emph{Proceedings of the 29th International Conference on Software Engineering}} \emph{(\bibinfo{series}{ICSE '07})}. \bibinfo{publisher}{IEEE Computer Society}, \bibinfo{address}{USA}, \bibinfo{pages}{354–363}.
\newblock
\showISBNx{0769528287}
\urldef\tempurl%
\url{https://doi.org/10.1109/ICSE.2007.87}
\showDOI{\tempurl}


\bibitem[Chong and Hurlbutt(2007b)]%
        {chong2007social}
\bibfield{author}{\bibinfo{person}{Jan Chong} {and} \bibinfo{person}{Tom Hurlbutt}.} \bibinfo{year}{2007}\natexlab{b}.
\newblock \showarticletitle{The social dynamics of pair programming}. In \bibinfo{booktitle}{\emph{29th International Conference on Software Engineering (ICSE'07)}}. IEEE, \bibinfo{pages}{354--363}.
\newblock


\bibitem[Cliburn(2003)]%
        {cliburn2003experiences}
\bibfield{author}{\bibinfo{person}{Daniel~C Cliburn}.} \bibinfo{year}{2003}\natexlab{}.
\newblock \showarticletitle{Experiences with pair programming at a small college}.
\newblock \bibinfo{journal}{\emph{Journal of Computing Sciences in Colleges}} \bibinfo{volume}{19}, \bibinfo{number}{1} (\bibinfo{year}{2003}), \bibinfo{pages}{20--29}.
\newblock


\bibitem[Collins and Bilge(2020)]%
        {collins2020intersectionality}
\bibfield{author}{\bibinfo{person}{Patricia~Hill Collins} {and} \bibinfo{person}{Sirma Bilge}.} \bibinfo{year}{2020}\natexlab{}.
\newblock \bibinfo{booktitle}{\emph{Intersectionality}}.
\newblock \bibinfo{publisher}{John Wiley \& Sons}.
\newblock


\bibitem[Compeau and Higgins(1995)]%
        {compeau1995computer}
\bibfield{author}{\bibinfo{person}{Deborah~R Compeau} {and} \bibinfo{person}{Christopher~A Higgins}.} \bibinfo{year}{1995}\natexlab{}.
\newblock \showarticletitle{Computer self-efficacy: Development of a measure and initial test}.
\newblock \bibinfo{journal}{\emph{MIS quarterly}} (\bibinfo{year}{1995}), \bibinfo{pages}{189--211}.
\newblock


\bibitem[Cooper et~al\mbox{.}(2020)]%
        {cooper2020exploring}
\bibfield{author}{\bibinfo{person}{Saraah Cooper}, \bibinfo{person}{Ben Clinkscale}, \bibinfo{person}{Briana Williams}, {and} \bibinfo{person}{Myles Lewis}.} \bibinfo{year}{2020}\natexlab{}.
\newblock \showarticletitle{Exploring the impact of exposing CS majors to programming concepts using IDE programming vs. non-IDE programming in the classroom}. In \bibinfo{booktitle}{\emph{Proceedings of the 51st ACM Technical Symposium on Computer Science Education}}. \bibinfo{pages}{1422--1422}.
\newblock


\bibitem[Crenshaw(1997)]%
        {crenshaw1997mapping}
\bibfield{author}{\bibinfo{person}{Kimberle Crenshaw}.} \bibinfo{year}{1997}\natexlab{}.
\newblock \showarticletitle{Mapping the margins: Intersectionality, identity politics, and violence against women of color}.
\newblock \bibinfo{journal}{\emph{The legal response to violence against women}}  \bibinfo{volume}{5} (\bibinfo{year}{1997}), \bibinfo{pages}{91}.
\newblock


\bibitem[Cuadrado et~al\mbox{.}(2012)]%
        {cuadrado2012gender}
\bibfield{author}{\bibinfo{person}{Isabel Cuadrado}, \bibinfo{person}{Marisol Navas}, \bibinfo{person}{Fernando Molero}, \bibinfo{person}{Emilio Ferrer}, {and} \bibinfo{person}{J~Francisco Morales}.} \bibinfo{year}{2012}\natexlab{}.
\newblock \showarticletitle{Gender differences in leadership styles as a function of leader and subordinates' sex and type of organization}.
\newblock \bibinfo{journal}{\emph{Journal of Applied Social Psychology}} \bibinfo{volume}{42}, \bibinfo{number}{12} (\bibinfo{year}{2012}), \bibinfo{pages}{3083--3113}.
\newblock


\bibitem[Cunningham et~al\mbox{.}(2023)]%
        {cunningham2023grounds}
\bibfield{author}{\bibinfo{person}{Jay Cunningham}, \bibinfo{person}{Gabrielle Benabdallah}, \bibinfo{person}{Daniela Rosner}, {and} \bibinfo{person}{Alex Taylor}.} \bibinfo{year}{2023}\natexlab{}.
\newblock \showarticletitle{On the grounds of solutionism: Ontologies of blackness and HCI}.
\newblock \bibinfo{journal}{\emph{ACM Transactions on Computer-Human Interaction}} \bibinfo{volume}{30}, \bibinfo{number}{2} (\bibinfo{year}{2023}), \bibinfo{pages}{1--17}.
\newblock


\bibitem[Dake(2022)]%
        {Dake_2022}
\bibfield{author}{\bibinfo{person}{Austen Dake}.} \bibinfo{year}{2022}\natexlab{}.
\newblock \bibinfo{title}{2020 state of salaries report: Salary benchmarks and talent preferences}.
\newblock
\newblock
\urldef\tempurl%
\url{https://hired.com/blog/highlights/2020-state-of-salaries-report/}
\showURL{%
\tempurl}


\bibitem[De~Dreu and Weingart(2003)]%
        {de2003task}
\bibfield{author}{\bibinfo{person}{Carsten~KW De~Dreu} {and} \bibinfo{person}{Laurie~R Weingart}.} \bibinfo{year}{2003}\natexlab{}.
\newblock \showarticletitle{Task versus relationship conflict, team performance, and team member satisfaction: a meta-analysis.}
\newblock \bibinfo{journal}{\emph{Journal of applied Psychology}} \bibinfo{volume}{88}, \bibinfo{number}{4} (\bibinfo{year}{2003}), \bibinfo{pages}{741}.
\newblock


\bibitem[de~Souza~Santos et~al\mbox{.}(2023)]%
        {santos2023benefits}
\bibfield{author}{\bibinfo{person}{R. de Souza~Santos}, \bibinfo{person}{C. Magalhaes}, {and} \bibinfo{person}{P. Ralph}.} \bibinfo{year}{2023}\natexlab{}.
\newblock \showarticletitle{Benefits and Limitations of Remote Work to LGBTQIA+ Software Professionals}. In \bibinfo{booktitle}{\emph{Proceedings of the 45th IEEE/ACM International Conference on Software Engineering (ICSE 2023)}}.
\newblock


\bibitem[Deci(1971)]%
        {deci1971effects}
\bibfield{author}{\bibinfo{person}{Edward~L Deci}.} \bibinfo{year}{1971}\natexlab{}.
\newblock \showarticletitle{Effects of externally mediated rewards on intrinsic motivation.}
\newblock \bibinfo{journal}{\emph{Journal of personality and Social Psychology}} \bibinfo{volume}{18}, \bibinfo{number}{1} (\bibinfo{year}{1971}), \bibinfo{pages}{105}.
\newblock


\bibitem[Delgado and Stefancic(2023)]%
        {delgado2023critical}
\bibfield{author}{\bibinfo{person}{Richard Delgado} {and} \bibinfo{person}{Jean Stefancic}.} \bibinfo{year}{2023}\natexlab{}.
\newblock \bibinfo{booktitle}{\emph{Critical race theory: An introduction}}. Vol.~\bibinfo{volume}{87}.
\newblock \bibinfo{publisher}{NyU press}.
\newblock


\bibitem[Denner et~al\mbox{.}(2014)]%
        {denner2014pair}
\bibfield{author}{\bibinfo{person}{Jill Denner}, \bibinfo{person}{Linda Werner}, \bibinfo{person}{Shannon Campe}, {and} \bibinfo{person}{Eloy Ortiz}.} \bibinfo{year}{2014}\natexlab{}.
\newblock \showarticletitle{Pair programming: Under what conditions is it advantageous for middle school students?}
\newblock \bibinfo{journal}{\emph{Journal of Research on Technology in Education}} \bibinfo{volume}{46}, \bibinfo{number}{3} (\bibinfo{year}{2014}), \bibinfo{pages}{277--296}.
\newblock


\bibitem[Dillahunt et~al\mbox{.}(2009)]%
        {dillahunt2009s}
\bibfield{author}{\bibinfo{person}{Tawanna Dillahunt}, \bibinfo{person}{Jennifer Mankoff}, \bibinfo{person}{Eric Paulos}, {and} \bibinfo{person}{Susan Fussell}.} \bibinfo{year}{2009}\natexlab{}.
\newblock \showarticletitle{It's not all about" Green" energy use in low-income communities}. In \bibinfo{booktitle}{\emph{Proceedings of the 11th international conference on Ubiquitous computing}}. \bibinfo{pages}{255--264}.
\newblock


\bibitem[Dipboye and Colella(2013)]%
        {dipboye2013discrimination}
\bibfield{author}{\bibinfo{person}{Robert~L Dipboye} {and} \bibinfo{person}{Adrienne Colella}.} \bibinfo{year}{2013}\natexlab{}.
\newblock \bibinfo{booktitle}{\emph{Discrimination at work: The psychological and organizational bases}}.
\newblock \bibinfo{publisher}{Psychology Press}.
\newblock


\bibitem[DiSessa(2000)]%
        {disessa2000changing}
\bibfield{author}{\bibinfo{person}{Andrea~A DiSessa}.} \bibinfo{year}{2000}\natexlab{}.
\newblock \bibinfo{booktitle}{\emph{Changing minds: Computers, learning, and literacy}}.
\newblock \bibinfo{publisher}{Mit Press}.
\newblock


\bibitem[Dovidio(2001)]%
        {dovidio2001nature}
\bibfield{author}{\bibinfo{person}{John~F Dovidio}.} \bibinfo{year}{2001}\natexlab{}.
\newblock \showarticletitle{On the nature of contemporary prejudice: The third wave}.
\newblock \bibinfo{journal}{\emph{Journal of social issues}} \bibinfo{volume}{57}, \bibinfo{number}{4} (\bibinfo{year}{2001}), \bibinfo{pages}{829--849}.
\newblock


\bibitem[Dutta et~al\mbox{.}(2023)]%
        {dutta2023diversity}
\bibfield{author}{\bibinfo{person}{Riya Dutta}, \bibinfo{person}{Diego~Elias Costa}, \bibinfo{person}{Emad Shihab}, {and} \bibinfo{person}{Tanja Tajmel}.} \bibinfo{year}{2023}\natexlab{}.
\newblock \showarticletitle{Diversity Awareness in Software Engineering Participant Research}.
\newblock \bibinfo{journal}{\emph{arXiv preprint arXiv:2302.00042}} (\bibinfo{year}{2023}).
\newblock


\bibitem[Dyb\r{a} et~al\mbox{.}(2007)]%
        {Dyba2007}
\bibfield{author}{\bibinfo{person}{Tore Dyb\r{a}}, \bibinfo{person}{Erik Arisholm}, \bibinfo{person}{Dag Sjøberg}, \bibinfo{person}{Jo Hannay}, {and} \bibinfo{person}{Forrest Shull}.} \bibinfo{year}{2007}\natexlab{}.
\newblock \showarticletitle{Are Two Heads Better than One? On the Effectiveness of Pair Programming}.
\newblock \bibinfo{journal}{\emph{Software, IEEE}}  \bibinfo{volume}{24} (\bibinfo{date}{12} \bibinfo{year}{2007}), \bibinfo{pages}{12 -- 15}.
\newblock
\urldef\tempurl%
\url{https://doi.org/10.1109/MS.2007.158}
\showDOI{\tempurl}


\bibitem[Eagly and Johnson(1990)]%
        {eagly1990gender}
\bibfield{author}{\bibinfo{person}{Alice~H Eagly} {and} \bibinfo{person}{Blair~T Johnson}.} \bibinfo{year}{1990}\natexlab{}.
\newblock \showarticletitle{Gender and leadership style: A meta-analysis.}
\newblock \bibinfo{journal}{\emph{Psychological bulletin}} \bibinfo{volume}{108}, \bibinfo{number}{2} (\bibinfo{year}{1990}), \bibinfo{pages}{233}.
\newblock


\bibitem[Edelman and Inc(2010)]%
        {Edelman2010}
\bibfield{author}{\bibinfo{person}{Berland Edelman} {and} \bibinfo{person}{Inc}.} \bibinfo{year}{2010}\natexlab{}.
\newblock \bibinfo{booktitle}{\emph{Creativity and education: Why it matters}}.
\newblock
\urldef\tempurl%
\url{http://www.adobe.com/aboutadobe/pressroom/pdfs/Adobe_Creativity_and_Education_Why_It_Matters_study.pdf}
\showURL{%
Retrieved September 18th, 2019 from \tempurl}


\bibitem[Edwards(2012)]%
        {edwards2012role}
\bibfield{author}{\bibinfo{person}{Anne Edwards}.} \bibinfo{year}{2012}\natexlab{}.
\newblock \showarticletitle{The role of common knowledge in achieving collaboration across practices}.
\newblock \bibinfo{journal}{\emph{Learning, Culture and Social Interaction}} \bibinfo{volume}{1}, \bibinfo{number}{1} (\bibinfo{year}{2012}), \bibinfo{pages}{22--32}.
\newblock


\bibitem[Edwards et~al\mbox{.}(2001)]%
        {edwards2001race}
\bibfield{author}{\bibinfo{person}{Christopher~L Edwards}, \bibinfo{person}{Roger~B Fillingim}, {and} \bibinfo{person}{Francis Keefe}.} \bibinfo{year}{2001}\natexlab{}.
\newblock \showarticletitle{Race, ethnicity and pain}.
\newblock \bibinfo{journal}{\emph{Pain}} \bibinfo{volume}{94}, \bibinfo{number}{2} (\bibinfo{year}{2001}), \bibinfo{pages}{133--137}.
\newblock


\bibitem[Emerson and Murphy(2015)]%
        {emerson2015company}
\bibfield{author}{\bibinfo{person}{Katherine~TU Emerson} {and} \bibinfo{person}{Mary~C Murphy}.} \bibinfo{year}{2015}\natexlab{}.
\newblock \showarticletitle{A company I can trust? Organizational lay theories moderate stereotype threat for women}.
\newblock \bibinfo{journal}{\emph{Personality and Social Psychology Bulletin}} \bibinfo{volume}{41}, \bibinfo{number}{2} (\bibinfo{year}{2015}), \bibinfo{pages}{295--307}.
\newblock


\bibitem[Enayati(2002)]%
        {enayati2002research}
\bibfield{author}{\bibinfo{person}{Jasmin Enayati}.} \bibinfo{year}{2002}\natexlab{}.
\newblock \showarticletitle{The research: Effective communication and decision-making in diverse groups}.
\newblock \bibinfo{journal}{\emph{Multi-stakeholder processes for governance and sustainability: Beyond deadlock and conflict}} (\bibinfo{year}{2002}), \bibinfo{pages}{73--95}.
\newblock


\bibitem[Erete et~al\mbox{.}(2018)]%
        {erete2018intersectional}
\bibfield{author}{\bibinfo{person}{Sheena Erete}, \bibinfo{person}{Aarti Israni}, {and} \bibinfo{person}{Tawanna Dillahunt}.} \bibinfo{year}{2018}\natexlab{}.
\newblock \showarticletitle{An intersectional approach to designing in the margins}.
\newblock \bibinfo{journal}{\emph{Interactions}} \bibinfo{volume}{25}, \bibinfo{number}{3} (\bibinfo{year}{2018}), \bibinfo{pages}{66--69}.
\newblock


\bibitem[Erete(2015)]%
        {erete2015engaging}
\bibfield{author}{\bibinfo{person}{Sheena~L Erete}.} \bibinfo{year}{2015}\natexlab{}.
\newblock \showarticletitle{Engaging around neighborhood issues: How online communication affects offline behavior}. In \bibinfo{booktitle}{\emph{Proceedings of the 18th ACM Conference on Computer Supported Cooperative Work \& Social Computing}}. \bibinfo{pages}{1590--1601}.
\newblock


\bibitem[et~al.(2020)]%
        {Robe2020}
\bibfield{author}{\bibinfo{person}{P.~{Robe} et al.}} \bibinfo{year}{2020}\natexlab{}.
\newblock \showarticletitle{Can Machine Learning Facilitate Remote Pair Programming? Challenges, Insights Implications}. In \bibinfo{booktitle}{\emph{VL/HCC}}. \bibinfo{pages}{1--11}.
\newblock


\bibitem[Feagin(1995)]%
        {feagin1995living}
\bibfield{author}{\bibinfo{person}{Joe~R Feagin}.} \bibinfo{year}{1995}\natexlab{}.
\newblock \bibinfo{booktitle}{\emph{Living with racism: The black middle-class experience}}.
\newblock \bibinfo{publisher}{Beacon Press}.
\newblock


\bibitem[Feagin and McKinney(2005)]%
        {feagin2005many}
\bibfield{author}{\bibinfo{person}{Joe~R Feagin} {and} \bibinfo{person}{Karyn~D McKinney}.} \bibinfo{year}{2005}\natexlab{}.
\newblock \bibinfo{booktitle}{\emph{The many costs of racism}}.
\newblock \bibinfo{publisher}{Rowman \& Littlefield Publishers}.
\newblock


\bibitem[Felix and Studios({[n.\,d.]})]%
        {Felix}
\bibfield{author}{\bibinfo{person}{Felix} {and} \bibinfo{person}{Paul Studios}.} \bibinfo{year}{[n.\,d.]}\natexlab{}.
\newblock \bibinfo{title}{Traveling while black on oculus quest 2}.
\newblock
\newblock
\urldef\tempurl%
\url{https://www.oculus.com/experiences/quest/2121787737926354/?utm_source=www.wilsoncenter.org&amp;utm_medium=oculusredirect&amp;store&amp;item_id=2121787737926354}
\showURL{%
\tempurl}


\bibitem[Filer et~al\mbox{.}(1991)]%
        {filer1991voting}
\bibfield{author}{\bibinfo{person}{John~E Filer}, \bibinfo{person}{Lawrence~W Kenny}, {and} \bibinfo{person}{Rebecca~B Morton}.} \bibinfo{year}{1991}\natexlab{}.
\newblock \showarticletitle{Voting laws, educational policies, and minority turnout}.
\newblock \bibinfo{journal}{\emph{The Journal of Law and Economics}} \bibinfo{volume}{34}, \bibinfo{number}{2, Part 1} (\bibinfo{year}{1991}), \bibinfo{pages}{371--393}.
\newblock


\bibitem[Fisher and Marshall(2009)]%
        {fisher2009understanding}
\bibfield{author}{\bibinfo{person}{Murray~J Fisher} {and} \bibinfo{person}{Andrea~P Marshall}.} \bibinfo{year}{2009}\natexlab{}.
\newblock \showarticletitle{Understanding descriptive statistics}.
\newblock \bibinfo{journal}{\emph{Australian critical care}} \bibinfo{volume}{22}, \bibinfo{number}{2} (\bibinfo{year}{2009}), \bibinfo{pages}{93--97}.
\newblock


\bibitem[Ford et~al\mbox{.}(2019)]%
        {ford2019remote}
\bibfield{author}{\bibinfo{person}{D. Ford}, \bibinfo{person}{R. Milewicz}, {and} \bibinfo{person}{A. Serebrenik}.} \bibinfo{year}{2019}\natexlab{}.
\newblock \showarticletitle{How Remote Work Can Foster a More Inclusive Environment for Transgender Developers}. In \bibinfo{booktitle}{\emph{2019 IEEE/ACM 2nd International Workshop on Gender Equality in Software Engineering (GE)}}. \bibinfo{pages}{9--12}.
\newblock


\bibitem[Galdo et~al\mbox{.}(2022)]%
        {galdo2022pair}
\bibfield{author}{\bibinfo{person}{Aisha~Chung Galdo}, \bibinfo{person}{Mehmet Celepkolu}, \bibinfo{person}{Nicholas Lytle}, {and} \bibinfo{person}{Kristy~Elizabeth Boyer}.} \bibinfo{year}{2022}\natexlab{}.
\newblock \showarticletitle{Pair programming in a pandemic: Understanding middle school students' remote collaboration experiences}. In \bibinfo{booktitle}{\emph{Proceedings of the 53rd ACM Technical Symposium on Computer Science Education-Volume 1}}. \bibinfo{pages}{335--341}.
\newblock


\bibitem[Gitnux(2023)]%
        {Gitnux_2023}
\bibfield{author}{\bibinfo{person}{Author: Gitnux}.} \bibinfo{year}{2023}\natexlab{}.
\newblock \bibinfo{title}{The most surprising diversity in tech statistics and Trends in 2023 • gitnux}.
\newblock
\newblock
\urldef\tempurl%
\url{https://blog.gitnux.com/diversity-in-tech-statistics/#:~:text=According%20to%20a%202020%20report,and%20Black%20employees%20(6.1%25).}
\showURL{%
\tempurl}


\bibitem[Goode-Cross and Grim(2016)]%
        {goode2016unspoken}
\bibfield{author}{\bibinfo{person}{David~T Goode-Cross} {and} \bibinfo{person}{Karen~Ann Grim}.} \bibinfo{year}{2016}\natexlab{}.
\newblock \showarticletitle{“An Unspoken Level of Comfort” Black Therapists’ Experiences Working With Black Clients}.
\newblock \bibinfo{journal}{\emph{Journal of Black Psychology}} \bibinfo{volume}{42}, \bibinfo{number}{1} (\bibinfo{year}{2016}), \bibinfo{pages}{29--53}.
\newblock


\bibitem[Gramlich(2019)]%
        {Gramlich_2019}
\bibfield{author}{\bibinfo{person}{John Gramlich}.} \bibinfo{year}{2019}\natexlab{}.
\newblock \bibinfo{title}{The gap between the number of blacks and whites in prison is shrinking}.
\newblock
\newblock
\urldef\tempurl%
\url{https://www.pewresearch.org/short-reads/2019/04/30/shrinking-gap-between-number-of-blacks-and-whites-in-prison/}
\showURL{%
\tempurl}


\bibitem[Gregory et~al\mbox{.}(2021)]%
        {gregory2021agile}
\bibfield{author}{\bibinfo{person}{Peggy Gregory}, \bibinfo{person}{Casper Lassenius}, \bibinfo{person}{Xiaofeng Wang}, {and} \bibinfo{person}{Philippe Kruchten}.} \bibinfo{year}{2021}\natexlab{}.
\newblock \bibinfo{booktitle}{\emph{Agile Processes in Software Engineering and Extreme Programming: 22nd International Conference on Agile Software Development, XP 2021, Virtual Event, June 14--18, 2021, Proceedings}}.
\newblock \bibinfo{publisher}{Springer Nature}.
\newblock


\bibitem[Grimes et~al\mbox{.}(2008)]%
        {grimes2008eatwell}
\bibfield{author}{\bibinfo{person}{Andrea Grimes}, \bibinfo{person}{Martin Bednar}, \bibinfo{person}{Jay~David Bolter}, {and} \bibinfo{person}{Rebecca~E Grinter}.} \bibinfo{year}{2008}\natexlab{}.
\newblock \showarticletitle{EatWell: sharing nutrition-related memories in a low-income community}. In \bibinfo{booktitle}{\emph{Proceedings of the 2008 ACM conference on Computer supported cooperative work}}. \bibinfo{pages}{87--96}.
\newblock


\bibitem[Guo et~al\mbox{.}(2014)]%
        {guo2014local}
\bibfield{author}{\bibinfo{person}{Jiong Guo}, \bibinfo{person}{Danny Hermelin}, {and} \bibinfo{person}{Christian Komusiewicz}.} \bibinfo{year}{2014}\natexlab{}.
\newblock \showarticletitle{Local search for string problems: Brute-force is essentially optimal}.
\newblock \bibinfo{journal}{\emph{Theoretical Computer Science}}  \bibinfo{volume}{525} (\bibinfo{year}{2014}), \bibinfo{pages}{30--41}.
\newblock


\bibitem[Gurin(1999)]%
        {gurin1999new}
\bibfield{author}{\bibinfo{person}{Patricia Gurin}.} \bibinfo{year}{1999}\natexlab{}.
\newblock \showarticletitle{New research on the benefits of diversity in college and beyond: An empirical analysis}.
\newblock \bibinfo{journal}{\emph{Diversity Digest}} \bibinfo{volume}{3}, \bibinfo{number}{3} (\bibinfo{year}{1999}), \bibinfo{pages}{5--15}.
\newblock


\bibitem[Gündemir et~al\mbox{.}(2014)]%
        {Gundemir2014}
\bibfield{author}{\bibinfo{person}{S. Gündemir}, \bibinfo{person}{A.~C. Homan}, \bibinfo{person}{C.~K. de Dreu}, {and} \bibinfo{person}{M. van Vugt}.} \bibinfo{year}{2014}\natexlab{}.
\newblock \showarticletitle{Think leader, think White? Capturing and weakening an implicit pro-White leadership bias}.
\newblock \bibinfo{journal}{\emph{PLoS One}}  \bibinfo{volume}{9} (\bibinfo{year}{2014}), \bibinfo{pages}{e83915}.
\newblock
Issue 1.
\urldef\tempurl%
\url{https://doi.org/10.1371/journal.pone.0083915}
\showDOI{\tempurl}


\bibitem[Han et~al\mbox{.}(2010)]%
        {Han2010}
\bibfield{author}{\bibinfo{person}{Keun-Woo Han}, \bibinfo{person}{EunKyoung Lee}, {and} \bibinfo{person}{Youngjun Lee}.} \bibinfo{year}{2010}\natexlab{}.
\newblock \showarticletitle{The Impact of a Peer-Learning Agent Based on Pair Programming in a Programming Course}.
\newblock \bibinfo{journal}{\emph{Education, IEEE Transactions on}}  \bibinfo{volume}{53} (\bibinfo{date}{06} \bibinfo{year}{2010}), \bibinfo{pages}{318 -- 327}.
\newblock
\urldef\tempurl%
\url{https://doi.org/10.1109/TE.2009.2019121}
\showDOI{\tempurl}


\bibitem[Hanks(2005)]%
        {hanks2005student}
\bibfield{author}{\bibinfo{person}{Brian Hanks}.} \bibinfo{year}{2005}\natexlab{}.
\newblock \showarticletitle{Student performance in CS1 with distributed pair programming}.
\newblock \bibinfo{journal}{\emph{ACM SIGCSE Bulletin}} \bibinfo{volume}{37}, \bibinfo{number}{3} (\bibinfo{year}{2005}), \bibinfo{pages}{316--320}.
\newblock


\bibitem[Hannay et~al\mbox{.}(2009)]%
        {hannay2009effects}
\bibfield{author}{\bibinfo{person}{Jo~E Hannay}, \bibinfo{person}{Erik Arisholm}, \bibinfo{person}{Harald Engvik}, {and} \bibinfo{person}{Dag~IK Sjoberg}.} \bibinfo{year}{2009}\natexlab{}.
\newblock \showarticletitle{Effects of personality on pair programming}.
\newblock \bibinfo{journal}{\emph{IEEE Transactions on Software Engineering}} \bibinfo{volume}{36}, \bibinfo{number}{1} (\bibinfo{year}{2009}), \bibinfo{pages}{61--80}.
\newblock


\bibitem[Haslam et~al\mbox{.}(2005)]%
        {haslam2005anxiety}
\bibfield{author}{\bibinfo{person}{C Haslam}, \bibinfo{person}{Sarah Atkinson}, \bibinfo{person}{SS Brown}, {and} \bibinfo{person}{RA Haslam}.} \bibinfo{year}{2005}\natexlab{}.
\newblock \showarticletitle{Anxiety and depression in the workplace: effects on the individual and organisation (a focus group investigation)}.
\newblock \bibinfo{journal}{\emph{Journal of affective disorders}} \bibinfo{volume}{88}, \bibinfo{number}{2} (\bibinfo{year}{2005}), \bibinfo{pages}{209--215}.
\newblock


\bibitem[Hebl and Dovidio(2005)]%
        {hebl2005promoting}
\bibfield{author}{\bibinfo{person}{Michelle~R Hebl} {and} \bibinfo{person}{John~F Dovidio}.} \bibinfo{year}{2005}\natexlab{}.
\newblock \showarticletitle{Promoting the “social” in the examination of social stigmas}.
\newblock \bibinfo{journal}{\emph{Personality and social psychology review}} \bibinfo{volume}{9}, \bibinfo{number}{2} (\bibinfo{year}{2005}), \bibinfo{pages}{156--182}.
\newblock


\bibitem[Higson et~al\mbox{.}(2019)]%
        {higson2019bayesian}
\bibfield{author}{\bibinfo{person}{Edward Higson}, \bibinfo{person}{Will Handley}, \bibinfo{person}{Michael Hobson}, {and} \bibinfo{person}{Anthony Lasenby}.} \bibinfo{year}{2019}\natexlab{}.
\newblock \showarticletitle{Bayesian sparse reconstruction: a brute-force approach to astronomical imaging and machine learning}.
\newblock \bibinfo{journal}{\emph{Monthly Notices of the Royal Astronomical Society}} \bibinfo{volume}{483}, \bibinfo{number}{4} (\bibinfo{year}{2019}), \bibinfo{pages}{4828--4846}.
\newblock


\bibitem[Hosanagar(2020)]%
        {hosanagar2020human}
\bibfield{author}{\bibinfo{person}{Kartik Hosanagar}.} \bibinfo{year}{2020}\natexlab{}.
\newblock \bibinfo{booktitle}{\emph{A human's guide to machine intelligence: how algorithms are shaping our lives and how we can stay in control}}.
\newblock \bibinfo{publisher}{Penguin}.
\newblock


\bibitem[Howard et~al\mbox{.}(2009)]%
        {Howard2009}
\bibfield{author}{\bibinfo{person}{Elizabeth Howard}, \bibinfo{person}{Donna Evans}, \bibinfo{person}{Jill Courte}, {and} \bibinfo{person}{Cathy Bishop-Clark}.} \bibinfo{year}{2009}\natexlab{}.
\newblock \showarticletitle{A qualitative look at Alice and pair-programming}.
\newblock \bibinfo{journal}{\emph{Number}}  \bibinfo{volume}{7} (\bibinfo{date}{08} \bibinfo{year}{2009}).
\newblock


\bibitem[Howard(2006)]%
        {howard2006attitudes}
\bibfield{author}{\bibinfo{person}{Elizabeth~V Howard}.} \bibinfo{year}{2006}\natexlab{}.
\newblock \showarticletitle{Attitudes on using pair-programming}.
\newblock \bibinfo{journal}{\emph{Journal of Educational Technology Systems}} \bibinfo{volume}{35}, \bibinfo{number}{1} (\bibinfo{year}{2006}), \bibinfo{pages}{89--103}.
\newblock


\bibitem[Hughes et~al\mbox{.}(2020)]%
        {hughes2020remote}
\bibfield{author}{\bibinfo{person}{Janet Hughes}, \bibinfo{person}{Ann Walshe}, \bibinfo{person}{Bobby Law}, {and} \bibinfo{person}{Brendan Murphy}.} \bibinfo{year}{2020}\natexlab{}.
\newblock \showarticletitle{Remote pair programming}. In \bibinfo{booktitle}{\emph{12th International Conference on Computer Supported Education}}. SciTePress, \bibinfo{pages}{476--483}.
\newblock


\bibitem[Ickes(1984)]%
        {ickes1984compositions}
\bibfield{author}{\bibinfo{person}{William Ickes}.} \bibinfo{year}{1984}\natexlab{}.
\newblock \showarticletitle{Compositions in Black and White: Determinants of interaction in interracial dyads.}
\newblock \bibinfo{journal}{\emph{Journal of Personality and Social Psychology}} \bibinfo{volume}{47}, \bibinfo{number}{2} (\bibinfo{year}{1984}), \bibinfo{pages}{330}.
\newblock


\bibitem[Ikeda and Shiramatsu(2017)]%
        {ikeda2017generating}
\bibfield{author}{\bibinfo{person}{Yuto Ikeda} {and} \bibinfo{person}{Shun Shiramatsu}.} \bibinfo{year}{2017}\natexlab{}.
\newblock \showarticletitle{Generating questions asked by facilitator agents using preceding context in web-based discussion}. In \bibinfo{booktitle}{\emph{2017 IEEE International conference on agents (ICA)}}. IEEE, \bibinfo{pages}{127--132}.
\newblock


\bibitem[Isajiw(1993)]%
        {isajiw1993definition}
\bibfield{author}{\bibinfo{person}{Wsevolod~W Isajiw}.} \bibinfo{year}{1993}\natexlab{}.
\newblock \showarticletitle{Definition and dimensions of ethnicity: A theoretical framework}.
\newblock \bibinfo{journal}{\emph{Challenges of measuring an ethnic world: Science, politics and reality}} (\bibinfo{year}{1993}), \bibinfo{pages}{407--427}.
\newblock


\bibitem[Isaksen and Treffinger(2004)]%
        {Isakesen2004}
\bibfield{author}{\bibinfo{person}{Scott~G. Isaksen} {and} \bibinfo{person}{Donald~J. Treffinger}.} \bibinfo{year}{2004}\natexlab{}.
\newblock \showarticletitle{{Celebrating 50 years of Reflective Practice: Versions of Creative Problem Solving}}.
\newblock \bibinfo{journal}{\emph{The Journal of Creative Behavior}} \bibinfo{volume}{38}, \bibinfo{number}{2} (\bibinfo{date}{June} \bibinfo{year}{2004}), \bibinfo{pages}{75--101}.
\newblock


\bibitem[Jaccard(1901)]%
        {Jaccard1901}
\bibfield{author}{\bibinfo{person}{Paul Jaccard}.} \bibinfo{year}{1901}\natexlab{}.
\newblock \showarticletitle{Etude de la distribution florale dans une portion des Alpes et du Jura}.
\newblock \bibinfo{journal}{\emph{Bulletin de la Societe Vaudoise des Sciences Naturelles}}  \bibinfo{volume}{37} (\bibinfo{date}{01} \bibinfo{year}{1901}), \bibinfo{pages}{547--579}.
\newblock
\urldef\tempurl%
\url{https://doi.org/10.5169/seals-266450}
\showDOI{\tempurl}


\bibitem[Jackson et~al\mbox{.}(1995)]%
        {jackson1995composition}
\bibfield{author}{\bibinfo{person}{Pamela~Braboy Jackson}, \bibinfo{person}{Peggy~A Thoits}, {and} \bibinfo{person}{Howard~F Taylor}.} \bibinfo{year}{1995}\natexlab{}.
\newblock \showarticletitle{Composition of the workplace and psychological well-being: The effects of tokenism on America's Black elite}.
\newblock \bibinfo{journal}{\emph{Social Forces}} \bibinfo{volume}{74}, \bibinfo{number}{2} (\bibinfo{year}{1995}), \bibinfo{pages}{543--557}.
\newblock


\bibitem[Jacobson(2014)]%
        {jacobson2014google}
\bibfield{author}{\bibinfo{person}{Murrey Jacobson}.} \bibinfo{year}{2014}\natexlab{}.
\newblock \showarticletitle{Google finally discloses its diversity record, and it’s not good}.
\newblock \bibinfo{journal}{\emph{PBS News Hour}} (\bibinfo{year}{2014}).
\newblock


\bibitem[Janarthanan(2012)]%
        {janarthanan2012serious}
\bibfield{author}{\bibinfo{person}{Vasudevan Janarthanan}.} \bibinfo{year}{2012}\natexlab{}.
\newblock \showarticletitle{Serious video games: Games for education and health}. In \bibinfo{booktitle}{\emph{2012 Ninth International Conference on Information Technology-New Generations}}. IEEE, \bibinfo{pages}{875--878}.
\newblock


\bibitem[Jensen(2022)]%
        {Jensen_2022}
\bibfield{author}{\bibinfo{person}{Eric Jensen}.} \bibinfo{year}{2022}\natexlab{}.
\newblock \bibinfo{title}{Measuring racial and ethnic diversity for the 2020 census}.
\newblock
\newblock
\urldef\tempurl%
\url{https://www.census.gov/newsroom/blogs/random-samplings/2021/08/measuring-racial-ethnic-diversity-2020-census.html}
\showURL{%
\tempurl}


\bibitem[Jiang and Pell(2015)]%
        {JIANG20159}
\bibfield{author}{\bibinfo{person}{Xiaoming Jiang} {and} \bibinfo{person}{Marc~D. Pell}.} \bibinfo{year}{2015}\natexlab{}.
\newblock \showarticletitle{On how the brain decodes vocal cues about speaker confidence}.
\newblock \bibinfo{journal}{\emph{Cortex}}  \bibinfo{volume}{66} (\bibinfo{year}{2015}), \bibinfo{pages}{9 -- 34}.
\newblock
\showISSN{0010-9452}
\urldef\tempurl%
\url{http://www.sciencedirect.com/science/article/pii/S0010945215000593}
\showURL{%
\tempurl}


\bibitem[Johns et~al\mbox{.}(2008)]%
        {johns2008stereotype}
\bibfield{author}{\bibinfo{person}{Michael Johns}, \bibinfo{person}{Michael Inzlicht}, {and} \bibinfo{person}{Toni Schmader}.} \bibinfo{year}{2008}\natexlab{}.
\newblock \showarticletitle{Stereotype threat and executive resource depletion: examining the influence of emotion regulation.}
\newblock \bibinfo{journal}{\emph{Journal of Experimental Psychology: General}} \bibinfo{volume}{137}, \bibinfo{number}{4} (\bibinfo{year}{2008}), \bibinfo{pages}{691}.
\newblock


\bibitem[Johnson-Bailey(1999)]%
        {johnson1999ties}
\bibfield{author}{\bibinfo{person}{Juanita Johnson-Bailey}.} \bibinfo{year}{1999}\natexlab{}.
\newblock \showarticletitle{The ties that bind and the shackles that separate: Race, gender, class, and color in a research process}.
\newblock \bibinfo{journal}{\emph{International Journal of Qualitative Studies in Education}} \bibinfo{volume}{12}, \bibinfo{number}{6} (\bibinfo{year}{1999}), \bibinfo{pages}{659--670}.
\newblock


\bibitem[Joshi and Roh(2009)]%
        {joshi2009role}
\bibfield{author}{\bibinfo{person}{Aparna Joshi} {and} \bibinfo{person}{Hyuntak Roh}.} \bibinfo{year}{2009}\natexlab{}.
\newblock \showarticletitle{The role of context in work team diversity research: A meta-analytic review}.
\newblock \bibinfo{journal}{\emph{Academy of management journal}} \bibinfo{volume}{52}, \bibinfo{number}{3} (\bibinfo{year}{2009}), \bibinfo{pages}{599--627}.
\newblock


\bibitem[Kaiser and Miller(2001)]%
        {kaiser2001stop}
\bibfield{author}{\bibinfo{person}{Cheryl~R Kaiser} {and} \bibinfo{person}{Carol~T Miller}.} \bibinfo{year}{2001}\natexlab{}.
\newblock \showarticletitle{Stop complaining! The social costs of making attributions to discrimination}.
\newblock \bibinfo{journal}{\emph{Personality and Social Psychology Bulletin}} \bibinfo{volume}{27}, \bibinfo{number}{2} (\bibinfo{year}{2001}), \bibinfo{pages}{254--263}.
\newblock


\bibitem[Katira et~al\mbox{.}(2005)]%
        {katira2005towards}
\bibfield{author}{\bibinfo{person}{Neha Katira}, \bibinfo{person}{Laurie Williams}, {and} \bibinfo{person}{Jason Osborne}.} \bibinfo{year}{2005}\natexlab{}.
\newblock \showarticletitle{Towards increasing the compatibility of student pair programmers}. In \bibinfo{booktitle}{\emph{Proceedings of the 27th international conference on Software engineering}}. \bibinfo{pages}{625--626}.
\newblock


\bibitem[{Kaur Kuttal} et~al\mbox{.}(2019)]%
        {Kuttal2019}
\bibfield{author}{\bibinfo{person}{S. {Kaur Kuttal}}, \bibinfo{person}{K. {Gerstner}}, {and} \bibinfo{person}{A. {Bejarano}}.} \bibinfo{year}{2019}\natexlab{}.
\newblock \showarticletitle{Remote Pair Programming in Online CS Education: Investigating through a Gender Lens}. In \bibinfo{booktitle}{\emph{2019 IEEE Symposium on Visual Languages and Human-Centric Computing (VL/HCC)}}. \bibinfo{pages}{75--85}.
\newblock


\bibitem[Kaushal(2014)]%
        {kaushal2014social}
\bibfield{author}{\bibinfo{person}{K Kaushal}.} \bibinfo{year}{2014}\natexlab{}.
\newblock \showarticletitle{Social desirability bias in face to face interviews}.
\newblock \bibinfo{journal}{\emph{Journal of postgraduate medicine}} \bibinfo{volume}{60}, \bibinfo{number}{4} (\bibinfo{year}{2014}), \bibinfo{pages}{415}.
\newblock


\bibitem[Kavitha and Ahmed(2013)]%
        {Kavitha2013}
\bibfield{author}{\bibinfo{person}{R.~K. Kavitha} {and} \bibinfo{person}{M.~S.~Irfan Ahmed}.} \bibinfo{year}{2013}\natexlab{}.
\newblock \showarticletitle{{Knowledge Sharing Through Pair Programming in Learning Environments: An empirical study}}.
\newblock \bibinfo{journal}{\emph{Education and Information Technologies}} \bibinfo{volume}{20}, \bibinfo{number}{2} (\bibinfo{date}{Oct.} \bibinfo{year}{2013}), \bibinfo{pages}{319--333}.
\newblock


\bibitem[King(2006)]%
        {king2006jim}
\bibfield{author}{\bibinfo{person}{Ryan~Scott King}.} \bibinfo{year}{2006}\natexlab{}.
\newblock \showarticletitle{Jim crow is alive and well in the 21st century: Felony disenfranchisement and the continuing struggle to silence the african-american voice}.
\newblock \bibinfo{journal}{\emph{Souls}} \bibinfo{volume}{8}, \bibinfo{number}{2} (\bibinfo{year}{2006}), \bibinfo{pages}{7--21}.
\newblock


\bibitem[Kogler~Hill and Gant(2000)]%
        {koglerhill2000}
\bibfield{author}{\bibinfo{person}{Susan~E Kogler~Hill} {and} \bibinfo{person}{George Gant}.} \bibinfo{year}{2000}\natexlab{}.
\newblock \showarticletitle{Mentoring by Minorities for Minorities: The Organizational Communications Support Program}.
\newblock \bibinfo{journal}{\emph{Review of Business}} \bibinfo{volume}{21}, \bibinfo{number}{1/2} (\bibinfo{year}{2000}), \bibinfo{pages}{53--57}.
\newblock
\newblock
\shownote{Entrepreneurship Database; ProQuest Central}.


\bibitem[{Kuttal} et~al\mbox{.}(2020)]%
        {Kuttal2020}
\bibfield{author}{\bibinfo{person}{S.~K. {Kuttal}}, \bibinfo{person}{J. {Myers}}, \bibinfo{person}{S. {Gurka}}, \bibinfo{person}{D. {Magar}}, \bibinfo{person}{D. {Piorkowski}}, {and} \bibinfo{person}{R. {Bellamy}}.} \bibinfo{year}{2020}\natexlab{}.
\newblock \showarticletitle{Towards Designing Conversational Agents for Pair Programming: Accounting for Creativity Strategies and Conversational Styles}. In \bibinfo{booktitle}{\emph{2020 IEEE Symposium on Visual Languages and Human-Centric Computing (VL/HCC)}}. \bibinfo{pages}{1--11}.
\newblock


\bibitem[Kuttal et~al\mbox{.}(2021)]%
        {Kuttal2021}
\bibfield{author}{\bibinfo{person}{Sandeep~Kaur Kuttal}, \bibinfo{person}{Bali Ong}, \bibinfo{person}{Kate Kwasny}, {and} \bibinfo{person}{Peter Robe}.} \bibinfo{year}{2021}\natexlab{}.
\newblock \showarticletitle{Trade-Offs for Substituting a Human with an Agent in a Pair Programming Context: The Good, the Bad, and the Ugly}. In \bibinfo{booktitle}{\emph{Proceedings of the 2021 CHI Conference on Human Factors in Computing Systems}} (Yokohama, Japan) \emph{(\bibinfo{series}{CHI '21})}. \bibinfo{publisher}{Association for Computing Machinery}, \bibinfo{address}{New York, NY, USA}, Article \bibinfo{articleno}{243}, \bibinfo{numpages}{20}~pages.
\newblock
\showISBNx{9781450380966}
\urldef\tempurl%
\url{https://doi.org/10.1145/3411764.3445659}
\showDOI{\tempurl}


\bibitem[L.~Jones and D.~Fleming(2013)]%
        {Jones2013}
\bibfield{author}{\bibinfo{person}{Danielle L.~Jones} {and} \bibinfo{person}{Scott D.~Fleming}.} \bibinfo{year}{2013}\natexlab{}.
\newblock \showarticletitle{What use is a backseat driver? A qualitative investigation of pair programming}.
\newblock \bibinfo{journal}{\emph{Proceedings of IEEE Symposium on Visual Languages and Human-Centric Computing, VL/HCC}}, \bibinfo{pages}{103--110}.
\newblock


\bibitem[Lebron(2023)]%
        {lebron2023making}
\bibfield{author}{\bibinfo{person}{Christopher~J Lebron}.} \bibinfo{year}{2023}\natexlab{}.
\newblock \bibinfo{booktitle}{\emph{The making of black lives matter: A brief history of an idea}}.
\newblock \bibinfo{publisher}{Oxford University Press}.
\newblock


\bibitem[Leman(2015)]%
        {leman2015groups}
\bibfield{author}{\bibinfo{person}{Patrick~J Leman}.} \bibinfo{year}{2015}\natexlab{}.
\newblock \showarticletitle{How do groups work? Age differences in performance and the social outcomes of peer collaboration}.
\newblock \bibinfo{journal}{\emph{Cognitive science}} \bibinfo{volume}{39}, \bibinfo{number}{4} (\bibinfo{year}{2015}), \bibinfo{pages}{804--820}.
\newblock


\bibitem[Leman and Oldham(2005)]%
        {leman2005children}
\bibfield{author}{\bibinfo{person}{Patrick~J Leman} {and} \bibinfo{person}{Zo{\"e} Oldham}.} \bibinfo{year}{2005}\natexlab{}.
\newblock \showarticletitle{Do children need to learn to collaborate?: The effect of age and age differences on collaborative recall}.
\newblock \bibinfo{journal}{\emph{Cognitive Development}} \bibinfo{volume}{20}, \bibinfo{number}{1} (\bibinfo{year}{2005}), \bibinfo{pages}{33--48}.
\newblock


\bibitem[Lemov(2010)]%
        {lemov2010teach}
\bibfield{author}{\bibinfo{person}{Doug Lemov}.} \bibinfo{year}{2010}\natexlab{}.
\newblock \bibinfo{booktitle}{\emph{Teach like a champion: 49 techniques that put students on the path to college (K-12)}}.
\newblock \bibinfo{publisher}{John Wiley \& Sons}.
\newblock


\bibitem[Lewis(1982)]%
        {Lewis1982}
\bibfield{author}{\bibinfo{person}{Clayton Lewis}.} \bibinfo{year}{1982}\natexlab{}.
\newblock \bibinfo{booktitle}{\emph{Using the "thinking-aloud" method in cognitive interface design}}.
\newblock \bibinfo{publisher}{IBM T.J. Watson Research Center}, \bibinfo{address}{Yorktown Heights, N.Y.}
\newblock


\bibitem[Liu and Schonwetter(2004)]%
        {Liu2004}
\bibfield{author}{\bibinfo{person}{Zhiqiang Liu} {and} \bibinfo{person}{Dieter~J Schonwetter}.} \bibinfo{year}{2004}\natexlab{}.
\newblock \showarticletitle{Teaching creativity in engineering}.
\newblock \bibinfo{journal}{\emph{International Journal of Engineering Education}} \bibinfo{volume}{20}, \bibinfo{number}{5} (\bibinfo{year}{2004}), \bibinfo{pages}{801--808}.
\newblock


\bibitem[Loeber et~al\mbox{.}(2000)]%
        {loeber2000stability}
\bibfield{author}{\bibinfo{person}{Rolf Loeber}, \bibinfo{person}{Matthew Drinkwater}, \bibinfo{person}{Yanming Yin}, \bibinfo{person}{Stewart~J Anderson}, \bibinfo{person}{Laura~C Schmidt}, {and} \bibinfo{person}{Anne Crawford}.} \bibinfo{year}{2000}\natexlab{}.
\newblock \showarticletitle{Stability of family interaction from ages 6 to 18}.
\newblock \bibinfo{journal}{\emph{Journal of abnormal child psychology}}  \bibinfo{volume}{28} (\bibinfo{year}{2000}), \bibinfo{pages}{353--369}.
\newblock


\bibitem[Lomotey(1993)]%
        {lomotey1993african}
\bibfield{author}{\bibinfo{person}{Kofi Lomotey}.} \bibinfo{year}{1993}\natexlab{}.
\newblock \showarticletitle{African-American principals: Bureaucrat/administrators and ethno-humanists}.
\newblock \bibinfo{journal}{\emph{Urban Education}} \bibinfo{volume}{27}, \bibinfo{number}{4} (\bibinfo{year}{1993}), \bibinfo{pages}{395--412}.
\newblock


\bibitem[Lopez et~al\mbox{.}(2019)]%
        {lopez2019investigating}
\bibfield{author}{\bibinfo{person}{Sarah Lopez}, \bibinfo{person}{Yi Yang}, \bibinfo{person}{Kevin Beltran}, \bibinfo{person}{Soo~Jung Kim}, \bibinfo{person}{Jennifer Cruz~Hernandez}, \bibinfo{person}{Chelsy Simran}, \bibinfo{person}{Bingkun Yang}, {and} \bibinfo{person}{Beste~F Yuksel}.} \bibinfo{year}{2019}\natexlab{}.
\newblock \showarticletitle{Investigating implicit gender bias and embodiment of white males in virtual reality with full body visuomotor synchrony}. In \bibinfo{booktitle}{\emph{Proceedings of the 2019 CHI Conference on human factors in computing systems}}. \bibinfo{pages}{1--12}.
\newblock


\bibitem[Lord and Saenz(1985)]%
        {lord1985memory}
\bibfield{author}{\bibinfo{person}{Charles~G Lord} {and} \bibinfo{person}{Delia~S Saenz}.} \bibinfo{year}{1985}\natexlab{}.
\newblock \showarticletitle{Memory deficits and memory surfeits: Differential cognitive consequences of tokenism for tokens and observers.}
\newblock \bibinfo{journal}{\emph{Journal of personality and social psychology}} \bibinfo{volume}{49}, \bibinfo{number}{4} (\bibinfo{year}{1985}), \bibinfo{pages}{918}.
\newblock


\bibitem[Lott et~al\mbox{.}(2021)]%
        {Lott2021}
\bibfield{author}{\bibinfo{person}{C. Lott}, \bibinfo{person}{A. McAuliffe}, {and} \bibinfo{person}{S. Kuttal}.} \bibinfo{year}{2021}\natexlab{}.
\newblock \showarticletitle{Remote Pair Collaborations of CS Students: Leaving Women Behind?}. In \bibinfo{booktitle}{\emph{Proceedings of Visual Languages and Human-Centric Computing}}.
\newblock


\bibitem[Lui and Chan(2006)]%
        {lui2006pair}
\bibfield{author}{\bibinfo{person}{Kim~Man Lui} {and} \bibinfo{person}{Keith~CC Chan}.} \bibinfo{year}{2006}\natexlab{}.
\newblock \showarticletitle{Pair programming productivity: Novice--novice vs. expert--expert}.
\newblock \bibinfo{journal}{\emph{International Journal of Human-computer studies}} \bibinfo{volume}{64}, \bibinfo{number}{9} (\bibinfo{year}{2006}), \bibinfo{pages}{915--925}.
\newblock


\bibitem[Major et~al\mbox{.}(2002)]%
        {major2002antecedents}
\bibfield{author}{\bibinfo{person}{Brenda Major}, \bibinfo{person}{Wendy~J Quinton}, {and} \bibinfo{person}{Shannon~K McCoy}.} \bibinfo{year}{2002}\natexlab{}.
\newblock \showarticletitle{Antecedents and consequences of attributions to discrimination: Theoretical and empirical advances}.
\newblock In \bibinfo{booktitle}{\emph{Advances in experimental social psychology}}. Vol.~\bibinfo{volume}{34}. \bibinfo{publisher}{Elsevier}, \bibinfo{pages}{251--330}.
\newblock


\bibitem[Margaret~Burnett(2018)]%
        {Burnett2018}
\bibfield{author}{\bibinfo{person}{Christopher Mendez Alannah Oleson Claudia Hilderbrand Zoe Steine-Hanson Andrew J.~Ko Margaret~Burnett, Anita~Sarma}.} \bibinfo{year}{2018}\natexlab{}.
\newblock \showarticletitle{Designing Technologies to Support Human Problem Solving}. In \bibinfo{booktitle}{\emph{Workshop at VL/HCC}}.
\newblock


\bibitem[Matthew and Denis({[n.\,d.]})]%
        {matthew19s}
\bibfield{author}{\bibinfo{person}{Clair Matthew} {and} \bibinfo{person}{J Denis}.} \bibinfo{year}{[n.\,d.]}\natexlab{}.
\newblock \showarticletitle{S.(2015). Sociology of racism}.
\newblock \bibinfo{journal}{\emph{International Encyclopedia of the Social \& Behavioral Sciences}} \bibinfo{volume}{19}, \bibinfo{number}{2} (\bibinfo{year}{[n.\,d.]}), \bibinfo{pages}{857--863}.
\newblock


\bibitem[Mcdowell et~al\mbox{.}(2002)]%
        {Mcdowell2002}
\bibfield{author}{\bibinfo{person}{Charles Mcdowell}, \bibinfo{person}{Linda Werner}, \bibinfo{person}{Heather Bullock}, {and} \bibinfo{person}{Julian Fernald}.} \bibinfo{year}{2002}\natexlab{}.
\newblock \showarticletitle{The effects of pair-programming on performance in an introductory programming course}.
\newblock \bibinfo{journal}{\emph{ACM SIGCSE Bulletin}}  \bibinfo{volume}{34}, \bibinfo{pages}{38--42}.
\newblock
\urldef\tempurl%
\url{https://doi.org/10.1145/563340.563353}
\showDOI{\tempurl}


\bibitem[Mcdowell et~al\mbox{.}(2003)]%
        {Mcdowell2003}
\bibfield{author}{\bibinfo{person}{Charles Mcdowell}, \bibinfo{person}{Linda Werner}, \bibinfo{person}{Heather Bullock}, {and} \bibinfo{person}{J. Fernald}.} \bibinfo{year}{2003}\natexlab{}.
\newblock \showarticletitle{The impact of pair programming on student performance, perception and persistence}. \bibinfo{pages}{602-- 607}.
\newblock
\showISBNx{0-7695-1877-X}
\urldef\tempurl%
\url{https://doi.org/10.1109/ICSE.2003.1201243}
\showDOI{\tempurl}


\bibitem[Mcdowell et~al\mbox{.}(2006)]%
        {McDowell2006}
\bibfield{author}{\bibinfo{person}{Charles Mcdowell}, \bibinfo{person}{Linda Werner}, \bibinfo{person}{Heather Bullock}, {and} \bibinfo{person}{Julian Fernald}.} \bibinfo{year}{2006}\natexlab{}.
\newblock \showarticletitle{Pair programming improves student retention, confidence, and program quality}.
\newblock \bibinfo{journal}{\emph{Commun. ACM}}  \bibinfo{volume}{49} (\bibinfo{date}{08} \bibinfo{year}{2006}), \bibinfo{pages}{90--95}.
\newblock
\urldef\tempurl%
\url{https://doi.org/10.1145/1145293}
\showDOI{\tempurl}


\bibitem[McDowell et~al\mbox{.}(2006)]%
        {mcdowell2006pair}
\bibfield{author}{\bibinfo{person}{Charlie McDowell}, \bibinfo{person}{Linda Werner}, \bibinfo{person}{Heather~E Bullock}, {and} \bibinfo{person}{Julian Fernald}.} \bibinfo{year}{2006}\natexlab{}.
\newblock \showarticletitle{Pair programming improves student retention, confidence, and program quality}.
\newblock \bibinfo{journal}{\emph{Commun. ACM}} \bibinfo{volume}{49}, \bibinfo{number}{8} (\bibinfo{year}{2006}), \bibinfo{pages}{90--95}.
\newblock


\bibitem[M'charek(2013)]%
        {m2013beyond}
\bibfield{author}{\bibinfo{person}{Amade M'charek}.} \bibinfo{year}{2013}\natexlab{}.
\newblock \showarticletitle{Beyond fact or fiction: On the materiality of race in practice}.
\newblock \bibinfo{journal}{\emph{Cultural anthropology}} \bibinfo{volume}{28}, \bibinfo{number}{3} (\bibinfo{year}{2013}), \bibinfo{pages}{420--442}.
\newblock


\bibitem[McLeod et~al\mbox{.}(1996)]%
        {mcleod1996ethnic}
\bibfield{author}{\bibinfo{person}{Poppy~Lauretta McLeod}, \bibinfo{person}{Sharon~Alisa Lobel}, {and} \bibinfo{person}{Taylor~H Cox~Jr}.} \bibinfo{year}{1996}\natexlab{}.
\newblock \showarticletitle{Ethnic diversity and creativity in small groups}.
\newblock \bibinfo{journal}{\emph{Small group research}} \bibinfo{volume}{27}, \bibinfo{number}{2} (\bibinfo{year}{1996}), \bibinfo{pages}{248--264}.
\newblock


\bibitem[Merriam et~al\mbox{.}(2001)]%
        {merriam2001power}
\bibfield{author}{\bibinfo{person}{Sharan~B Merriam}, \bibinfo{person}{Juanita Johnson-Bailey}, \bibinfo{person}{Ming-Yeh Lee}, \bibinfo{person}{Youngwha Kee}, \bibinfo{person}{Gabo Ntseane}, {and} \bibinfo{person}{Mazanah Muhamad}.} \bibinfo{year}{2001}\natexlab{}.
\newblock \showarticletitle{Power and positionality: Negotiating insider/outsider status within and across cultures}.
\newblock \bibinfo{journal}{\emph{International journal of lifelong education}} \bibinfo{volume}{20}, \bibinfo{number}{5} (\bibinfo{year}{2001}), \bibinfo{pages}{405--416}.
\newblock


\bibitem[Michie(1989)]%
        {michie1989brute}
\bibfield{author}{\bibinfo{person}{Donald Michie}.} \bibinfo{year}{1989}\natexlab{}.
\newblock \showarticletitle{Brute force in chess and science}.
\newblock \bibinfo{journal}{\emph{ICGA Journal}} \bibinfo{volume}{12}, \bibinfo{number}{3} (\bibinfo{year}{1989}), \bibinfo{pages}{127--143}.
\newblock


\bibitem[Miller(2004)]%
        {miller2004expectancy}
\bibfield{author}{\bibinfo{person}{Chelli A~Leutschaft Miller}.} \bibinfo{year}{2004}\natexlab{}.
\newblock \emph{\bibinfo{title}{Expectancy bias: An exploration of practitioner effectiveness in the clinical diagnostic process}}.
\newblock \bibinfo{thesistype}{Ph.\,D. Dissertation}. \bibinfo{school}{Capella University}.
\newblock


\bibitem[Moreland et~al\mbox{.}(2018)]%
        {moreland2018creating}
\bibfield{author}{\bibinfo{person}{Richard~L Moreland}, \bibinfo{person}{John~M Levine}, {and} \bibinfo{person}{Melissa~L Wingert}.} \bibinfo{year}{2018}\natexlab{}.
\newblock \showarticletitle{Creating the ideal group: Composition effects at work}.
\newblock In \bibinfo{booktitle}{\emph{Understanding group behavior}}. \bibinfo{publisher}{Psychology Press}, \bibinfo{pages}{11--35}.
\newblock


\bibitem[Muller and Padberg(2004)]%
        {muller2004empirical}
\bibfield{author}{\bibinfo{person}{Matthias~M Muller} {and} \bibinfo{person}{Frank Padberg}.} \bibinfo{year}{2004}\natexlab{}.
\newblock \showarticletitle{An empirical study about the feelgood factor in pair programming}. In \bibinfo{booktitle}{\emph{10th International Symposium on Software Metrics, 2004. Proceedings.}} IEEE, \bibinfo{pages}{151--158}.
\newblock


\bibitem[Murphy and Steele(2009)]%
        {murphy2009importance}
\bibfield{author}{\bibinfo{person}{MC Murphy} {and} \bibinfo{person}{CM Steele}.} \bibinfo{year}{2009}\natexlab{}.
\newblock \showarticletitle{The importance of context: Understanding the effects of situational cues on perceived identity contingencies and sense of belonging}.
\newblock \bibinfo{journal}{\emph{Unpublished manuscript}} (\bibinfo{year}{2009}).
\newblock


\bibitem[Murphy-Hill et~al\mbox{.}(2022)]%
        {murphyhill2022pushback}
\bibfield{author}{\bibinfo{person}{Emerson~R. Murphy-Hill}, \bibinfo{person}{Ciera Jaspan}, \bibinfo{person}{Carolyn~D. Egelman}, {and} \bibinfo{person}{Lan Cheng}.} \bibinfo{year}{2022}\natexlab{}.
\newblock \showarticletitle{The Pushback Effects of Race, Ethnicity, Gender, and Age in Code Review}.
\newblock \bibinfo{journal}{\emph{Commun. ACM}} \bibinfo{volume}{65}, \bibinfo{number}{3} (\bibinfo{year}{2022}), \bibinfo{pages}{52--57}.
\newblock


\bibitem[Nagappan et~al\mbox{.}(2003a)]%
        {Nagappan2003}
\bibfield{author}{\bibinfo{person}{Nachiappan Nagappan}, \bibinfo{person}{Laurie Williams}, \bibinfo{person}{Miriam Ferzli}, \bibinfo{person}{Eric Wiebe}, \bibinfo{person}{Kai Yang}, \bibinfo{person}{Carol Miller}, {and} \bibinfo{person}{Suzanne Balik}.} \bibinfo{year}{2003}\natexlab{a}.
\newblock \showarticletitle{Improving the CS1 experience with pair programming}.
\newblock \bibinfo{journal}{\emph{ACM Sigcse Bulletin}}  \bibinfo{volume}{35}, \bibinfo{pages}{359--362}.
\newblock
\urldef\tempurl%
\url{https://doi.org/10.1145/792548.612006}
\showDOI{\tempurl}


\bibitem[Nagappan et~al\mbox{.}(2003b)]%
        {nagappan2003improving}
\bibfield{author}{\bibinfo{person}{Nachiappan Nagappan}, \bibinfo{person}{Laurie Williams}, \bibinfo{person}{Miriam Ferzli}, \bibinfo{person}{Eric Wiebe}, \bibinfo{person}{Kai Yang}, \bibinfo{person}{Carol Miller}, {and} \bibinfo{person}{Suzanne Balik}.} \bibinfo{year}{2003}\natexlab{b}.
\newblock \showarticletitle{Improving the CS1 experience with pair programming}.
\newblock \bibinfo{journal}{\emph{ACM Sigcse Bulletin}} \bibinfo{volume}{35}, \bibinfo{number}{1} (\bibinfo{year}{2003}), \bibinfo{pages}{359--362}.
\newblock


\bibitem[ncwit({[n.\,d.]})]%
        {ncwit}
ncwit \bibinfo{year}{[n.\,d.]}\natexlab{}.
\newblock \bibinfo{title}{{National Center for Women \& Information Technology}}.
\newblock
\newblock
\urldef\tempurl%
\url{https://www.ncwit.org/}
\showURL{%
\tempurl}


\bibitem[Ngounou and Guti{\'e}rrez(2019)]%
        {ngounou2019value}
\bibfield{author}{\bibinfo{person}{Gislaine~N Ngounou} {and} \bibinfo{person}{Nancy~B Guti{\'e}rrez}.} \bibinfo{year}{2019}\natexlab{}.
\newblock \showarticletitle{The value of interracial facilitation of racial equity training}.
\newblock \bibinfo{journal}{\emph{Phi Delta Kappan}} \bibinfo{volume}{100}, \bibinfo{number}{8} (\bibinfo{year}{2019}), \bibinfo{pages}{56--61}.
\newblock


\bibitem[Noble(2018)]%
        {noble2018algorithms}
\bibfield{author}{\bibinfo{person}{Safiya~Umoja Noble}.} \bibinfo{year}{2018}\natexlab{}.
\newblock \showarticletitle{Algorithms of oppression}.
\newblock In \bibinfo{booktitle}{\emph{Algorithms of oppression}}. \bibinfo{publisher}{New York university press}.
\newblock


\bibitem[Nosek(1998)]%
        {Nosek1998}
\bibfield{author}{\bibinfo{person}{John Nosek}.} \bibinfo{year}{1998}\natexlab{}.
\newblock \showarticletitle{The Case for Collaborative Programming}.
\newblock \bibinfo{journal}{\emph{Commun. ACM}}  \bibinfo{volume}{41} (\bibinfo{date}{03} \bibinfo{year}{1998}).
\newblock
\urldef\tempurl%
\url{https://doi.org/10.1145/272287.272333}
\showDOI{\tempurl}


\bibitem[Oberai and Anand(2018)]%
        {oberai2018unconscious}
\bibfield{author}{\bibinfo{person}{Himani Oberai} {and} \bibinfo{person}{Ila~Mehrotra Anand}.} \bibinfo{year}{2018}\natexlab{}.
\newblock \showarticletitle{Unconscious bias: thinking without thinking}.
\newblock \bibinfo{journal}{\emph{Human Resource Management International Digest}} \bibinfo{volume}{26}, \bibinfo{number}{6} (\bibinfo{year}{2018}), \bibinfo{pages}{14--17}.
\newblock


\bibitem[Ogbonnaya-Ogburu et~al\mbox{.}(2020)]%
        {ogbonnaya2020critical}
\bibfield{author}{\bibinfo{person}{Ihudiya~Finda Ogbonnaya-Ogburu}, \bibinfo{person}{Angela~DR Smith}, \bibinfo{person}{Alexandra To}, {and} \bibinfo{person}{Kentaro Toyama}.} \bibinfo{year}{2020}\natexlab{}.
\newblock \showarticletitle{Critical race theory for HCI}. In \bibinfo{booktitle}{\emph{Proceedings of the 2020 CHI conference on human factors in computing systems}}. \bibinfo{pages}{1--16}.
\newblock


\bibitem[Osborn(1957)]%
        {Osborn1957}
\bibfield{author}{\bibinfo{person}{A.F. Osborn}.} \bibinfo{year}{1957}\natexlab{}.
\newblock \bibinfo{booktitle}{\emph{{Applied Imagination: Principles and Procedures of Creative Thinking}}}.
\newblock \bibinfo{publisher}{Charles Scribner's Sons}.
\newblock
\showLCCN{57007589}


\bibitem[Ospina and Foldy(2009a)]%
        {Leadership2009}
\bibfield{author}{\bibinfo{person}{Sonia Ospina} {and} \bibinfo{person}{Erica Foldy}.} \bibinfo{year}{2009}\natexlab{a}.
\newblock \showarticletitle{A critical review of race and ethnicity in the leadership literature: Surfacing context, power and the collective dimensions of leadership}.
\newblock \bibinfo{journal}{\emph{The Leadership Quarterly}} \bibinfo{volume}{20}, \bibinfo{number}{6} (\bibinfo{year}{2009}), \bibinfo{pages}{876--896}.
\newblock
\showISSN{1048-9843}
\urldef\tempurl%
\url{https://doi.org/10.1016/j.leaqua.2009.09.005}
\showDOI{\tempurl}
\newblock
\shownote{The Leadership Quarterly Yearly Review of Leadership}.


\bibitem[Ospina and Foldy(2009b)]%
        {ospina2009critical}
\bibfield{author}{\bibinfo{person}{Sonia Ospina} {and} \bibinfo{person}{Erica Foldy}.} \bibinfo{year}{2009}\natexlab{b}.
\newblock \showarticletitle{A critical review of race and ethnicity in the leadership literature: Surfacing context, power and the collective dimensions of leadership}.
\newblock \bibinfo{journal}{\emph{The leadership quarterly}} \bibinfo{volume}{20}, \bibinfo{number}{6} (\bibinfo{year}{2009}), \bibinfo{pages}{876--896}.
\newblock


\bibitem[Oviatt and Cohen(2000)]%
        {Oviatt2000}
\bibfield{author}{\bibinfo{person}{Sharon Oviatt} {and} \bibinfo{person}{Philip Cohen}.} \bibinfo{year}{2000}\natexlab{}.
\newblock \showarticletitle{Perceptual User Interfaces: Multimodal Interfaces That Process What Comes Naturally}.
\newblock \bibinfo{journal}{\emph{Commun. ACM}} \bibinfo{volume}{43}, \bibinfo{number}{3} (\bibinfo{date}{March} \bibinfo{year}{2000}), \bibinfo{pages}{45–53}.
\newblock
\showISSN{0001-0782}
\urldef\tempurl%
\url{https://doi.org/10.1145/330534.330538}
\showDOI{\tempurl}


\bibitem[Pace(2008)]%
        {pace2008can}
\bibfield{author}{\bibinfo{person}{Tyler Pace}.} \bibinfo{year}{2008}\natexlab{}.
\newblock \showarticletitle{Can an orc catch a cab in stormwind? Cybertype preference in the World of Warcraft character creation interface}.
\newblock In \bibinfo{booktitle}{\emph{CHI'08 Extended Abstracts on Human Factors in Computing Systems}}. \bibinfo{pages}{2493--2502}.
\newblock


\bibitem[Page(2007)]%
        {page2007making}
\bibfield{author}{\bibinfo{person}{Scott~E Page}.} \bibinfo{year}{2007}\natexlab{}.
\newblock \showarticletitle{Making the difference: Applying a logic of diversity}.
\newblock \bibinfo{journal}{\emph{Academy of Management Perspectives}} \bibinfo{volume}{21}, \bibinfo{number}{4} (\bibinfo{year}{2007}), \bibinfo{pages}{6--20}.
\newblock


\bibitem[Page-Jones(1988)]%
        {Page1988}
\bibfield{author}{\bibinfo{person}{M. Page-Jones}.} \bibinfo{year}{1988}\natexlab{}.
\newblock \bibinfo{booktitle}{\emph{{The Practical Guide to Structured Systems Design}}}.
\newblock \bibinfo{publisher}{Prentice Hall}.
\newblock
\showISBNx{9780136907695}
\showLCCN{87034000}


\bibitem[Peck et~al\mbox{.}(2021)]%
        {peck2021evidence}
\bibfield{author}{\bibinfo{person}{Tabitha~C Peck}, \bibinfo{person}{Jessica~J Good}, {and} \bibinfo{person}{Katharina Seitz}.} \bibinfo{year}{2021}\natexlab{}.
\newblock \showarticletitle{Evidence of racial bias using immersive virtual reality: Analysis of head and hand motions during shooting decisions}.
\newblock \bibinfo{journal}{\emph{IEEE Transactions on Visualization and Computer Graphics}} \bibinfo{volume}{27}, \bibinfo{number}{5} (\bibinfo{year}{2021}), \bibinfo{pages}{2502--2512}.
\newblock


\bibitem[Pinel(1999)]%
        {pinel1999stigma}
\bibfield{author}{\bibinfo{person}{Elizabeth~C Pinel}.} \bibinfo{year}{1999}\natexlab{}.
\newblock \showarticletitle{Stigma consciousness: the psychological legacy of social stereotypes.}
\newblock \bibinfo{journal}{\emph{Journal of personality and social psychology}} \bibinfo{volume}{76}, \bibinfo{number}{1} (\bibinfo{year}{1999}), \bibinfo{pages}{114}.
\newblock


\bibitem[Plonka et~al\mbox{.}(2012)]%
        {plonka2012disengagement}
\bibfield{author}{\bibinfo{person}{Laura Plonka}, \bibinfo{person}{Helen Sharp}, {and} \bibinfo{person}{Janet Van Der~Linden}.} \bibinfo{year}{2012}\natexlab{}.
\newblock \showarticletitle{Disengagement in pair programming: Does it matter?}. In \bibinfo{booktitle}{\emph{2012 34th international conference on software engineering (ICSE)}}. IEEE, \bibinfo{pages}{496--506}.
\newblock


\bibitem[Pomales et~al\mbox{.}(1986)]%
        {pomales1986effects}
\bibfield{author}{\bibinfo{person}{Jay Pomales}, \bibinfo{person}{Charles~D Claiborn}, {and} \bibinfo{person}{Teresa~D LaFromboise}.} \bibinfo{year}{1986}\natexlab{}.
\newblock \showarticletitle{Effects of Black students' racial identity on perceptions of White counselors varying in cultural sensitivity.}
\newblock \bibinfo{journal}{\emph{Journal of Counseling Psychology}} \bibinfo{volume}{33}, \bibinfo{number}{1} (\bibinfo{year}{1986}), \bibinfo{pages}{57}.
\newblock


\bibitem[Purdie-Vaughns et~al\mbox{.}(2008)]%
        {purdie2008social}
\bibfield{author}{\bibinfo{person}{Valerie Purdie-Vaughns}, \bibinfo{person}{Claude~M Steele}, \bibinfo{person}{Paul~G Davies}, \bibinfo{person}{Ruth Ditlmann}, {and} \bibinfo{person}{Jennifer~Randall Crosby}.} \bibinfo{year}{2008}\natexlab{}.
\newblock \showarticletitle{Social identity contingencies: how diversity cues signal threat or safety for African Americans in mainstream institutions.}
\newblock \bibinfo{journal}{\emph{Journal of personality and social psychology}} \bibinfo{volume}{94}, \bibinfo{number}{4} (\bibinfo{year}{2008}), \bibinfo{pages}{615}.
\newblock


\bibitem[Rader et~al\mbox{.}(2011)]%
        {rader2011brick}
\bibfield{author}{\bibinfo{person}{Emilee Rader}, \bibinfo{person}{Margaret Echelbarger}, {and} \bibinfo{person}{Justine Cassell}.} \bibinfo{year}{2011}\natexlab{}.
\newblock \showarticletitle{Brick by brick: iterating interventions to bridge the achievement gap with virtual peers}. In \bibinfo{booktitle}{\emph{Proceedings of the SIGCHI Conference on Human Factors in Computing Systems}}. \bibinfo{pages}{2971--2974}.
\newblock


\bibitem[Raghavan et~al\mbox{.}(2020)]%
        {raghavan2020mitigating}
\bibfield{author}{\bibinfo{person}{Manish Raghavan}, \bibinfo{person}{Solon Barocas}, \bibinfo{person}{Jon Kleinberg}, {and} \bibinfo{person}{Karen Levy}.} \bibinfo{year}{2020}\natexlab{}.
\newblock \showarticletitle{Mitigating bias in algorithmic hiring: Evaluating claims and practices}. In \bibinfo{booktitle}{\emph{Proceedings of the 2020 conference on fairness, accountability, and transparency}}. \bibinfo{pages}{469--481}.
\newblock


\bibitem[Raji and Buolamwini(2019)]%
        {raji2019actionable}
\bibfield{author}{\bibinfo{person}{Inioluwa~Deborah Raji} {and} \bibinfo{person}{Joy Buolamwini}.} \bibinfo{year}{2019}\natexlab{}.
\newblock \showarticletitle{Actionable auditing: Investigating the impact of publicly naming biased performance results of commercial ai products}. In \bibinfo{booktitle}{\emph{Proceedings of the 2019 AAAI/ACM Conference on AI, Ethics, and Society}}. \bibinfo{pages}{429--435}.
\newblock


\bibitem[Ralph et~al\mbox{.}(2020)]%
        {ralph2020pandemic}
\bibfield{author}{\bibinfo{person}{Paul Ralph}, \bibinfo{person}{Sebastian Baltes}, \bibinfo{person}{Gilda Adisaputri}, \bibinfo{person}{Richard Torkar}, \bibinfo{person}{Vladimir Kovalenko}, \bibinfo{person}{Marcos Kalinowski}, {and} \bibinfo{person}{Razan Alkadhi}.} \bibinfo{year}{2020}\natexlab{}.
\newblock \showarticletitle{Pandemic Programming: How {COVID-19} Affects Software Developers and How Their Organizations Can Help}.
\newblock \bibinfo{journal}{\emph{Empirical Software Engineering}}  \bibinfo{volume}{25} (\bibinfo{year}{2020}), \bibinfo{pages}{4927--4961}.
\newblock
\urldef\tempurl%
\url{https://doi.org/10.1007/s10664-020-09816-2}
\showDOI{\tempurl}


\bibitem[Ramalingam and Wiedenbeck(1998)]%
        {ramalingam1998development}
\bibfield{author}{\bibinfo{person}{Vennila Ramalingam} {and} \bibinfo{person}{Susan Wiedenbeck}.} \bibinfo{year}{1998}\natexlab{}.
\newblock \showarticletitle{Development and validation of scores on a computer programming self-efficacy scale and group analyses of novice programmer self-efficacy}.
\newblock \bibinfo{journal}{\emph{Journal of Educational Computing Research}} \bibinfo{volume}{19}, \bibinfo{number}{4} (\bibinfo{year}{1998}), \bibinfo{pages}{367--381}.
\newblock


\bibitem[Randolph-Seng et~al\mbox{.}(2016)]%
        {randolph2016diversity}
\bibfield{author}{\bibinfo{person}{Brandon Randolph-Seng}, \bibinfo{person}{Claudia~C Cogliser}, \bibinfo{person}{Angela~F Randolph}, \bibinfo{person}{Terri~A Scandura}, \bibinfo{person}{Carliss~D Miller}, {and} \bibinfo{person}{Rachelle Smith-Genth{\^o}s}.} \bibinfo{year}{2016}\natexlab{}.
\newblock \showarticletitle{Diversity in leadership: Race in leader-member exchanges}.
\newblock \bibinfo{journal}{\emph{Leadership \& Organization Development Journal}} \bibinfo{volume}{37}, \bibinfo{number}{6} (\bibinfo{year}{2016}), \bibinfo{pages}{750--773}.
\newblock


\bibitem[Rankin(2023)]%
        {Rankin_2023}
\bibfield{author}{\bibinfo{person}{Joy~Lisi Rankin}.} \bibinfo{year}{2023}\natexlab{}.
\newblock \bibinfo{title}{Learning to code isn’t enough}.
\newblock
\newblock
\urldef\tempurl%
\url{https://www.technologyreview.com/2023/04/20/1071291/learn-to-code-legacy-new-projects-education/amp/}
\showURL{%
\tempurl}


\bibitem[Rankin and Irish(2020)]%
        {rankin2020seat}
\bibfield{author}{\bibinfo{person}{Yolanda~A Rankin} {and} \bibinfo{person}{India Irish}.} \bibinfo{year}{2020}\natexlab{}.
\newblock \showarticletitle{A seat at the table: Black feminist thought as a critical framework for inclusive game design}.
\newblock \bibinfo{journal}{\emph{Proceedings of the ACM on Human-Computer Interaction}} \bibinfo{volume}{4}, \bibinfo{number}{CSCW2} (\bibinfo{year}{2020}), \bibinfo{pages}{1--26}.
\newblock


\bibitem[Rankin and Thomas(2019)]%
        {rankin2019straighten}
\bibfield{author}{\bibinfo{person}{Yolanda~A Rankin} {and} \bibinfo{person}{Jakita~O Thomas}.} \bibinfo{year}{2019}\natexlab{}.
\newblock \showarticletitle{Straighten up and fly right: Rethinking intersectionality in HCI research}.
\newblock \bibinfo{journal}{\emph{Interactions}} \bibinfo{volume}{26}, \bibinfo{number}{6} (\bibinfo{year}{2019}), \bibinfo{pages}{64--68}.
\newblock


\bibitem[Riccucci(2021)]%
        {riccucci2021managing}
\bibfield{author}{\bibinfo{person}{Norma~M Riccucci}.} \bibinfo{year}{2021}\natexlab{}.
\newblock \bibinfo{booktitle}{\emph{Managing diversity in public sector workforces}}.
\newblock \bibinfo{publisher}{Routledge}.
\newblock


\bibitem[Robe and Kuttal(2022)]%
        {Robe2022}
\bibfield{author}{\bibinfo{person}{Peter Robe} {and} \bibinfo{person}{Sandeep~Kaur Kuttal}.} \bibinfo{year}{2022}\natexlab{}.
\newblock \showarticletitle{Designing PairBuddy—A Conversational Agent for Pair Programming}.
\newblock \bibinfo{journal}{\emph{ACM Trans. Comput.-Hum. Interact.}} \bibinfo{volume}{29}, \bibinfo{number}{4}, Article \bibinfo{articleno}{34} (\bibinfo{date}{may} \bibinfo{year}{2022}), \bibinfo{numpages}{44}~pages.
\newblock
\showISSN{1073-0516}
\urldef\tempurl%
\url{https://doi.org/10.1145/3498326}
\showDOI{\tempurl}


\bibitem[Roberson and Block(2001)]%
        {roberson20016}
\bibfield{author}{\bibinfo{person}{Loriann Roberson} {and} \bibinfo{person}{Caryn~J Block}.} \bibinfo{year}{2001}\natexlab{}.
\newblock \showarticletitle{6. Racioethnicity and job performance: A review and critique of theoretical perspectives on the causes of group differences}.
\newblock \bibinfo{journal}{\emph{Research in organizational behavior}}  \bibinfo{volume}{23} (\bibinfo{year}{2001}), \bibinfo{pages}{247--325}.
\newblock


\bibitem[Roberson et~al\mbox{.}(2003)]%
        {roberson2003stereotype}
\bibfield{author}{\bibinfo{person}{Loriann Roberson}, \bibinfo{person}{Elizabeth~A Deitch}, \bibinfo{person}{Arthur~P Brief}, {and} \bibinfo{person}{Caryn~J Block}.} \bibinfo{year}{2003}\natexlab{}.
\newblock \showarticletitle{Stereotype threat and feedback seeking in the workplace}.
\newblock \bibinfo{journal}{\emph{Journal of Vocational Behavior}} \bibinfo{volume}{62}, \bibinfo{number}{1} (\bibinfo{year}{2003}), \bibinfo{pages}{176--188}.
\newblock


\bibitem[Roth(2016)]%
        {roth2016multiple}
\bibfield{author}{\bibinfo{person}{Wendy~D Roth}.} \bibinfo{year}{2016}\natexlab{}.
\newblock \showarticletitle{The multiple dimensions of race}.
\newblock \bibinfo{journal}{\emph{Ethnic and Racial Studies}} \bibinfo{volume}{39}, \bibinfo{number}{8} (\bibinfo{year}{2016}), \bibinfo{pages}{1310--1338}.
\newblock


\bibitem[Ruvalcaba et~al\mbox{.}(2016a)]%
        {Ruvalcaba2016}
\bibfield{author}{\bibinfo{person}{Omar Ruvalcaba}, \bibinfo{person}{Linda Werner}, {and} \bibinfo{person}{Jill Denner}.} \bibinfo{year}{2016}\natexlab{a}.
\newblock \showarticletitle{Observations of Pair Programming: Variations in Collaboration Across Demographic Groups}. \bibinfo{pages}{90--95}.
\newblock
\urldef\tempurl%
\url{https://doi.org/10.1145/2839509.2844558}
\showDOI{\tempurl}


\bibitem[Ruvalcaba et~al\mbox{.}(2016b)]%
        {ruvalcaba2016observations}
\bibfield{author}{\bibinfo{person}{Omar Ruvalcaba}, \bibinfo{person}{Linda Werner}, {and} \bibinfo{person}{Jill Denner}.} \bibinfo{year}{2016}\natexlab{b}.
\newblock \showarticletitle{Observations of pair programming: Variations in collaboration across demographic groups}. In \bibinfo{booktitle}{\emph{Proceedings of the 47th ACM technical symposium on computing science education}}. \bibinfo{pages}{90--95}.
\newblock


\bibitem[Santos et~al\mbox{.}(2020)]%
        {santos2020increasing}
\bibfield{author}{\bibinfo{person}{Adrian Santos}, \bibinfo{person}{Sira Vegas}, \bibinfo{person}{Fernando Uyaguari}, \bibinfo{person}{Oscar Dieste}, \bibinfo{person}{Burak Turhan}, {and} \bibinfo{person}{Natalia Juristo}.} \bibinfo{year}{2020}\natexlab{}.
\newblock \showarticletitle{Increasing validity through replication: an illustrative TDD case}.
\newblock \bibinfo{journal}{\emph{Software Quality Journal}}  \bibinfo{volume}{28} (\bibinfo{year}{2020}), \bibinfo{pages}{371--395}.
\newblock


\bibitem[Santos and Ralph(2022)]%
        {santos2022grounded}
\bibfield{author}{\bibinfo{person}{R.~E. Santos} {and} \bibinfo{person}{P. Ralph}.} \bibinfo{year}{2022}\natexlab{}.
\newblock \showarticletitle{A Grounded Theory of Coordination in Remote-First and Hybrid Software Teams}. In \bibinfo{booktitle}{\emph{44th IEEE/ACM International Conference on Software Engineering (ICSE 2022)}}. \bibinfo{pages}{25--35}.
\newblock


\bibitem[Scandura et~al\mbox{.}(1986)]%
        {scandura1986managers}
\bibfield{author}{\bibinfo{person}{Terri~A Scandura}, \bibinfo{person}{George~B Graen}, {and} \bibinfo{person}{Michael~A Novak}.} \bibinfo{year}{1986}\natexlab{}.
\newblock \showarticletitle{When managers decide not to decide autocratically: An investigation of leader--member exchange and decision influence.}
\newblock \bibinfo{journal}{\emph{Journal of applied psychology}} \bibinfo{volume}{71}, \bibinfo{number}{4} (\bibinfo{year}{1986}), \bibinfo{pages}{579}.
\newblock


\bibitem[Schlesinger et~al\mbox{.}(2017)]%
        {schlesinger2017intersectional}
\bibfield{author}{\bibinfo{person}{Ari Schlesinger}, \bibinfo{person}{W~Keith Edwards}, {and} \bibinfo{person}{Rebecca~E Grinter}.} \bibinfo{year}{2017}\natexlab{}.
\newblock \showarticletitle{Intersectional HCI: Engaging identity through gender, race, and class}. In \bibinfo{booktitle}{\emph{Proceedings of the 2017 CHI conference on human factors in computing systems}}. \bibinfo{pages}{5412--5427}.
\newblock


\bibitem[Schroeder and Gefenas(2009)]%
        {schroeder2009vulnerability}
\bibfield{author}{\bibinfo{person}{Doris Schroeder} {and} \bibinfo{person}{Eugenijus Gefenas}.} \bibinfo{year}{2009}\natexlab{}.
\newblock \showarticletitle{Vulnerability: too vague and too broad?}
\newblock \bibinfo{journal}{\emph{Cambridge Quarterly of Healthcare Ethics}} \bibinfo{volume}{18}, \bibinfo{number}{2} (\bibinfo{year}{2009}), \bibinfo{pages}{113--121}.
\newblock


\bibitem[Schyns et~al\mbox{.}(2009)]%
        {Schyns2009}
\bibfield{author}{\bibinfo{person}{Philippe~G. Schyns}, \bibinfo{person}{Lucy~S. Petro}, {and} \bibinfo{person}{Marie~L. Smith}.} \bibinfo{year}{2009}\natexlab{}.
\newblock \showarticletitle{Transmission of Facial Expressions of Emotion Co-Evolved with Their Efficient Decoding in the Brain: Behavioral and Brain Evidence}.
\newblock \bibinfo{journal}{\emph{PLOS ONE}} \bibinfo{volume}{4}, \bibinfo{number}{5} (\bibinfo{date}{05} \bibinfo{year}{2009}), \bibinfo{pages}{1--16}.
\newblock
\urldef\tempurl%
\url{https://doi.org/10.1371/journal.pone.0005625}
\showDOI{\tempurl}


\bibitem[Seaman(1999)]%
        {Seaman1999}
\bibfield{author}{\bibinfo{person}{C.~B. Seaman}.} \bibinfo{year}{1999}\natexlab{}.
\newblock \showarticletitle{"Qualitative Methods in Empirical Studies of Software Engineering"}. In \bibinfo{booktitle}{\emph{IEEE Transactions on Software Engineering}}, Vol.~\bibinfo{volume}{25}. \bibinfo{pages}{557--572}.
\newblock


\bibitem[Seo and Kim(2016)]%
        {Seo2016}
\bibfield{author}{\bibinfo{person}{Young-Ho Seo} {and} \bibinfo{person}{Jong-Hoon Kim}.} \bibinfo{year}{2016}\natexlab{}.
\newblock \showarticletitle{Analyzing the Effects of Coding Education through Pair Programming for the Computational Thinking and Creativity of Elementary School Students}.
\newblock \bibinfo{journal}{\emph{Indian Journal of Science and Technology}}  \bibinfo{volume}{9} (\bibinfo{date}{12} \bibinfo{year}{2016}).
\newblock
\urldef\tempurl%
\url{https://doi.org/10.17485/ijst/2016/v9i46/107837}
\showDOI{\tempurl}


\bibitem[Sesko and Biernat(2010)]%
        {SESKO2010356}
\bibfield{author}{\bibinfo{person}{Amanda~K. Sesko} {and} \bibinfo{person}{Monica Biernat}.} \bibinfo{year}{2010}\natexlab{}.
\newblock \showarticletitle{Prototypes of race and gender: The invisibility of Black women}.
\newblock \bibinfo{journal}{\emph{Journal of Experimental Social Psychology}} \bibinfo{volume}{46}, \bibinfo{number}{2} (\bibinfo{year}{2010}), \bibinfo{pages}{356--360}.
\newblock
\showISSN{0022-1031}
\urldef\tempurl%
\url{https://doi.org/10.1016/j.jesp.2009.10.016}
\showDOI{\tempurl}


\bibitem[Shelton(2003)]%
        {shelton2003interpersonal}
\bibfield{author}{\bibinfo{person}{J~Nicole Shelton}.} \bibinfo{year}{2003}\natexlab{}.
\newblock \showarticletitle{Interpersonal concerns in social encounters between majority and minority group members}.
\newblock \bibinfo{journal}{\emph{Group Processes \& Intergroup Relations}} \bibinfo{volume}{6}, \bibinfo{number}{2} (\bibinfo{year}{2003}), \bibinfo{pages}{171--185}.
\newblock


\bibitem[Shelton and Richeson(2006)]%
        {shelton2006interracial}
\bibfield{author}{\bibinfo{person}{J~Nicole Shelton} {and} \bibinfo{person}{Jennifer~A Richeson}.} \bibinfo{year}{2006}\natexlab{}.
\newblock \showarticletitle{Interracial interactions: A relational approach}.
\newblock \bibinfo{journal}{\emph{Advances in experimental social psychology}}  \bibinfo{volume}{38} (\bibinfo{year}{2006}), \bibinfo{pages}{121--181}.
\newblock


\bibitem[Sices et~al\mbox{.}(2009)]%
        {sices2009sugar}
\bibfield{author}{\bibinfo{person}{Laura Sices}, \bibinfo{person}{Lucia Egbert}, {and} \bibinfo{person}{Mary~Beth Mercer}.} \bibinfo{year}{2009}\natexlab{}.
\newblock \showarticletitle{Sugar-coaters and straight talkers: communicating about developmental delays in primary care}.
\newblock \bibinfo{journal}{\emph{Pediatrics}} \bibinfo{volume}{124}, \bibinfo{number}{4} (\bibinfo{year}{2009}), \bibinfo{pages}{e705--e713}.
\newblock


\bibitem[Simon and Hanks(2008)]%
        {simon2008first}
\bibfield{author}{\bibinfo{person}{Beth Simon} {and} \bibinfo{person}{Brian Hanks}.} \bibinfo{year}{2008}\natexlab{}.
\newblock \showarticletitle{First-year students' impressions of pair programming in CS1}.
\newblock \bibinfo{journal}{\emph{Journal on Educational Resources in Computing (JERIC)}} \bibinfo{volume}{7}, \bibinfo{number}{4} (\bibinfo{year}{2008}), \bibinfo{pages}{1--28}.
\newblock


\bibitem[Smith et~al\mbox{.}(2020)]%
        {smith2020s}
\bibfield{author}{\bibinfo{person}{Angela~DR Smith}, \bibinfo{person}{Alex~A Ahmed}, \bibinfo{person}{Adriana Alvarado~Garcia}, \bibinfo{person}{Bryan Dosono}, \bibinfo{person}{Ihudiya Ogbonnaya-Ogburu}, \bibinfo{person}{Yolanda Rankin}, \bibinfo{person}{Alexandra To}, {and} \bibinfo{person}{Kentaro Toyama}.} \bibinfo{year}{2020}\natexlab{}.
\newblock \showarticletitle{What's race got to do with it? Engaging in race in HCI}. In \bibinfo{booktitle}{\emph{Extended abstracts of the 2020 CHI conference on human factors in computing systems}}. \bibinfo{pages}{1--8}.
\newblock


\bibitem[Smith and Schonfeld(2000)]%
        {smith2000benefits}
\bibfield{author}{\bibinfo{person}{Daryl~G Smith} {and} \bibinfo{person}{Natalie~B Schonfeld}.} \bibinfo{year}{2000}\natexlab{}.
\newblock \showarticletitle{The benefits of diversity what the research tells us}.
\newblock \bibinfo{journal}{\emph{About campus}} \bibinfo{volume}{5}, \bibinfo{number}{5} (\bibinfo{year}{2000}), \bibinfo{pages}{16--23}.
\newblock


\bibitem[Sol{\'o}rzano and Yosso(2002)]%
        {solorzano2002critical}
\bibfield{author}{\bibinfo{person}{Daniel~G Sol{\'o}rzano} {and} \bibinfo{person}{Tara~J Yosso}.} \bibinfo{year}{2002}\natexlab{}.
\newblock \showarticletitle{Critical race methodology: Counter-storytelling as an analytical framework for education research}.
\newblock \bibinfo{journal}{\emph{Qualitative inquiry}} \bibinfo{volume}{8}, \bibinfo{number}{1} (\bibinfo{year}{2002}), \bibinfo{pages}{23--44}.
\newblock


\bibitem[Solyst et~al\mbox{.}(2023)]%
        {solyst2023would}
\bibfield{author}{\bibinfo{person}{Jaemarie Solyst}, \bibinfo{person}{Shixian Xie}, \bibinfo{person}{Ellia Yang}, \bibinfo{person}{Angela~EB Stewart}, \bibinfo{person}{Motahhare Eslami}, \bibinfo{person}{Jessica Hammer}, {and} \bibinfo{person}{Amy Ogan}.} \bibinfo{year}{2023}\natexlab{}.
\newblock \showarticletitle{“I Would Like to Design”: Black Girls Analyzing and Ideating Fair and Accountable AI}. In \bibinfo{booktitle}{\emph{Proceedings of the 2023 CHI Conference on Human Factors in Computing Systems}}. \bibinfo{pages}{1--14}.
\newblock


\bibitem[Sommers(2006)]%
        {sommers2006racial}
\bibfield{author}{\bibinfo{person}{Samuel~R Sommers}.} \bibinfo{year}{2006}\natexlab{}.
\newblock \showarticletitle{On racial diversity and group decision making: identifying multiple effects of racial composition on jury deliberations.}
\newblock \bibinfo{journal}{\emph{Journal of personality and social psychology}} \bibinfo{volume}{90}, \bibinfo{number}{4} (\bibinfo{year}{2006}), \bibinfo{pages}{597}.
\newblock


\bibitem[Sommerville(2010)]%
        {Sommerville10}
\bibfield{author}{\bibinfo{person}{Ian Sommerville}.} \bibinfo{year}{2010}\natexlab{}.
\newblock \bibinfo{booktitle}{\emph{Software Engineering} (\bibinfo{edition}{9} ed.)}.
\newblock \bibinfo{publisher}{Addison-Wesley}, \bibinfo{address}{Harlow, England}.
\newblock
\showISBNx{978-0-13-703515-1}


\bibitem[Spencer et~al\mbox{.}(1999)]%
        {spencer1999stereotype}
\bibfield{author}{\bibinfo{person}{Steven~J Spencer}, \bibinfo{person}{Claude~M Steele}, {and} \bibinfo{person}{Diane~M Quinn}.} \bibinfo{year}{1999}\natexlab{}.
\newblock \showarticletitle{Stereotype threat and women's math performance}.
\newblock \bibinfo{journal}{\emph{Journal of experimental social psychology}} \bibinfo{volume}{35}, \bibinfo{number}{1} (\bibinfo{year}{1999}), \bibinfo{pages}{4--28}.
\newblock


\bibitem[Spiers et~al\mbox{.}(2017)]%
        {Spiers2017_Prejudice}
\bibfield{author}{\bibinfo{person}{Hugo~J. Spiers}, \bibinfo{person}{Bradley~C. Love}, \bibinfo{person}{Mike~E. Le~Pelley}, \bibinfo{person}{Charlotte~E. Gibb}, {and} \bibinfo{person}{Robin~A. Murphy}.} \bibinfo{year}{2017}\natexlab{}.
\newblock \showarticletitle{{Anterior Temporal Lobe Tracks the Formation of Prejudice}}.
\newblock \bibinfo{journal}{\emph{Journal of Cognitive Neuroscience}} \bibinfo{volume}{29}, \bibinfo{number}{3} (\bibinfo{date}{03} \bibinfo{year}{2017}), \bibinfo{pages}{530--544}.
\newblock
\showISSN{0898-929X}
\urldef\tempurl%
\url{https://doi.org/10.1162/jocn_a_01056}
\showDOI{\tempurl}
\showeprint{https://direct.mit.edu/jocn/article-pdf/29/3/530/1952357/jocn\_a\_01056.pdf}


\bibitem[Squire(2008)]%
        {squire2008video}
\bibfield{author}{\bibinfo{person}{Kurt~D Squire}.} \bibinfo{year}{2008}\natexlab{}.
\newblock \showarticletitle{Video games and education: Designing learning systems for an interactive age}.
\newblock \bibinfo{journal}{\emph{Educational technology}} (\bibinfo{year}{2008}), \bibinfo{pages}{17--26}.
\newblock


\bibitem[Steele(1997)]%
        {steele1997threat}
\bibfield{author}{\bibinfo{person}{Claude~M Steele}.} \bibinfo{year}{1997}\natexlab{}.
\newblock \showarticletitle{A threat in the air: How stereotypes shape intellectual identity and performance.}
\newblock \bibinfo{journal}{\emph{American psychologist}} \bibinfo{volume}{52}, \bibinfo{number}{6} (\bibinfo{year}{1997}), \bibinfo{pages}{613}.
\newblock


\bibitem[Steele(2011)]%
        {steele2011whistling}
\bibfield{author}{\bibinfo{person}{Claude~M Steele}.} \bibinfo{year}{2011}\natexlab{}.
\newblock \bibinfo{booktitle}{\emph{Whistling Vivaldi: How stereotypes affect us and what we can do}}.
\newblock \bibinfo{publisher}{WW Norton \& Company}.
\newblock


\bibitem[Steele et~al\mbox{.}(2002)]%
        {steele2002contending}
\bibfield{author}{\bibinfo{person}{Claude~M Steele}, \bibinfo{person}{Steven~J Spencer}, {and} \bibinfo{person}{Joshua Aronson}.} \bibinfo{year}{2002}\natexlab{}.
\newblock \showarticletitle{Contending with group image: The psychology of stereotype and social identity threat}.
\newblock In \bibinfo{booktitle}{\emph{Advances in experimental social psychology}}. Vol.~\bibinfo{volume}{34}. \bibinfo{publisher}{Elsevier}, \bibinfo{pages}{379--440}.
\newblock


\bibitem[Stevens and Levi(2023)]%
        {stevens2023introduction}
\bibfield{author}{\bibinfo{person}{Dannelle~D Stevens} {and} \bibinfo{person}{Antonia~J Levi}.} \bibinfo{year}{2023}\natexlab{}.
\newblock \bibinfo{booktitle}{\emph{Introduction to rubrics: An assessment tool to save grading time, convey effective feedback, and promote student learning}}.
\newblock \bibinfo{publisher}{Routledge}.
\newblock


\bibitem[Sun et~al\mbox{.}(2015)]%
        {sun2015effectiveness}
\bibfield{author}{\bibinfo{person}{Wenying Sun}, \bibinfo{person}{George Marakas}, {and} \bibinfo{person}{Miguel Aguirre-Urreta}.} \bibinfo{year}{2015}\natexlab{}.
\newblock \showarticletitle{The effectiveness of pair programming: Software professionals' perceptions}.
\newblock \bibinfo{journal}{\emph{IEEE Software}} \bibinfo{volume}{33}, \bibinfo{number}{4} (\bibinfo{year}{2015}), \bibinfo{pages}{72--79}.
\newblock


\bibitem[Tee et~al\mbox{.}(2009)]%
        {tee2009exploring}
\bibfield{author}{\bibinfo{person}{Kimberly Tee}, \bibinfo{person}{AJ~Bernheim Brush}, {and} \bibinfo{person}{Kori~M Inkpen}.} \bibinfo{year}{2009}\natexlab{}.
\newblock \showarticletitle{Exploring communication and sharing between extended families}.
\newblock \bibinfo{journal}{\emph{International Journal of Human-Computer Studies}} \bibinfo{volume}{67}, \bibinfo{number}{2} (\bibinfo{year}{2009}), \bibinfo{pages}{128--138}.
\newblock


\bibitem[Thomas et~al\mbox{.}(2003)]%
        {thomas2003code}
\bibfield{author}{\bibinfo{person}{Lynda Thomas}, \bibinfo{person}{Mark Ratcliffe}, {and} \bibinfo{person}{Ann Robertson}.} \bibinfo{year}{2003}\natexlab{}.
\newblock \showarticletitle{Code warriors and code-a-phobes: a study in attitude and pair programming}.
\newblock \bibinfo{journal}{\emph{ACM SIGCSE Bulletin}} \bibinfo{volume}{35}, \bibinfo{number}{1} (\bibinfo{year}{2003}), \bibinfo{pages}{363--367}.
\newblock


\bibitem[Thompson(2006)]%
        {thompson2006development}
\bibfield{author}{\bibinfo{person}{Ross~A Thompson}.} \bibinfo{year}{2006}\natexlab{}.
\newblock \showarticletitle{The development of the person: social understanding, relationships, conscience, self.}
\newblock  (\bibinfo{year}{2006}).
\newblock


\bibitem[Thériault et~al\mbox{.}(2021)]%
        {theriault2021bodyswapping}
\bibfield{author}{\bibinfo{person}{Rachel Thériault}, \bibinfo{person}{Jay~A. Olson}, \bibinfo{person}{Sabrina~A. Krol}, {and} \bibinfo{person}{Amir Raz}.} \bibinfo{year}{2021}\natexlab{}.
\newblock \showarticletitle{Body Swapping with a Black Person Boosts Empathy: Using Virtual Reality to Embody Another}.
\newblock \bibinfo{journal}{\emph{Quarterly Journal of Experimental Psychology (2006)}} \bibinfo{volume}{74}, \bibinfo{number}{12} (\bibinfo{year}{2021}), \bibinfo{pages}{2057--2074}.
\newblock
\urldef\tempurl%
\url{https://doi.org/10.1177/17470218211024826}
\showDOI{\tempurl}


\bibitem[Tsai et~al\mbox{.}(2019)]%
        {tsai2019developing}
\bibfield{author}{\bibinfo{person}{Meng-Jung Tsai}, \bibinfo{person}{Ching-Yeh Wang}, {and} \bibinfo{person}{Po-Fen Hsu}.} \bibinfo{year}{2019}\natexlab{}.
\newblock \showarticletitle{Developing the computer programming self-efficacy scale for computer literacy education}.
\newblock \bibinfo{journal}{\emph{Journal of Educational Computing Research}} \bibinfo{volume}{56}, \bibinfo{number}{8} (\bibinfo{year}{2019}), \bibinfo{pages}{1345--1360}.
\newblock


\bibitem[Tsompanoudi et~al\mbox{.}(2019)]%
        {Tsompanoudi2019}
\bibfield{author}{\bibinfo{person}{Despina Tsompanoudi}, \bibinfo{person}{Maya Satratzemi}, \bibinfo{person}{Stelios Xinogalos}, {and} \bibinfo{person}{Leonidas Karamitopoulos}.} \bibinfo{year}{2019}\natexlab{}.
\newblock \bibinfo{title}{An Empirical Study on Factors related to Distributed Pair Programming}.  (\bibinfo{date}{04} \bibinfo{year}{2019}).
\newblock
\urldef\tempurl%
\url{https://www.learntechlib.org/p/208576}
\showURL{%
\tempurl}


\bibitem[VanDeGrift(2004)]%
        {vandegrift2004coupling}
\bibfield{author}{\bibinfo{person}{Tammy VanDeGrift}.} \bibinfo{year}{2004}\natexlab{}.
\newblock \showarticletitle{Coupling pair programming and writing: learning about students' perceptions and processes}. In \bibinfo{booktitle}{\emph{Proceedings of the 35th SIGCSE technical symposium on Computer science education}}. \bibinfo{pages}{2--6}.
\newblock


\bibitem[Voss et~al\mbox{.}(2011)]%
        {voss2011hippocampal}
\bibfield{author}{\bibinfo{person}{Joel~L Voss}, \bibinfo{person}{Brian~D Gonsalves}, \bibinfo{person}{Kara~D Federmeier}, \bibinfo{person}{Daniel Tranel}, {and} \bibinfo{person}{Neal~J Cohen}.} \bibinfo{year}{2011}\natexlab{}.
\newblock \showarticletitle{Hippocampal brain-network coordination during volitional exploratory behavior enhances learning}.
\newblock \bibinfo{journal}{\emph{Nature neuroscience}} \bibinfo{volume}{14}, \bibinfo{number}{1} (\bibinfo{year}{2011}), \bibinfo{pages}{115--120}.
\newblock


\bibitem[Wagner and Compton(2012)]%
        {wagner2012}
\bibfield{author}{\bibinfo{person}{Tony Wagner} {and} \bibinfo{person}{Robert~A Compton}.} \bibinfo{year}{2012}\natexlab{}.
\newblock \bibinfo{booktitle}{\emph{Creating innovators: The making of young people who will change the world}}.
\newblock \bibinfo{publisher}{Simon and Schuster}.
\newblock


\bibitem[Walton and Cohen(2007)]%
        {walton2007question}
\bibfield{author}{\bibinfo{person}{Gregory~M Walton} {and} \bibinfo{person}{Geoffrey~L Cohen}.} \bibinfo{year}{2007}\natexlab{}.
\newblock \showarticletitle{A question of belonging: race, social fit, and achievement.}
\newblock \bibinfo{journal}{\emph{Journal of personality and social psychology}} \bibinfo{volume}{92}, \bibinfo{number}{1} (\bibinfo{year}{2007}), \bibinfo{pages}{82}.
\newblock


\bibitem[Washington(2020)]%
        {washington2020twice}
\bibfield{author}{\bibinfo{person}{Alicia~Nicki Washington}.} \bibinfo{year}{2020}\natexlab{}.
\newblock \showarticletitle{When twice as good isn't enough: The case for cultural competence in computing}. In \bibinfo{booktitle}{\emph{Proceedings of the 51st ACM technical symposium on computer science education}}. \bibinfo{pages}{213--219}.
\newblock


\bibitem[Wellman(1993)]%
        {wellman1993portraits}
\bibfield{author}{\bibinfo{person}{David~T Wellman}.} \bibinfo{year}{1993}\natexlab{}.
\newblock \bibinfo{booktitle}{\emph{Portraits of white racism}}.
\newblock \bibinfo{publisher}{Cambridge University Press}.
\newblock


\bibitem[Werner and Denning(2009)]%
        {werner2009pair}
\bibfield{author}{\bibinfo{person}{Linda Werner} {and} \bibinfo{person}{Jill Denning}.} \bibinfo{year}{2009}\natexlab{}.
\newblock \showarticletitle{Pair programming in middle school: What does it look like?}
\newblock \bibinfo{journal}{\emph{Journal of Research on Technology in Education}} \bibinfo{volume}{42}, \bibinfo{number}{1} (\bibinfo{year}{2009}), \bibinfo{pages}{29--49}.
\newblock


\bibitem[Werner et~al\mbox{.}(2004a)]%
        {Werner2004}
\bibfield{author}{\bibinfo{person}{Linda~L. Werner}, \bibinfo{person}{Brian Hanks}, {and} \bibinfo{person}{Charlie McDowell}.} \bibinfo{year}{2004}\natexlab{a}.
\newblock \showarticletitle{Pair-Programming Helps Female Computer Science Students}.
\newblock \bibinfo{journal}{\emph{J. Educ. Resour. Comput.}} \bibinfo{volume}{4}, \bibinfo{number}{1} (\bibinfo{date}{March} \bibinfo{year}{2004}), \bibinfo{pages}{4–es}.
\newblock
\showISSN{1531-4278}
\urldef\tempurl%
\url{https://doi.org/10.1145/1060071.1060075}
\showDOI{\tempurl}


\bibitem[Werner et~al\mbox{.}(2004b)]%
        {werner2004pair}
\bibfield{author}{\bibinfo{person}{Linda~L Werner}, \bibinfo{person}{Brian Hanks}, {and} \bibinfo{person}{Charlie McDowell}.} \bibinfo{year}{2004}\natexlab{b}.
\newblock \showarticletitle{Pair-programming helps female computer science students}.
\newblock \bibinfo{journal}{\emph{Journal on Educational Resources in Computing (JERIC)}} \bibinfo{volume}{4}, \bibinfo{number}{1} (\bibinfo{year}{2004}), \bibinfo{pages}{4--es}.
\newblock


\bibitem[Williams and Kessler(2002)]%
        {Williams2002a}
\bibfield{author}{\bibinfo{person}{Laurie Williams} {and} \bibinfo{person}{Robert Kessler}.} \bibinfo{year}{2002}\natexlab{}.
\newblock \bibinfo{booktitle}{\emph{Pair Programming Illuminated}}.
\newblock \bibinfo{publisher}{Addison-Wesley Longman Publishing Co., Inc.}, \bibinfo{address}{USA}.
\newblock
\showISBNx{0201745763}


\bibitem[Williams et~al\mbox{.}(2002a)]%
        {Williams2002b}
\bibfield{author}{\bibinfo{person}{Laurie Williams}, \bibinfo{person}{Eric Wiebe}, \bibinfo{person}{Kai Yang}, \bibinfo{person}{Miriam Ferzli}, {and} \bibinfo{person}{Carol Miller}.} \bibinfo{year}{2002}\natexlab{a}.
\newblock \showarticletitle{In Support of Pair Programming in the Introductory Computer Science Course}.
\newblock \bibinfo{journal}{\emph{Computer Science Education}} \bibinfo{volume}{12}, \bibinfo{number}{3} (\bibinfo{year}{2002}), \bibinfo{pages}{197--212}.
\newblock
\urldef\tempurl%
\url{https://doi.org/10.1076/csed.12.3.197.8618}
\showDOI{\tempurl}


\bibitem[Williams et~al\mbox{.}(2002b)]%
        {williams2002support}
\bibfield{author}{\bibinfo{person}{Laurie Williams}, \bibinfo{person}{Eric Wiebe}, \bibinfo{person}{Kai Yang}, \bibinfo{person}{Miriam Ferzli}, {and} \bibinfo{person}{Carol Miller}.} \bibinfo{year}{2002}\natexlab{b}.
\newblock \showarticletitle{In support of pair programming in the introductory computer science course}.
\newblock \bibinfo{journal}{\emph{Computer Science Education}} \bibinfo{volume}{12}, \bibinfo{number}{3} (\bibinfo{year}{2002}), \bibinfo{pages}{197--212}.
\newblock


\bibitem[Williams and Kessler(2000)]%
        {Williams2000a}
\bibfield{author}{\bibinfo{person}{Laurie~A. Williams} {and} \bibinfo{person}{Robert~R. Kessler}.} \bibinfo{year}{2000}\natexlab{}.
\newblock \showarticletitle{All I Really Need to Know about Pair Programming I Learned in Kindergarten}.
\newblock \bibinfo{journal}{\emph{Commun. ACM}} \bibinfo{volume}{43}, \bibinfo{number}{5} (\bibinfo{date}{May} \bibinfo{year}{2000}), \bibinfo{pages}{108–114}.
\newblock
\showISSN{0001-0782}
\urldef\tempurl%
\url{https://doi.org/10.1145/332833.332848}
\showDOI{\tempurl}


\bibitem[Wood(2011)]%
        {wood2011lynching}
\bibfield{author}{\bibinfo{person}{Amy~Louise Wood}.} \bibinfo{year}{2011}\natexlab{}.
\newblock \bibinfo{booktitle}{\emph{Lynching and spectacle: Witnessing racial violence in America, 1890-1940}}.
\newblock \bibinfo{publisher}{Univ of North Carolina Press}.
\newblock


\bibitem[Wray(2009)]%
        {wray2009pair}
\bibfield{author}{\bibinfo{person}{Stuart Wray}.} \bibinfo{year}{2009}\natexlab{}.
\newblock \showarticletitle{How pair programming really works}.
\newblock \bibinfo{journal}{\emph{IEEE software}} \bibinfo{volume}{27}, \bibinfo{number}{1} (\bibinfo{year}{2009}), \bibinfo{pages}{50--55}.
\newblock


\bibitem[Yanow(Year)]%
        {yanow}
\bibfield{author}{\bibinfo{person}{Dvora Yanow}.} \bibinfo{year}{Year}\natexlab{}.
\newblock \bibinfo{booktitle}{\emph{Constructing Race and Ethnicity in America: Category-making in Public Policy and Administration} (\bibinfo{edition}{1st} ed.)}.
\newblock \bibinfo{publisher}{Taylor \& Francis}.
\newblock


\bibitem[Ying et~al\mbox{.}(2019)]%
        {ying2019their}
\bibfield{author}{\bibinfo{person}{Kimberly~Michelle Ying}, \bibinfo{person}{Lydia~G Pezzullo}, \bibinfo{person}{Mohona Ahmed}, \bibinfo{person}{Kassandra Crompton}, \bibinfo{person}{Jeremiah Blanchard}, {and} \bibinfo{person}{Kristy~Elizabeth Boyer}.} \bibinfo{year}{2019}\natexlab{}.
\newblock \showarticletitle{In their own words: Gender differences in student perceptions of pair programming}. In \bibinfo{booktitle}{\emph{Proceedings of the 50th ACM Technical Symposium on Computer Science Education}}. \bibinfo{pages}{1053--1059}.
\newblock


\bibitem[Yukselturk and Altiok(2017)]%
        {yukselturk2017investigation}
\bibfield{author}{\bibinfo{person}{Erman Yukselturk} {and} \bibinfo{person}{Serhat Altiok}.} \bibinfo{year}{2017}\natexlab{}.
\newblock \showarticletitle{An investigation of the effects of programming with Scratch on the preservice IT teachers’ self-efficacy perceptions and attitudes towards computer programming}.
\newblock \bibinfo{journal}{\emph{British Journal of Educational Technology}} \bibinfo{volume}{48}, \bibinfo{number}{3} (\bibinfo{year}{2017}), \bibinfo{pages}{789--801}.
\newblock


\bibitem[Zhao(2012)]%
        {zhao2012}
\bibfield{author}{\bibinfo{person}{Yong Zhao}.} \bibinfo{year}{2012}\natexlab{}.
\newblock \bibinfo{booktitle}{\emph{World class learners: Educating creative and entrepreneurial students}}.
\newblock \bibinfo{publisher}{Corwin Press}.
\newblock


\bibitem[Zhong et~al\mbox{.}(2016)]%
        {zhong2016impact}
\bibfield{author}{\bibinfo{person}{Baichang Zhong}, \bibinfo{person}{Qiyun Wang}, {and} \bibinfo{person}{Jie Chen}.} \bibinfo{year}{2016}\natexlab{}.
\newblock \showarticletitle{The impact of social factors on pair programming in a primary school}.
\newblock \bibinfo{journal}{\emph{Computers in Human Behavior}}  \bibinfo{volume}{64} (\bibinfo{year}{2016}), \bibinfo{pages}{423--431}.
\newblock


\bibitem[Zhong et~al\mbox{.}(2017)]%
        {zhong2017investigating}
\bibfield{author}{\bibinfo{person}{Baichang Zhong}, \bibinfo{person}{Qiyun Wang}, \bibinfo{person}{Jie Chen}, {and} \bibinfo{person}{Yi Li}.} \bibinfo{year}{2017}\natexlab{}.
\newblock \showarticletitle{Investigating the period of switching roles in pair programming in a primary school}.
\newblock \bibinfo{journal}{\emph{Journal of Educational Technology \& Society}} \bibinfo{volume}{20}, \bibinfo{number}{3} (\bibinfo{year}{2017}), \bibinfo{pages}{220--233}.
\newblock


\bibitem[Zieris and Prechelt(2020)]%
        {Zieris2020}
\bibfield{author}{\bibinfo{person}{Franz Zieris} {and} \bibinfo{person}{Lutz Prechelt}.} \bibinfo{year}{2020}\natexlab{}.
\newblock \showarticletitle{Explaining pair programming session dynamics from knowledge gaps}. In \bibinfo{booktitle}{\emph{Proc. 42nd Int’l. Conf. on Software Engineering (ICSE’20)}}.
\newblock


\end{thebibliography}
